\documentclass{article}
  \usepackage[margin=1in]{geometry}
  \usepackage[utf8]{inputenc}
  \usepackage{amsfonts}
  \usepackage{amssymb}
  \usepackage{amsmath}
  \usepackage{mathrsfs}
  \usepackage[dvipsnames]{xcolor}
  \usepackage{dirtytalk}
  \usepackage{physics}
  \usepackage{rotating}
  \usepackage{multicol}
  \usepackage{setspace}
  \usepackage{hyperref}
  \usepackage{cleveref}
  \crefname{section}{§}{§§}
  \Crefname{section}{§}{§§}
  \usepackage{graphicx} % Required for inserting images
  \DeclareFontFamily{U}{mathx}{}
  \DeclareFontShape{U}{mathx}{m}{n}{<-> mathx10}{}
  \DeclareSymbolFont{mathx}{U}{mathx}{m}{n}
  \DeclareMathAccent{\widehat}{0}{mathx}{"70}
  \DeclareMathAccent{\widecheck}{0}{mathx}{"71}
  \newenvironment{acknowledgements}{\vspace{2em}}{}

  \newcommand{\uppi}{\pi}
  
  \newcommand{\Real}{\mathrm{Re}} % \Real: Real part of
  \newcommand{\Imag}{\mathrm{Im}} % \Imag: Imaginary part of
  
  \DeclareMathAlphabet\mathsfbi{OT1}{cmss}{m}{sl}
  \newcommand{\doi}[1]{doi:~\href{https://doi.org/#1}{#1}}
  
  \newcommand{\mat}[1]{\mathsf{#1}} % \mat: Matrices
  \newcommand{\gmat}[1]{\mathsf{#1}} % \gmat: Greek matrices
  \newcommand{\pvect}[1]{\mathbf{#1}} % \pvect: Physical ectors
  \newcommand{\mvect}[1]{\boldsymbol{#1}} % \mvect: Modal vectors

  \title{\vspace{-0.5in}\textbf{The Weakly-Nonlinear Admittance at Open Ends of Two- and Three-Dimensional Acoustic Waveguides}}
  \author{Freddie Jensen$^1$, Harry Turnbull$^1$ and Edward James Brambley$^{1,2}$\footnote{Corresponding author: \href{mailto:E.J.Brambley@warwick.ac.uk}{E.J.Brambley@warwick.ac.uk}}\\[1ex]
  $^1$~Mathematics Institute, University of Warwick, Coventry CV4 7AL, UK\\
  $^2$~WMG, University of Warwick, Coventry CV4 7AL, UK}
\date{October 12, 2026\\[1em]\small{Published in the Journal of Fluid Mechanics \textbf{1044}, A45 (2026), \doi{10.1017/jfm.2026.12139}}}
  
\usepackage{hyperref}
\usepackage{subcaption}
\usepackage{natbib}
\usepackage{mathtools}
\hypersetup{
    colorlinks = true,
    urlcolor   = blue,
    citecolor  = blue,
    linkcolor   = blue,}

\usepackage{textcomp,gensymb}

\def\clap#1{\hbox to 0pt{\hss#1\hss}}

\def\mathclap{\mathpalette\mathclapinternal}

\def\mathclapinternal#1#2{%
\clap{$\mathsurround=0pt#1{#2}$}}

\newcommand{\I}{\mathrm{i}} % \I: $\sqrt{-1}$
\newcommand{\e}{\mathrm{e}} % \e: $\exp{1}$
\newcommand{\J}{\mathrm{J}} % \J: Bessel's function $J_m(z)$
\newcommand{\Ib}{\mathrm{I}} % \Ib: Bessel's function $I_m(z)$
\newcommand{\K}{\mathrm{K}} % \K: Bessel's function $K_m(z)$
\newcommand{\intd}{\mathrm{d}} % \intd : Roman d for integrals, eg. \int\ldots\,\intd \theta
\newcommand{\T}{\mathrm{T}} % \T: Transpose

\newcommand{\sgn}{\mathrm{sgn}}
\newcommand{\res}{\mathrm{res}}

\begin{document}
\maketitle

\begin{abstract}
We formulate a weakly-nonlinear exit condition for open ends of acoustic waveguides; to our knowledge this is the first time a general acoustic open end has been analysed outside the linear regime. The formulation neglects mean flow, vortex separation, and viscous effects.  The resulting admittance boundary condition, and its weakly-nonlinear counterpart, extend recent weakly-nonlinear modelling of curved ducts~(\citealp{jensenbrambley2026} J.~Fluid Mech.\ 1030 A19) to include open ends. We approximate free space by considering the open-ended duct to be enclosed within a much larger hard-walled concentric duct; within the larger duct, the smaller duct exit is modelled as an acoustic discontinuity.  Importantly, the superposition principle is unneeded, allowing the model to be applied in the nonlinear regime. The exit condition can be calculated without needing to solve the full problem in either the outer or inner ducts, making it numerically efficient. The exit condition is validated in the linear regime by comparison to Wiener--Hopf solutions of the duct end correction, and a novel nonlinear end correction is proposed; we find that both non-plane waves and nonlinearity cause the end correction to vary significantly from Rayleigh's classical $0.6$ radii result.  A number of numerical illustrations are then discussed, demonstrating nonlinear effects, sound radiating from curved ducts, sound radiating from an exponential horn (representative of brass instrument bells), and the harmonic effects of the open end on in-duct resonances. The model has potential applications to sound in woodwind and brass instruments. \textsc{Matlab} source code is provided in the supplementary material.
\end{abstract}

\section{Introduction}
The duct exit is an important component of a model of duct acoustics. An accurate picture of wave reflection and transmission at a duct outlet will inform modelling of both the duct's interior acoustics and its exterior radiative properties. For example, the fundamental resonance of a cylindrical duct closed at one end and open at the other is a quarter wavelength, with, at the closed end, a velocity node giving a pressure anti-node, and a pressure node located a short distance beyond the open end; this distance, known as the \emph{end-correction}, was postulated by \citet{rayleigh1871} to be about $0.6R$, where $R$ is the duct radius.  Subsequently, \citet{levine1948radiation} made great progress in using the complex-analytic Wiener--Hopf technique to calculate the acoustic radiation from an unflanged cylindrical pipe, enabling them to calculate an end correction of $0.6133R$. Subsequent work has attempted to extend their low-frequency calculation to higher spatial modes and frequencies \citep[e.g.][]{weinstein1948rigorous,snakowska2011}; however, all analytical work to date in this area has been \emph{linear}.

Linearity is a reasonable assumption for many applications of exit-condition modelling, e.g. wind instruments, certain types of organ pipe, and industrial exhaust ventilation. However, it has been experimentally proven that shock formation occurs in brass instruments \citep{hirschberg,pandya2003}, and in fact these instruments are understood to sound `brassy' due to the nonlinearity of their internal acoustics \citep{arnold2012}. Analytical models of this phenomenon support this condition \citep{gilbert2008,fernando} but do not include consideration of the duct's open end. Thus, a model combining nonlinearity with the open end has both novelty and an immediate application.

Another weakness of the preceding analytical work is its inability to capture the effects of brass instruments' complex geometries on the shocks they produce. Similarly to the exit condition, modelling acoustics in complex geometries has largely been restricted to the linear case \citep{felix1,felix2}.  Recently, however, \citet{mctavish+brambley-2019} combined weak nonlinearity with complex duct geometry in two dimensions, and this has more recently been expanded to three dimensions~\citep{jensenbrambley2026}, allowing the interaction between nonlinearity and duct curvature to be explored.  Both of these studies, however, ignored the duct outlet by assuming a totally absorbing \emph{characteristic admittance} as an outlet condition: this is essentially a mathematical contrivance, and corresponds physically to the duct outlet continuing in an infinite straight line with internal forward propagation only.  This lack of a model of the open duct end is the hole that the exit condition proposed here is intended to fill.

Although the viscosity of air is very small, viscosity can still play an important role in strongly-nonlinear acoustics in ducts.  Viscosity leads to both linear wall friction losses~\citep[e.g.][]{zinn1970,rendon2010} and nonlinear energy loss through vortex ejection at a sharp duct exit~\citep[e.g.][]{cummings1983,disselhorst1980}.  Both of these effects are known to contribute to the damping effect of Helmholtz resonator-style acoustic linings~\citep[e.g.][]{zinn1970,rienstra2018}, with a significant area reduction in a resonator neck~\citep[e.g.][]{zinn1970} or at a nozzle exit~\citep[e.g.][]{cummings1983} causing extremely high acoustic amplitudes.  Concerning the duct exit, \citet[][figure~21]{disselhorst1980} map out the parameter regions where end losses are important, and~\citet[][figure~5, $\log K \to \infty$]{cummings1983} show that without a significant nozzle area restriction, nonlinear viscous end effects are small.  Moreover, \citet{auregan2008} experimentally verify that, for a ceramic liner with an 80\% open area (and therefore no significant area reduction) the liner is well modelled without nonlinear viscous dissipative effects.  Due to the complexity of the problem, the modelling of nonlinear viscous effects from a sharp-cornered orifice usually relies on a plane-wave assumption~\citep[e.g.][]{zinn1970,cummings1983} together with ad-hoc assumptions such as different simplified governing equations for the inlet and outlet phases of an oscillation.  Experimentally, schlieren imaging of a trumpet played sufficiently loudly to produce shock waves~\citep[][figures~2, 4~and~5]{pandya2003} shows no evidence of the sharp-tip vortices related to nonlinear viscous dissipation~\citep{disselhorst1980}, and \citet{hirschberg} comment that they ``could not even observe the nonlinearities which certainly occur, for low pitches at fortissimo levels as a result of vortex shedding at the pipe end'' for ``sound from the pipe to the listener in the case of a clarinet'', ``even though the oscillation amplitudes were comparable to those found in the trombone''.
Whilst viscous effects, both linear in the duct and nonlinear at the duct exit, may therefore be of interest when modelling saturation amplitudes in weakly-nonlinear situations such as in brass instruments~\citep[e.g.][]{rendon2010}, we are inspired by physical experiments~\citep[e.g.][]{hirschberg,pandya2003} and simplicity to neglect viscosity here.  This is the same assumption made for the in-duct inviscid modelling mentioned above.

This paper aims to create an analytical model of the duct exit that is compatible with the inviscid weak nonlinearity of \citet{jensenbrambley2026}. To do so, we take inspiration from the linear duct exit studies of \citet{kemp}, \citet{felix2018modeling}, and~\citet{mangin2023modelling} in their use of another, larger duct, concentric with the original duct, as the `room' into which radiation is emitted; this is intended to approximate free space as the outer duct's walls are widened, and reflections are expected to be of comparable magnitude to reflections from the actual room in which an instrument is played.  The inspiration from \citet{kemp} in particular extends to our use of a rigid external waveguide (in contrast to the other two papers' PML), though we do not follow them in taking an average of their `baffle' and `dipole' exit conditions, since this relies on the superposition principle and therefore cannot be generalised to the nonlinear regime.  Instead, we follow \citet{felix2018modeling} and~\citet{mangin2023modelling} in using a mode-matching condition, although here we will include weakly-nonlinear effects that were neglected in their linear modelling.

After a brief summary in section \ref{sec:recap} of the mathematical framework from \citet{jensenbrambley2026} which will be used here, the derivation of the outlet condition is carried out in section \ref{sec:exitcondition}. This outlet condition is used in section \ref{sec:endcorrection} to derive the end-correction and its generalisation to include nonlinearity.  Results are presented in section~\ref{sec:results}, starting with a validation of the exit condition in the linear regime in section \ref{sec:linendcorr} through comparison of its calculated end correction with that derived from the Wiener--Hopf technique.  The influence of nonlinearity on the end correction is then investigated in section~\ref{sec:nonlinendcorr}, and the effect of the exit condition on duct resonances is investigated in section~\ref{sec:resonances}. Sections \ref{sec:2Dsource}-\ref{sec:2Dexphorn} then give numerical examples of various geometries in two and three dimensions, focussing predominantly but not exclusively on the linear case. The paper's findings are summarised in section \ref{sec:conc} alongside a general discussion, including possibilities for future work. \textsc{Matlab} source code numerically implementing the model to produce these results is included in the supplementary material, along with animations of some of the figures.

\section{Weakly-nonlinear multi-modal duct acoustics}\label{sec:recap}
In this section, we give a condensed version of the method outlined in \citet{jensenbrambley2026}.

We nondimensionalise the equations of adiabatic gas dynamics, and expand them for small perturbation Mach number $M$ about a zero-mean-flow base state.  We include terms up to second order in $M$, making the model \emph{weakly} nonlinear. After the elimination of the density, the acoustic variables are subject to the first of two series expansions, namely a Fourier series expansion in time (justified by the fact that musical instruments and resonant ducts emit discrete spectra)
\begin{align}
    & p'(\pvect{x},t) = \sum_{a = -\infty}^{\infty}P^a(\pvect{x})\e^{-\I a\omega t}, & & \pvect{u}'(\pvect{x},t) = \sum_{a = -\infty}^{\infty}\pvect{U}^a(\pvect{x})\e^{-\I a\omega t}.
\end{align}
A duct geometry is introduced, as shown in figure~\ref{fig:outerductdrawings}(a) below.  This may be in two dimensions (abbreviated hereinafter to 2D) or three dimensions (hereinafter 3D), with the same analysis used except for the different mode shapes described below.  The geometry is defined by (i) a centreline with curvature $\kappa$, and additionally with torsion $\tau$ in 3D, and (ii) a width $X(s)$ in 2D or radius $R(s)$ in 3D. Each of these may vary along the duct with the arclength variable $s$. In 2D the wall width may vary either side of the centreline according to functions $X_\pm(s)$, with $X = X_+ - X_-$; in 3D we maintain a circular cross-section, as this applies to most brass instruments. The velocity is expressed in a coordinate system moving with the centreline, and transverse components of the velocity are eliminated, leaving two acoustic variables: the pressure $P^a(\pvect{x})$ and the longitudinal velocity $U^a(\pvect{x})$.

These two acoustic variables are then subject to the second of the two series expansions: they are projected onto an orthogonal basis of spatial modes which are eigenfunctions of the straight-duct problem:
\begin{align}
    P^a(\pvect{x}) &= \sum_{\alpha = 0}^\infty P_\alpha^a(s)\psi_\alpha(\pvect{x}),&
    U^a(\pvect{x}) &= \sum_{\alpha = 0}^\infty U_\alpha^a(s)\psi_\alpha(\pvect{x}).
\end{align}
Modes in the 2D coordinate system $\pvect{x} = (s,x)$, where $x$ is the transverse coordinate, are given by
\begin{align}
\label{2Dmodedef}
    &\psi_\alpha = \frac{C_\alpha}{\sqrt{X}} \cos\left[\frac{\lambda_\alpha (x - X_-)}{X}\right],&& \lambda_\alpha = \alpha\uppi,
\end{align}
for $\alpha \in \mathbb{N}_0$, with the scaling constant $C_\alpha = \sqrt{2 - \delta_{\alpha0}}$ ensuring orthonormality. In the 3D system $\pvect{x} = (s,r,\theta)$, the modes are
\begin{align}
\label{3Dmodedef}
    &\psi_\alpha = \frac{C_\alpha}{\sqrt{\uppi}R} \J_{m_\alpha}(\lambda_\alpha r/R)\cos\left(m_\alpha(\theta-\theta_0(s)) - \frac{\xi_\alpha\uppi}{2}\right),&&\lambda_\alpha = j_{m_\alpha n_\alpha}',
\end{align}
for $m_\alpha \in \mathbb{N}_0$ the azimuthal modenumber, $\xi_\alpha \in \{0,1\}$ the symmetry index (giving either cosine or sine solutions), $j_{m_\alpha n_\alpha}'$ the $n_\alpha$-th zero of the Bessel function of order $m_\alpha$, and $\theta_0(s)$ an extra coordinate which twists the system in line with torsion ensuring an orthogonal metric ~\citep[see][for details]{jensenbrambley2026}. The scaling constant is
\begin{align}
    C_\alpha = 
    \begin{cases}
        \big|\J_{0}(\lambda_\alpha)\big|^{-1},&m_\alpha = 0,\\
        \bigg(\sqrt{\frac{1}{2}\left[1 - \frac{{m_\alpha}^2}{{\lambda_\alpha}^2}\right]} ~\big|\J_{m_\alpha}(\lambda_\alpha)\big|\bigg)^{-1},&m_\alpha \neq 0.
    \end{cases}
\end{align}
We will sometimes refer to modes by their index $\alpha$, and sometimes by the equivalent triple $(m_\alpha,n_\alpha,\xi_\alpha)$, depending on context. Because of the inclusion of weakly-nonlinear terms, most of the expressions used here will involve a linear matrix-like component and a quadratic convolution-like component (convolution being the form of the nonlinear terms' expression as Fourier coefficients).  For example, the weakly-nonlinear generalisation of the admittance, relating the axial velocity to the pressure, is given by
\begin{align}&
\label{eq:admdef}
    U_\alpha^a = \sum_{\beta=0}^\infty\mat{Y}_{\alpha\beta}^aP_\beta^a + \sum_{b = -\infty}^\infty\sum_{\beta,\gamma=0}^\infty\mathcal{Y}_{\alpha\beta\gamma}^{ab}P_\beta^{a - b}P_\gamma^b,
    &&\text{or equivalently}&&
    \mvect{u} = \mat{Y}\mvect{p} + \mathcal{Y}\langle\mvect{p},\mvect{p}\rangle,
\end{align}
where $\mat{Y}^a_{\alpha\beta}$ is a third-rank tensor representing a set of matrices $\mat{Y}^a$ for each frequency harmonic $a\in\mathbb{Z}$, and $\mathcal{Y}^{ab}_{\alpha\beta\gamma}$ is a fifth-rank tensor representing the weakly-nonlinear convolution.  We write this using the shorthand $\mvect{u} = \mat{Y}\mvect{p} + \mathcal{Y}\langle\mvect{p},\mvect{p}\rangle$, where $\mvect{u}$ and $\mvect{p}$ are vectors of velocity and pressure modal coefficients $U^a_\alpha$ and $P^a_\alpha$, $\mat{Y}$ is a block-diagonal matrix performing matrix multiplication on each frequency harmonic independently, and angle brackets denote the quadratic convolution operation
\begin{align}
    \mvect{q} &= \mathcal{N}\langle\mvect{v},\mvect{w}\rangle 
    &&\iff&
    q^a_\alpha &= \sum_{b=-\infty}^\infty\sum_{\beta,\gamma=0}^\infty \! \mathcal{N}^{ab}_{\alpha\beta\gamma}\, v^{a-b}_\beta\,w^b_\gamma. 
    \end{align}
To avoid repetition of complicated arguments, we will also use the shorthand notation $\mathcal{N}\langle\mvect{v},\mvect{v}\rangle \equiv \mathcal{N}\langle\mvect{v},\cdots\rangle$.
Using this notation, \citet{jensenbrambley2026} derived a system of evolution equations along the duct axis given by
\begin{equation}
    \label{LandNintro}
        \frac{\intd }{\intd s}\begin{pmatrix}
            \mvect{u} \\ \mvect{p}
        \end{pmatrix} = \mat{L}\begin{pmatrix}
            \mvect{u} \\ \mvect{p}
        \end{pmatrix} + \mathcal{N}\left\langle\begin{pmatrix}
            \mvect{u} \\ \mvect{p}
        \end{pmatrix},\,\cdots\right\rangle,
    \end{equation}
The operators $\mat{L}$ and $\mathcal{N}$ are known complicated expressions given by~\citet{jensenbrambley2026}; for example, in 3D, the matrix $\mat{L}^a$ is given by
\begin{equation}
 \mat{L}^a = \begin{pmatrix}
        -\frac{R'}{R}\mat{W} - \tau\mat{H} & ia\omega\left[\left(\mat{I} - \frac{\gmat{\Lambda}^2}{a^2\omega^2R^2}\right)\left(\mat{I} - \kappa R\mat{A}\right) - \frac{\kappa\widetilde{\mat{A}}}{a^2\omega^2R}\right] \\
        i a\omega\bigg(\mat{I} - \kappa R\mat{A}\bigg) &\frac{R'}{R}\mat{W}^\T + \tau\mat{H}^\T
    \end{pmatrix},   
\end{equation}
where $R' = \intd R/\intd s$, $\mat{W}$, $\mat{H}$, $\mat{A}$ and $\Tilde{\mat{A}}$ are constant matrices set by the modal basis, and $\mat{I}$ is the identity.
These equations are solved by introducing the admittance~\eqref{eq:admdef}, which allows for the prescription of a radiation condition. Note that, since we are working to order $O(M^2)$ and so terms of order $O(M^3)$ may be neglected, these quantities have inverses $\mat{Z} = \mat{Y}^{-1}$ and $\mathcal{Z} = -\mat{Z}\mathcal{Y}\langle\mat{Z},\mat{Z}\rangle$; here and elsewhere we use the notation that $\mathcal{A} = \mathcal{B}\langle\mat{C},\mat{D}\rangle$ defines a nonlinear operator such that $\mathcal{A}\langle\mvect{v}, \mvect{w}\rangle = \mathcal{B}\langle\mat{C}\mvect{v}, \mat{D}\mvect{w}\rangle$. In \citet{jensenbrambley2026}, first-order Riccati-style equations for these admittances are then derived and solved, subject to a boundary condition consisting of a known value of $\mat{Y}$ and $\mathcal{Y}$ at one given point in the duct; this was chosen, for convenience, to be at the duct outlet, and corresponds to forward propagation in an infinite, self-similar extension of the duct. Following this, a pressure condition (typically a plane piston source) is prescribed at the inlet and the pressure field is computed according to the duct's radiative properties, which are now known from the admittance. In the next section we derive a more physically realistic boundary condition for the admittances, corresponding to an open duct end. 

\section{Formulation of the exit condition and solution in the outer duct}\label{sec:exitcondition}
\subsection{The duct outlet as an acoustic source}\label{sec:outletintro}
\begin{figure}
\begin{subfigure}{\textwidth}
    \label{fig:first}
    \centering
    \includegraphics[trim=0cm 2.5cm 0cm 0cm, clip]{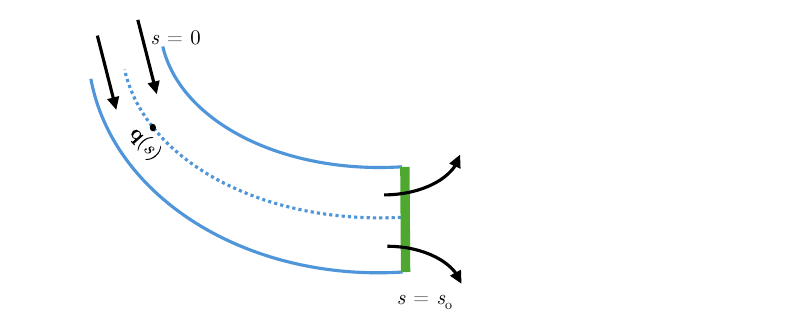}
    \subcaption{ }
\end{subfigure}
\hfill
\begin{subfigure}{\textwidth}
    \label{fig:second}
    \centering
    \includegraphics[trim=0cm 1.3cm 0cm 0.5cm, clip]{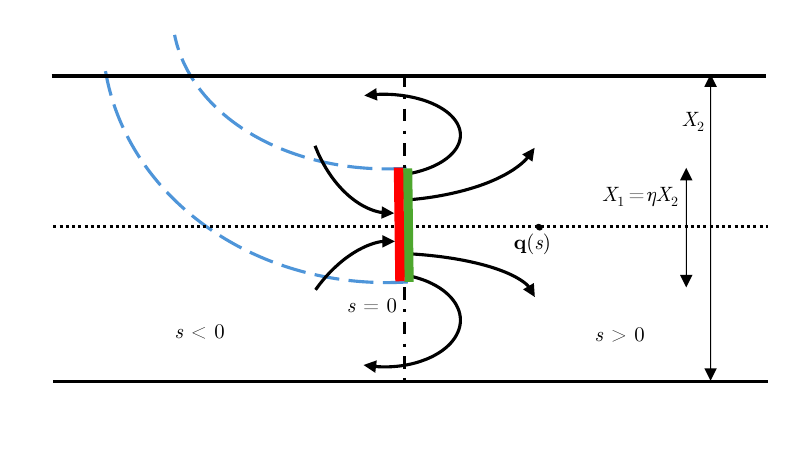}
    \subcaption{ }
\end{subfigure}
\caption{A schematic depicting the inner and outer ducts.
(a)~The inner duct, with centreline $\pvect{q}(s)$, inlet ($s = 0$), and the outlet boundary/surface of emission ($s = s_\text{o}$) depicted in green.
(b)~The outer duct, with centreline $\pvect{q}(s)$, relative duct widths $X_1$ and $X_2$, the acoustic source/surface of emission ($s = 0+$) depicted in green, the absorbing boundary or `acoustic wormhole' ($s=0-$) depicted in red, and the two domains of propagation $s > 0$ and $s < 0$, with acoustic variables continuous across the dash-dotted line. Note that the outer duct domain $s<0$ intersects with the inner duct domain, shown by the blue dashed lines, but this intersection is ignored in this model. Outer and inner widths $X_2$ and $X_1$ use the 2D notation.}
\label{fig:outerductdrawings}
\end{figure}

Here we wish to model the propagation of sound from a duct outlet, in order to inform our choice of radiation condition for the duct itself. Rather than a duct radiating sound into free space, we consider an inner duct positioned within an outer duct several times larger, i.e. `approximately free' space.  A schematic of this model in depicted in figure~\ref{fig:outerductdrawings}.
The inner duct exit, shown in green in figure~\ref{fig:outerductdrawings}, is modelled as an acoustic source within the outer-duct at arclength $s = 0$, with its width being a fraction $\eta < 1$ of the outer duct's width. The right (positive $s$-facing) side of the source then emanates sound into the outer duct, while the left side (shown in red in figure~\ref{fig:outerductdrawings}) will have an acoustic absorption condition to be specified later: since sound entering this surface departs forever from the domain of computation, we term this the `acoustic wormhole'. This is obviously a somewhat idealised situation, as the body of the inner duct is not present in the outer domain, but the expectation is as follows: of the sound leaving the bell of a trumpet, only a relatively small proportion will emanate backwards; a yet smaller proportion of this will reflect off the body of the trumpet; and this small fraction of reflected sound's contribution is unlikely to have a significant influence on the inner duct's acoustics.  While there are many possible boundary conditions that could be applied to the outer duct, here we will use a rigid wall boundary condition, as is used for the inner duct.  This assumption will be justified and discussed further in section~\ref{sec:convergence}. As an idea of the physical parameters typically involved, a brass instrument has a bell that is typically around 10cm or so in diameter. That means, with an $\eta = 1/40$ ratio (used for end-correction calculations later on) the outer duct is 4m in diameter. Clearly, unless the player is outside, reflections from the room the player is in would be comparable to the reflections from the ``fake'' outer duct employed here, and so are unlikely to affect the inner duct resonances. The curved arrows in figure \ref{fig:outerductdrawings} demonstrate that sound (1) leaves the inner-duct domain via the green surface, (2) passes into the right-hand side of the outer duct, (3) can also enter the left-hand side of the outer duct, and (4) may be absorbed by the `acoustic wormhole' surface shown in red.

We define $p^R(\pvect{x},t)$ and $u^R(\pvect{x},t)$ to be the pressure and longitudinal velocity just to the right of the source  ($s=0^+$), while $p^L(\pvect{x},t)$ and $u^L(\pvect{x},t)$ are the equivalent quantities just to the left of the source  ($s=0^-)$. These quantities are all expanded in the same way as described in section~\ref{sec:recap}, although being within the outer duct they are all expanded in terms of the outer-duct modes: taking $p^R(\pvect{x},t)$ as an example, we have
\begin{equation}
    p^R(\pvect{x},t) = \sum_{a = -\infty}^\infty\sum_{\alpha = 0}^{\infty}\e^{-\I a\omega t}\psi_\alpha^2(\pvect{x})P_{\alpha,2}^{a,R},
\end{equation}
where $\psi_\alpha^2(\pvect{x})$ are the outer-duct modes given by~\eqref{2Dmodedef} in 2D and~\eqref{3Dmodedef} in 3D (using the outer-duct width and $s = 0$ in these equations). Note that the definition of the coefficient $P_{\alpha,2}^{a,R}$ is
\begin{equation}
    P_{\alpha,2}^{a,R} = \frac{\omega}{2\pi}\int_{t = 0}^{2\pi}\iint_{S_2}p^R(\pvect{x},t)\e^{\I a\omega t}\psi_\alpha^2(\pvect{x})\mathrm{d}S\mathrm{d}t,
\end{equation}
where $S_2$ is the outer-duct cross section at  0.

We then treat $P_{\alpha,2}^{a,R}$ as a set of vectors with $\alpha$-coefficients, $\{\mvect{p}_2^{a,R}:\, a\in\mathbb{Z}\}$, which we refer to with the shorthand notation $\mvect{p}_2^R$ (and similarly with $\mvect{u}_2^R$, $\mvect{p}_2^L$ and $\mvect{u}_2^L$, an equivalent shorthand to the one seen in equation \ref{eq:admdef}). We can then introduce tensorial \emph{linear and nonlinear admittances} to the right of the source  $(s = 0^+)$, $\mat{Y}_{\alpha\beta}^{a,R}$ and $\mathcal{Y}_{\alpha\beta\gamma}^{ab,R}$, defined as in equation \eqref{eq:admdef}, and similarly for  $s = 0^-$, giving $\mat{Y}_{\alpha\beta}^{a,L}$ and $\mathcal{Y}_{\alpha\beta\gamma}^{ab,L}$. Equipped with these radiative quantities, we then stipulate that sound to the right of the source must only travel rightward in the outer duct, and sound to the left must only travel leftward; equivalently, we are taking the outer duct admittances ($\mat{Y}^R_2$, $\mathcal{Y}^R_2$) to be the \emph{positive characteristic} ($\overline{\mat{Y}}_2$, $\overline{\mathcal{Y}}_2$) for $s > 0$, and $\mat{Y}^L_2$, $\mathcal{Y}^L_2$ to be the \emph{negative characteristic} ($-\overline{\mat{Y}}_2$, $-\overline{\mathcal{Y}}_2$) for $s < 0$. Note that the impedances in these regions will then be ($\overline{\mat{Z}}_2,\overline{\mathcal{Z}}_2$) and ($-\overline{\mat{Z}}_2,\overline{\mathcal{Z}}_2$)\footnote{%
The non-matching signs here are due to reflection in $s=0$ giving a sign change of longitudinal velocity $u'$ but not of pressure $p'$, and the linear and weakly-nonlinear parts of the impedance being either linear or quadratic in this sign change.}
respectively, where $\overline{\mat{Z}}_2 = {\overline{\mat{Y}}_2}^{-1}$ and $\overline{\mathcal{Z}}_2 = -\overline{\mat{Z}}_2\overline{\mathcal{Y}}_2\langle\overline{\mat{Z}}_2,\overline{\mat{Z}}_2\rangle$. This notation is taken from \citet{jensenbrambley2026}, where explicit formulae and a derivation of the characteristic admittance is given.

\subsection{Restriction and zero-extension}\label{sec:trunczero}
Before we derive the outlet condition, it will first prove helpful to consider how modal expansions using the outer-duct modes $\psi_\alpha^2(\pvect{x})$ are related to modal expansions using the inner-duct modes; inner duct modes are denoted here as $\psi_\alpha^1(\pvect{x})$, and are similarly given by~\eqref{2Dmodedef} using the inner-duct width $X_1$ in 2D and by~\eqref{3Dmodedef} using the inner duct radius $R_1$ in 3D.  Expanding using the inner-duct modes would result in a different set of coefficients $\mvect{p}_1^R$, $\mvect{u}_1^R$, $\mvect{p}_1^L$ and $\mvect{u}_1^L$, and is valid only for $|x| < X_1/2$ in 2D and $r < R_1$ in 3D. We now relate the outer-duct coefficients to the inner-duct ones.  For example,
\begin{equation}
    \sum_{\beta = 0}^\infty\psi_\beta^1(\pvect{x})P_{\beta,1}^{a,R} = \sum_{\beta = 0}^\infty\psi_\beta^2(\pvect{x})P_{\beta,2}^{a,R}
    \qquad\text{provided}\quad \pvect{x}\in S_1.
\end{equation}
where $S_1$ is the inner-duct cross section at $s = 0$. Multiplying both sides by $\psi_\alpha^1(\pvect{x})$ and integrating both sides over an inner-duct cross section  gives (by orthogonality of the modes)
\begin{equation}
    P_{\alpha,1}^{a,R} = \sum_{\beta = 0}^\infty\int_{S_1}\psi_\alpha^1(\pvect{x})\psi_\beta^2(\pvect{x})\mathrm{d}S~P_{\beta,2}^{a,R} =: \sum_\beta \mat{F}_{\alpha\beta} P_{\beta,2}^{a,R},
\end{equation}
where we term $\mat{F}$ the \emph{restriction operator}. We utilise the vector notation here and write $\mvect{p}_1^R = \mat{F}\mvect{p}_2^R$.

Since restriction is a one-way operation (in that information for $\pvect{x}\in S_2\setminus S_1$ is lost), we cannot define an inverse operation, but we can still map from the inner duct to the outer duct if we think carefully about what this lost information can be replaced with.  While we expect a jump in the pressure and velocity as we move from one side of the source to the other, we do not expect this discontinuity to hold  around the source. We therefore define acoustic jump variables $u^*(\pvect{x},t)$ and $p^*(\pvect{x},t)$ as
\begin{align}\label{equ:jump}
    u^*(\pvect{x},t) = u^R(\pvect{x},t)|_{s = 0^+} - u^L(\pvect{x},t)|_{s = 0^-}, && p^*(\pvect{x},t) = p^R(\pvect{x},t)|_{s = 0^+} - p^L(\pvect{x},t)|_{s = 0^-},
\end{align}
in the expectation that these will be nonzero only on the region $S_1$. Finding the outer-duct coefficients of one of these jump variables, we have
\begin{equation}
    U_{\alpha,2}^{a,*} =  \frac{\omega}{2\pi}\int_{t = 0}^{2\pi}\iint_{S_2}u^*(\pvect{x},t)\e^{\I a\omega t}\psi_\alpha^2(\pvect{x})\mathrm{d}S\mathrm{d}t,
\end{equation}
but the integral over $S_2$ here may equivalently be performed over $S_1$, since $u^*$ is zero outside $S_1$. Making this change, and expanding $u^*(\pvect{x},t)$ in terms of the inner-duct modes, we get
\begin{equation}
    U_{\alpha,2}^{a,*} = \iint_{S_1}\psi_\alpha^2(\pvect{x})\psi_\beta^1(\pvect{x})\mathrm{d}S~U_{\beta,1}^{a,*},
\end{equation}
or in other words
\begin{align}\label{equ:zero-extend-star}
    \mvect{u}_2^* = \mat{F}^\T\!\mvect{u}_1^*, &&\text{and by the same property,} &&\mvect{p}_2^* = \mat{F}^\T\!\mvect{p}_1^*.
\end{align}
From this, we deduce that taking the transpose of the restriction operator results in a \emph{zero-extension} operator. Note that while $\mat{FF}^\T\! = \mat{I}$ (because zero-extending and then restricting returns the original), in general $\mat{F}^\T\!\mat{F}\! \neq \mat{I}$ (since restricting and then zero-extending essentially sets the original to zero for $\pvect{x}\in S_2\setminus S_1$).

For the mode definitions \eqref{2Dmodedef} and \eqref{3Dmodedef}, the restriction operator is given in 2D by
\begin{equation}
    \mat{F}_{\alpha\beta} = \begin{cases}
        \frac{C_\alpha C_\beta}{\pi}\frac{\eta^{3/2}\beta}{\alpha^2 - \eta^2\beta^2}\left[\sin\left(\frac{1 - \eta}{2}\beta\pi\right) - (-1)^\alpha\sin\left(\frac{1 + \eta}{2}\beta\pi\right)\right] &\text{if }\alpha \neq \eta\beta, \\
        \frac{C_\alpha C_\beta}{2}\sqrt{\eta}\cos\left(\frac{\alpha\pi}{2}\left[1 - \frac{1}{\eta}\right]\right) &\text{if }\alpha = \eta\beta \neq 0, \\
        C_\alpha C_\beta \sqrt{\eta} &\text{if }\alpha = \eta\beta = 0,
    \end{cases}
\end{equation}
and in 3D by
\begin{equation}
    \mat{F}_{\alpha\beta} = \begin{cases}
        \parbox{6cm}{$(1+\delta_{m_\alpha 0})\eta C_\alpha C_\beta\dfrac{\lambda_\alpha \J_{m_\alpha - 1}(\lambda_\alpha)\J_{m_\alpha}(\eta\lambda_\beta) - \eta\lambda_\beta\J_{m_\alpha}(\lambda_\alpha)\J_{m_\alpha - 1}(\eta\lambda_\beta)}{\eta^2\lambda_\beta^2 - \lambda_\alpha^2}$\hspace*{-6cm}}& \\
        &\text{if }m_\alpha = m_\beta,~\xi_\alpha = \xi_\beta,~\eta\lambda_\beta \neq \lambda_\alpha,\\
        \eta &\text{if }m_\alpha = m_\beta,~\xi_\alpha = \xi_\beta,~\eta\lambda_\beta = \lambda_\alpha, \\
        0 &\text{otherwise}.
    \end{cases}
\end{equation}

\subsection{Derivation of the outlet admittance}\label{sec:outletadm}
We now go on to derive the open-end admittance boundary condition, using the restriction and zero-extension operators $\mat{F}$ and $\mat{F}^\T$ from the previous section, applied at $s = 0$.   Specifically, we aim to derive a condition relating pressure and axial velocity at the exit of the smaller duct as it enters the larger duct (shown in green in figures~\ref{fig:outerductdrawings}(a) and~(b)); i.e.\ a relation between $\mvect{p}_1^R$ and $\mvect{u}_1^R$ at $s=0$.  To do so, we will assume: (i) a known relation between $\mvect{p}_1^L$ and $\mvect{u}_1^L$ (i.e.\ the condition on the red `acoustic wormhole' surface in figure~\ref{fig:outerductdrawings}(b)); (ii) that waves are only outgoing in the larger duct for both $s>0$ and $s<0$; and (iii) that the acoustic variables are continuous at the duct exit (across $s=0$) in the larger duct outside the smaller duct (for $\pvect{x}\in S_2\setminus S_1$). This last condition means that the jump in velocity and pressure when crossing $s=0$, given by $u^*$ and $p^*$ in~\eqref{equ:jump}, is only nonzero for $\pvect{x}\in S_1$.  The entirety of this section is evaluated within the outer duct at $s=0$. We start by using the acoustic jump variables in the outer duct to eliminate $\mvect{u}_2^L = \mvect{u}_2^R - \mvect{u}_2^*$ and $\mvect{p}_2^L = \mvect{p}_2^R - \mvect{p}_2^*$; in particular, the outer-left-duct admittance equation becomes
\begin{subequations}\label{equ:u2r}\begin{align}
    \mvect{u}_2^R - \mvect{u}_2^* &= -\overline{\mat{Y}}_2\big(\mvect{p}_2^R - \mvect{p}_2^*\big) - \overline{\mathcal{Y}}_2\big\langle\mvect{p}_2^R - \mvect{p}_2^*,\mvect{p}_2^R - \mvect{p}_2^*\big\rangle,
    \\ \text{or equivalently,}\quad
        \mvect{p}_2^R - \mvect{p}_2^* &= -\overline{\mat{Z}}_2\big(\mvect{u}_2^R - \mvect{u}_2^*\big) + \overline{\mathcal{Z}}_2\big\langle\mvect{u}_2^R - \mvect{u}_2^*,\mvect{u}_2^R - \mvect{u}_2^*\big\rangle.
\end{align}\end{subequations}
At linear order, this may be rearranged for $\mvect{u}_2^R$ using the identities $\mvect{u}_2^R = \overline{\mat{Y}}_2\mvect{p}_2^R$ and $\mvect{p}_2^R = \overline{\mat{Z}}_2\mvect{u}_2^R$ to give
\begin{align}
    &\mvect{u}_2^R = \frac{1}{2}\!\left(\mvect{u}_2^* + \overline{\mat{Y}}_2\mvect{p}_2^*\right) +O(M^2),&&\text{or equivalently}&&\mvect{p}_2^R = \frac{1}{2}\!\left(\mvect{p}_2^* + \overline{\mat{Z}}_2\mvect{u}_2^*\right)+O(M^2).
\end{align}
These two expressions may be substituted into the nonlinear terms in~\eqref{equ:u2r} (recall we are neglecting $O(M^3)$ and smaller terms), resulting in
\begin{subequations}
    \begin{align}
    \label{u2R}
        \mvect{u}_2^R &= \frac{1}{2}\!\left(\mvect{u}_2^* + \overline{\mat{Y}}_2\mvect{p}_2^*\right) + \frac{1}{8}\overline{\mathcal{Y}}_2\Big(\big\langle\mvect{p}_2^* + \overline{\mat{Z}}_2\mvect{u}_2^*,\mvect{p}_2^* + \overline{\mat{Z}}_2\mvect{u}_2^*\big\rangle - \big\langle\mvect{p}_2^* - \overline{\mat{Z}}_2\mvect{u}_2^*,\mvect{p}_2^* - \overline{\mat{Z}}_2\mvect{u}_2^*\big\rangle\Big),
        \\
    \label{p2R}
        \mvect{p}_2^R &= \frac{1}{2}\!\left(\mvect{p}_2^* + \overline{\mat{Z}}_2\mvect{u}_2^*\right) + \frac{1}{8}\overline{\mathcal{Z}}_2\Big(\big\langle\mvect{u}_2^* + \overline{\mat{Y}}_2\mvect{p}_2^*,\mvect{u}_2^* + \overline{\mat{Y}}_2\mvect{p}_2^*\big\rangle + \big\langle\mvect{u}_2^* - \overline{\mat{Y}}_2\mvect{p}_2^*,\mvect{u}_2^* - \overline{\mat{Y}}_2\mvect{p}_2^*\big\rangle\Big).
    \end{align}
\end{subequations}
By outer-duct continuity at $s=0$, both $u^*$ and $p^*$ are zero outside $\pvect{x}\in S_1$, and so we can use~\eqref{equ:zero-extend-star} to replace all outer-duct quantities on the right-hand sides with inner-duct ones. Applying $\mat{F}$ to both sides then gives
\begin{subequations}
    \begin{align}
    \label{u1R}
            \mvect{u}_1^R = &\frac{1}{2}\Big(\mvect{u}_{1}^* + \mat{F}\overline{\mat{Y}}_2\mat{F}^\T\!\!\mvect{p}_{1}^*\Big) 
            + \frac{1}{8}\mat{F}\overline{\mathcal{Y}}_2\Big(\big\langle\mat{F}^\T\!\!\mvect{p}1^* + \overline{\mat{Z}}_2\mat{F}^\T\!\mvect{u}_1^*,\mat{F}^\T\!\!\mvect{p}_1^* + \overline{\mat{Z}}_2\mat{F}^\T\!\mvect{u}_1^*\big\rangle - \big\langle\mat{F}^\T\!\!\mvect{p}_1^* - \overline{\mat{Z}}_2\mat{F}^\T\!\mvect{u}_1^*,\mat{F}^\T\!\!\mvect{p}_1^* - \overline{\mat{Z}}_2\mat{F}^\T\!\mvect{u}_1^*\big\rangle\Big),
\\[1ex]
                \label{p1R}
            \mvect{p}_1^R = &\frac{1}{2}\Big(\mvect{p}_1^* + \mat{F}\overline{\mat{Z}}_2\mat{F}^\T\!\mvect{u}_1^*\Big)
            + \frac{1}{8}\mat{F}\overline{\mathcal{Z}}_2\Big(\big\langle\mat{F}^\T\!\mvect{u}_1^* + \overline{\mat{Y}}_2\mat{F}^\T\!\!\mvect{p}_1^*,\mat{F}^\T\!\mvect{u}_1^* + \overline{\mat{Y}}_2\mat{F}^\T\!\!\mvect{p}_1^*\big\rangle + \big\langle\mat{F}^\T\!\mvect{u}_{1}^* - \overline{\mat{Y}}_2\mat{F}^\T\!\!\mvect{p}_1^*,\mat{F}^\T\!\mvect{u}_1^* - \overline{\mat{Y}}_2\mat{F}^\T\!\!\mvect{p}_1^*\big\rangle\Big).
        \end{align}
\end{subequations}
The inner-duct admittances to the right of the source, $\mat{Y}_1^R$ and $\mathcal{Y}_1^R$, are what we wish to determine; those to the left of the source, $\mat{Y}_1^L$ and $\mathcal{Y}_1^L$, will be prescribed. For the remainder of this section, we will suppress indices on radiative quantities and (unless otherwise specified) take $(\mat{Y}^R,\mathcal{Y}^R)$ to be $(\mat{Y}_1^R,\mathcal{Y}_1^R)$, $(\mat{Y}^L,\mathcal{Y}^L)$ to be $(\mat{Y}_1^L,\mathcal{Y}_1^L)$, and $(\overline{\mat{Y}},\overline{\mathcal{Y}})$ to be $(\overline{\mat{Y}}_2,\overline{\mathcal{Y}}_2)$ (with similar shorthand for all impedances). We then have the acoustic wormhole's impedance equation in terms of the jump and outlet quantities
\begin{equation}
\label{innerleftimp}
    \mvect{p}_1^* - \mvect{p}_1^R = \mat{Z}^L\Big(\mvect{u}_1^* - \mvect{u}_1^R\Big) - \mathcal{Z}^L\Big\langle\mvect{u}_1^* - \mvect{u}_1^R,\mvect{u}_1^* - \mvect{u}_1^R\Big\rangle.
\end{equation}
Substituting $\mvect{p}_1^*$ from this expression into \eqref{u1R}, at linear order we may rearrange to find
\begin{equation}
        \mvect{u}_1^* = (\mat{I} + \mat{Q})\mvect{u}_1^R - (\mat{I} - \mat{Q})\mat{Y}^L\mvect{p}_1^R + O\big(M^2\big),
\end{equation}
where $\mat{Q} := \big(\mat{I} + \mat{F}\overline{\mat{Y}}\mat{F}^\T\!\mat{Z}^L\big)^{-1}$ is a shorthand quantity.  This expression can be substituted back into~\eqref{innerleftimp} to gain an expression for $\mvect{p}_1^*$, again at linear order
\begin{equation}
        \mvect{p}_1^* = \mat{Z}^L\mat{Q}\Big(\mvect{u}_1^R + \mat{Y}^L\!\mvect{p}_1^R\Big) + O\big(M^2\big).
\end{equation}
We may now perform the same two steps, but including the quadratic quantities, making use of these linear identities within the nonlinear terms (since the error introduced by doing so is $O\big(M^3\big)$, i.e. comparable with the magnitude of other neglected terms).  This results in
\begin{subequations}\begin{align}
\label{u1*}
        \mvect{u}_1^* = (\mat{I} + \mat{Q})\mvect{u}_1^R - &(\mat{I} - \mat{Q})\mat{Y}^L\mvect{p}_1^R + (\mat{I} - \mat{Q})\mat{Y}^L\mathcal{Z}^L\bigg\langle\mat{Q}\mvect{u}_1^R - (\mat{I} - \mat{Q})\mat{Y}^L\mvect{p}_1^R,\cdots\bigg\rangle \\\notag
        + \frac{1}{4}\mat{Q}\mat{F}\overline{\mat{Y}}\overline{\mathcal{Z}}\bigg(- &\,\bigg\langle\overline{\mat{Y}}\mat{F}^\T\!\mat{Z}^L\mat{Q}\left(\mvect{u}_1^R + \mat{Y}^L\mvect{p}_1^R\right) - \mat{F}^\T\big[(\mat{I} + \mat{Q})\mvect{u}_1^R - (\mat{I} - \mat{Q})\mat{Y}^L\mvect{p}_1^R\big],\cdots\bigg\rangle\\\notag
        +&\,\bigg\langle\overline{\mat{Y}}\mat{F}^\T\!\mat{Z}^L\mat{Q}\left(\mvect{u}_1^R + \mat{Y}^L\mvect{p}_1^R\right) + \mat{F}^\T\big[(\mat{I} + \mat{Q})\mvect{u}_1^R - (\mat{I} - \mat{Q})\mat{Y}^L\mvect{p}_1^R\big],\cdots\bigg\rangle\bigg),
\\
\label{p1*}
        \mvect{p}_1^* = \mat{Z}^L\mat{Q}\left(\mvect{u}_1^R + \mat{Y}^L\right.&\left.\!\!\!\mvect{p}_1^R\right) - \mat{Z}^L\mat{Q}\mat{Y}^L\mathcal{Z}^L\bigg\langle\mat{Q}\mvect{u}_1^R - (\mat{I} - \mat{Q})\mat{Y}^L\mvect{p}_1^R,\cdots\bigg\rangle \\\notag
        + \frac{1}{4}\mat{Z}^L\mat{Q}\mat{F}\overline{\mat{Y}}\overline{\mathcal{Z}}\bigg(- &\,\bigg\langle\overline{\mat{Y}}\mat{F}^\T\!\mat{Z}^L\mat{Q}\left(\mvect{u}_1^R + \mat{Y}^L\mvect{p}_1^R\right) - \mat{F}^\T\big[(\mat{I} + \mat{Q})\mvect{u}_1^R - (\mat{I} - \mat{Q})\mat{Y}^L\mvect{p}_1^R\big],\cdots\bigg\rangle\\\notag
        +&\,\bigg\langle\overline{\mat{Y}}\mat{F}^\T\!\mat{Z}^L\mat{Q}\left(\mvect{u}_1^R + \mat{Y}^L\mvect{p}_1^R\right) + \mat{F}^\T\big[(\mat{I} + \mat{Q})\mvect{u}_1^R - (\mat{I} - \mat{Q})\mat{Y}^L\mvect{p}_1^R\big],\cdots\bigg\rangle\bigg),
    \end{align}\end{subequations}
where `$\cdots$' indicates the left argument of the $\langle\bullet,\bullet\rangle$ operation being repeated for the right. Since we have not yet used \eqref{p1R}, we may now do so, and so we substitute \eqref{p1*} and \eqref{u1*} into it. At linear order, this results in
\begin{equation}
    \mvect{p}_1^R = \frac{1}{2}\big[\mat{F}\overline{\mat{Z}}\mat{F}^\T\!(\mat{I} + \mat{Q}) + \mat{Z}^L\mat{Q}\big]\mvect{u}_1^R + \frac{1}{2}\big[-\mat{F}\overline{\mat{Z}}\mat{F}^\T\!(\mat{I} - \mat{Q}) + \mat{Z}^L\mat{Q}\big]\mat{Y}^L\mvect{p}_1^R + O\big(M^2\big).
\end{equation}
This may be rearranged, using another shorthand quantity  $\widetilde{\mat{Q}} := \big(\mat{I} + \mat{Y}^L\mat{F}\overline{\mat{Z}}\mat{F}^\T\big)^{-1}$, into the form $\mvect{p}_1^R$ = $\mat{Z}^R\mvect{u}_1^R + O\big(M^2\big)$, where
\begin{equation}\label{equ:ZRlin}
    \mat{Z}^R = -\mat{Z}^L\big[\mat{I} + \widetilde{\mat{Q}}^{-1}(\mat{I} - \mat{Q})\big]^{-1}\big[\mat{I} - \widetilde{\mat{Q}}^{-1}(\mat{I} + \mat{Q})\big].
\end{equation}
As before, we now carry out the same steps, but including the quadratic quantities, and making use of linear identities within the nonlinear terms.  The full nonlinear substitution of \eqref{p1*} and \eqref{u1*} into \eqref{p1R} is then $\mvect{p}_1^R = \mat{Z}^R\mvect{u}_1^R + \mathcal{Z}^R\big\langle\mvect{u}_1^R,\mvect{u}_1^R\big\rangle$, where
    \begin{align}\label{equ:ZRnonlin}
        \mathcal{Z}^R = \mat{Z}^L\big[\mat{I} + \widetilde{\mat{Q}}^{-1}(\mat{I} - \mat{Q})\big]^{-1}&\bigg\{-\big[\mat{I} - \widetilde{\mat{Q}}^{-1}(\mat{I} - \mat{Q})\big]\mat{Y}^L\mathcal{Z}^L\Big\langle\mat{Q} - (\mat{I} - \mat{Q})\mat{Y}^L\mat{Z}^R,\cdots\!\Big\rangle \\\notag
        + \frac{1}{4}\Big(\!\mat{Y}^L\mat{F} - \widetilde{\mat{Q}}^{-1}\mat{Q}\mat{F}\overline{\mat{Y}}\Big)\overline{\mathcal{Z}}&\Big\langle\overline{\mat{Y}}\mat{F}^\T\mat{Z}^L\mat{Q}\big(\mat{I} + \!\mat{Y}^L\mat{Z}^R\big) - \mat{F}^\T\big[(\mat{I} +\! \mat{Q}) - (\mat{I} - \mat{Q})\mat{Y}^L\mat{Z}^R\big],\cdots\!\Big\rangle \\\notag
        + \frac{1}{4}\Big(\!\mat{Y}^L\mat{F} + \widetilde{\mat{Q}}^{-1}\mat{Q}\mat{F}\overline{\mat{Y}}\Big)\overline{\mathcal{Z}}&\Big\langle\overline{\mat{Y}}\mat{F}^\T\mat{Z}^L\mat{Q}\big(\mat{I} + \!\mat{Y}^L\mat{Z}^R\big) + \mat{F}^\T\big[(\mat{I} +\! \mat{Q}) - (\mat{I} - \mat{Q})\mat{Y}^L\mat{Z}^R\big],\cdots\!\Big\rangle\bigg\},
    \end{align}
and we recall the notation that $\mathcal{A} = \mathcal{B}\langle\mat{C},\mat{D}\rangle$ defines a nonlinear operator such that $\mathcal{A}\langle\mvect{v}, \mvect{w}\rangle = \mathcal{B}\langle\mat{C}\mvect{v}, \mat{D}\mvect{w}\rangle$.
Using the nonlinear inversion rule from section \ref{sec:recap}, we may invert the impedances $\mat{Z}^R$ and $\mathcal{Z}^R$ to get the admittances
\begin{subequations}
    \begin{align}\label{equ:YRlin}
        \mat{Y}^R =& -\big[\mat{I} - \widetilde{\mat{Q}}^{-1}(\mat{I} + \mat{Q})\big]^{-1}\big[\mat{I} + \widetilde{\mat{Q}}^{-1}(\mat{I} - \mat{Q})\big]\mat{Y}^L,
\\\label{equ:YRnonlin}
            \mathcal{Y}^R =& \big[\mat{I} - \widetilde{\mat{Q}}^{-1}(\mat{I} + \mat{Q})\big]^{-1}\bigg\{\big[\mat{I} - \widetilde{\mat{Q}}^{-1}(\mat{I} - \mat{Q})\big]\mathcal{Y}^L\Big\langle\mat{I} - \mat{Z}^L\mat{Q}\left(\mat{Y}^L + \mat{Y}^R\right),\cdots\Big\rangle \\\notag&
            - \frac{1}{4}\Big(\!\mat{Y}^L\mat{F}\overline{\mat{Z}} - \widetilde{\mat{Q}}^{-1}\mat{Q}\mat{F}\Big)\overline{\mathcal{Y}}\Big\langle\mat{F}^\T\mat{Z}^L\mat{Q}\big(\mat{Y}^L + \!\mat{Y}^R\big) - \overline{\mat{Z}}\mat{F}^\T\big[(\mat{I} + \mat{Q})\mat{Y}^R - (\mat{I} - \mat{Q})\mat{Y}^L\big],\cdots\Big\rangle \\\notag&
            - \frac{1}{4}\Big(\!\mat{Y}^L\mat{F}\overline{\mat{Z}} + \widetilde{\mat{Q}}^{-1}\mat{Q}\mat{F}\Big)\overline{\mathcal{Y}}\Big\langle\mat{F}^\T\mat{Z}^L\mat{Q}\big(\mat{Y}^L + \!\mat{Y}^R\big) + \overline{\mat{Z}}\mat{F}^\T\big[(\mat{I} + \mat{Q})\mat{Y}^R - (\mat{I} - \mat{Q})\mat{Y}^L\big],\cdots\Big\rangle\bigg\}.
        \end{align}
\end{subequations}

It finally remains to choose the boundary condition to apply on the acoustic wormhole (left of the discontinuity) by specifying the admittance $\mat{Y}^L$ and $\mathcal{Y}^L$.  This boundary is somewhat artificial, as we have here neglected the outer walls of the inner duct, and this boundary sits within the region that should be inaccessible to sound from outside the inner duct (see figure~\ref{fig:outerductdrawings}(b) for a corresponding schematic).  In order to avoid spurious reflections from this artificial boundary, we here choose the left-hand side admittances to be perfectly absorbing, i.e.\ $\mat{Y}^L = \overline{\mat{Y}}_1$ and $\mathcal{Y}^L = \overline{\mathcal{Y}}_1$, and hence sound may enter the `acoustic wormhole' but no sound leaves from it. This, in combination with~\eqref{equ:YRlin} and~\eqref{equ:YRnonlin}, fully determines the admittance at the inner duct's open end, which may be used as an initial condition to solve the Riccati-style system in $(\mat{Y}_1(s),\mathcal{Y}_1(s))$ within the inner duct from arclength $s_\text{o}$ to $0$ \citep[as described in][]{jensenbrambley2026}, so that the admittance is then known everywhere throughout the system, and can then be used to calculate acoustic variables such as the pressure.

\subsection{Pressure continuity conditions}\label{sec:pressurecont}
Once the admittance has been calculated in the inner duct, the pressure in the inner duct $\mvect{p}_1^R(s)$ may be computed (subject to an inlet condition).  Details of this calculation are the subject of~\citet{jensenbrambley2026}.  For the examples that follow, we will specify the inlet condition in terms of the incoming (forward-going) acoustics only; that is, we assume that acoustics travelling backwards along the duct away from the open end may leave the duct at the inlet unimpeded.  While this is a helpful inlet condition to use for studying duct resonances, as we will do later, we note in passing that it is equally possible to use this framework for any inlet condition, or indeed for a duct open at both ends with a suitable sound source located in its interior (which would involve the decomposition of the problem into two ducts either side of the source, with the two sides potentially communicating with one another across it).

In this section, we derive the relations needed for transforming the computed pressure at the outlet of the inner duct into a boundary condition for computation of the pressure \emph{field} in the outer duct. We start by imposing continuity of acoustic velocity at $s=0$ outside the smaller duct, so that the velocity discontinuity is zero-extended from the outlet,
\begin{equation}
    \mvect{u}_2^R - \mvect{u}_2^L = \mat{F}^\T\!\big(\mvect{u}_1^R - \mvect{u}_1^L\big).
\end{equation}
In terms of the pressure, using the outer-duct admittance gives
\begin{equation}
    \overline{\mat{Y}}\left(\mvect{p}_2^R + \mvect{p}_2^L\right) + \overline{\mathcal{Y}}\left(\langle\mvect{p}_2^R,\mvect{p}_2^R\rangle + \langle\mvect{p}_2^L,\mvect{p}_2^L\rangle\right)
    = \mat{F}^\T\!\left(\mat{Y}^R\mvect{p}_1^R - \mat{Y}^L\mvect{p}_1^L + \mathcal{Y}^R\langle\mvect{p}_1^R,\mvect{p}_1^R\rangle - \mathcal{Y}^L\langle\mvect{p}_1^L,\mvect{p}_1^L\rangle\right),
\end{equation}
from which $\mvect{p}_2^L$ may be eliminated by imposing continuity of pressure in the outer duct away from the outlet; that is, by using the zero-extension of the pressure discontinuity,
\begin{equation}\label{equ:p2cont}
    \mvect{p}_2^L = \mvect{p}_2^R + \mat{F}^\T\!\big(\mvect{p}_1^L - \mvect{p}_1^R\big),
\end{equation}
resulting in the equation for $\mvect{p}_2^R$ in terms of $\mvect{p}_1^R$ and $\mvect{p}_1^L$,
\begin{align}\label{equ:p2R}
    \mvect{p}_2^R = &\frac{1}{2}\overline{\mat{Z}}\mat{F}^\T\!\left(\mat{Y}^R\mvect{p}_1^R - \mat{Y}^L\mvect{p}_1^L\right) + \frac{1}{2}\mat{F}^\T\!\left(\mvect{p}_1^R - \mvect{p}_1^L\right) 
    + \frac{1}{2}\overline{\mat{Z}}\Bigg\{\mat{F}^\T\!\left(\mathcal{Y}^R\big\langle\mvect{p}_1^R,\mvect{p}_1^R\big\rangle - \mathcal{Y}^L\big\langle\mvect{p}_1^L,\mvect{p}_1^L\big\rangle\right) \\\notag
    & -\frac{1}{4}\overline{\mathcal{Y}}\bigg(\Big\langle\overline{\mat{Z}}\mat{F}^\T\!\big(\mat{Y}^R\mvect{p}_1^R - \mat{Y}^L\mvect{p}_1^L\big) + \mat{F}^\T\big(\mvect{p}_1^R - \mvect{p}_1^L\big),\ldots\Big\rangle
     + \Big\langle\overline{\mat{Z}}\mat{F}^\T\!\big(\mat{Y}^R\mvect{p}_1^R - \mat{Y}^L\mvect{p}_1^L\big) - \mat{F}^\T\big(\mvect{p}_1^R - \mvect{p}_1^L\big),\ldots\Big\rangle\bigg)\Bigg\}
\end{align}
Taking the restriction of both sides maps $\mvect{p}_2^R$ to $\mvect{p}_1^R$; we then rearrange for $\mvect{p}_1^L$, getting
\begin{equation}
    \mvect{p}_1^L = \mat{C}\mvect{p}_1^R + \big(\mat{I} + \mat{F}\overline{\mat{Z}}\mat{F}^\T\!\mat{Y}^L\big)^{-1}\mat{F}\mathcal{C}\langle\mvect{p}_1^R,\mvect{p}_1^R\rangle,
\end{equation}
where $\mat{C}$ and $\mathcal{C}$ are defined as
\begin{subequations}
    \begin{gather}
        \mat{C} = -\big(\mat{I} + \mat{F}\overline{\mat{Z}}\mat{F}^\T\!\mat{Y}^L\big)^{-1}\Big(\mat{I} - \mat{F}\overline{\mat{Z}}\mat{F}^\T\!\mat{Y}^R\Big),
    \\
        \mathcal{C} = \overline{\mat{Z}}\Big[\mat{F}^\T\big(\mathcal{Y}^R - \mathcal{Y}^L\langle\mat{C},\mat{C}\rangle\big)
        - \frac{1}{2}\overline{\mathcal{Y}}\Big(\big\langle\overline{\mat{Z}}\mat{F}^\T\!\big(\mat{Y}^R - \mat{Y}^L\mat{C}\big),\cdots\big\rangle + \big\langle\mat{F}^\T\!\big(\mat{I} - \mat{C}\big),\cdots\big\rangle\Big)\Big].
    \end{gather}
\end{subequations}
Substituting this into the original equation~\eqref{equ:p2R} for $\mvect{p}_2^R$, we get
\begin{multline}\label{equ:final-p2R}
        \mvect{p}_2^R = \frac{1}{2}\Big[\big(\overline{\mat{Z}}\mat{F}^\T\!\mat{Y}^R + \mat{F}^\T\big) - \big(\overline{\mat{Z}}\mat{F}^\T\!\mat{Y}^L + \mat{F}^\T\big)\mat{C}\Big]\mvect{p}_1^R
        + \frac{1}{2}\Big[\mat{I} - \big(\overline{\mat{Z}}\mat{F}^\T\!\mat{Y}^L + \mat{F}^\T\big)\big(\mat{I} + \mat{F}\overline{\mat{Z}}\mat{F}^\T\!\mat{Y}^L\big)^{-1}\!\mat{F}\Big]\mathcal{C}\langle\mvect{p}_1^R,\mvect{p}_1^R\rangle,
\end{multline}
and then using~\eqref{equ:p2cont} gives
\begin{multline}\label{equ:final-p2L}
        \mvect{p}_2^L = \frac{1}{2}\Big[\big(\overline{\mat{Z}}\mat{F}^\T\!\mat{Y}^R - \mat{F}^\T\big) - \big(\overline{\mat{Z}}\mat{F}^\T\!\mat{Y}^L - \mat{F}^\T\big)\mat{C}\Big]\mvect{p}_1^R
        + \frac{1}{2}\Big[\mat{I} - \big(\overline{\mat{Z}}\mat{F}^\T\!\mat{Y}^L - \mat{F}^\T\big)\big(\mat{I} + \mat{F}\overline{\mat{Z}}\mat{F}^\T\!\mat{Y}^L\big)^{-1}\!\mat{F}\Big]\mathcal{C}\langle\mvect{p}_1^R,\mvect{p}_1^R\rangle.
\end{multline}
Equations~\eqref{equ:final-p2R} and~\eqref{equ:final-p2L} give the pressure at $s=0^+$ and $s=0^-$ as boundary conditions from which the pressure within the outer duct for $s>0$ and $s<0$ can be solved, respectively.  The method of solution is the same as for the inner duct~\citep[again as described by][]{jensenbrambley2026}.  This allows the acoustic field within the outer duct, modelling free space, to be plotted and visualized. Examples of this will be given below.

\subsection{Numerical Method}
In this section we describe details of our numerical implementation of the open end exit calculation outlined in section~\ref{sec:outletadm}, and the subsequent pressure matching described in section~\ref{sec:pressurecont}. \textsc{Matlab} source code numerically implementing this model is included in the supplementary material;  this code was used to produce the results that follow in section~\ref{sec:results}. For details of the numerical method for calculating propagation along either the inner or outer ducts, and a demonstration of its convergence as more modes are added, the reader is referred to \citet{jensenbrambley2026}.

When calculating the duct exit condition the infinite series of modes must be truncated during numerical implementation. Regarding truncation, it should be noted that the decision (taken in section \ref{sec:outletadm}) to set $\mat{Y}_1^L = \overline{\mat{Y}}_1$ and $\mathcal{Y}_1^L = \overline{\mathcal{Y}}_1$ on the acoustic wormhole has a numerical justification as well as a physical one. The natural alternative to characteristic admittances to the left of the outlet would be to take a zero admittance (corresponding to a hard wall and allowing no sound to propagate through): this causes numerical difficulties with the definitions of $\mat{Q}$ and $\widetilde{\mat{Q}}$, and even when the equations are reformulated to avoid these problems, numerical difficulties still arise as a result of the non-invertibility of $\mat{F}\overline{\mat{Z}}\mat{F}^\T$ and $\mat{F}\overline{\mat{Y}}\mat{F}^\T$ when truncated. In contrast, for the perfectly-absorbing characteristic admittance used here, no such numerical difficulties arise.

\subsubsection{Choosing modenumbers and the width ratio}\label{sec:modenumberchoice}
When truncating, we generally follow the rule of thumb set by \citet{mctavphd} that $\alpha_\text{max}^2 \geq \alpha_\text{max}^1/\eta$, where $\alpha_\text{max}^1$ and $\alpha_\text{max}^2$ are the maximum modenumbers in the inner and outer ducts respectively. As well as choosing modenumbers carefully, it is also important that $\eta$ be small enough to approximate free space; there is therefore a trade-off between needing small $\eta$, large $\alpha_\text{max}^2$, and manageable runtime.  The computations we perform with this framework are not always of the same complexity, since we will focus on different parts of the solution depending on what we wish to calculate. While the number of modes in the inner duct $\alpha_\text{max}^1$ always remains around 10, $\alpha_\text{max}^2$ and $\eta$ can vary much more.

When only a single computation of the linear-regime exit condition at a given location is required (for example, when calculating the linear end correction described in section~\ref{sec:endcorrection}, or when calculating resonances in section \ref{sec:resonances}), with no need to solve any ODEs in the outer duct, large numbers of modes can be used. In such cases, we will typically take on the order of a thousand modes for $\alpha_\text{max}^2$, meaning that we can take $\eta$ down to $1/40$ without under-resolving the duct exit. Such computations will only require tens of megabytes of memory in the linear case. These memory requirements apply in either 2D or 3D, but the calculation can take much longer in 3D than 2D, since the 3D calculation involves integration of millions of Bessel function products: a 3D calculation will often last hours on a standard desktop computer using a single core, whereas a 2D one will be done in seconds. Figures \ref{3Dendcorrectionplot.pdf} and \ref{2Dendcorrectionplot.pdf} provide a means of calibrating $\eta$ and $\alpha_\text{max}^2$ by validating against results from the Wiener--Hopf technique, as discussed further in section~\ref{sec:convergence} below. We may still solve ODEs in the inner duct (i.e. in section \ref{sec:resonances}) using this large value of $\alpha^2_\mathrm{max}$, since the number of modes in the inner duct $\alpha^1_\mathrm{max}$ remains quite small, and linearity requires only a single temporal mode, so this does not present a time or memory constraint.

When nonlinearity becomes involved, memory requirements are a greater concern due to the presence of third-rank tensors. Using thousands of modes for a nonlinear calculation of the exit condition at a single location would require more than $128\,\mathrm{GB}$ of memory, so the number of modes was reduced in this case. It is demonstrated for the linear regime in section \ref{sec:convergence} that the parameters $(\eta,\alpha_\text{max}^1,\alpha_\text{max}^2) = (1/10,10,200)$ still provide a reasonable approximation to the semi-analytical Wiener--Hopf solution: this requires on the order of $10\,\mathrm{MB}$ of memory even in the nonlinear case. 

Computational time is also more of a limiting factor in the nonlinear regime (such as in sections \ref{sec:nonlinendcorr} and \ref{sec:2Dnonlin}), especially since we will see in section~\ref{sec:nonlin-end-theory} that the nonlinear exit condition cannot be considered in isolation without also solving for the acoustic field within the inner duct.
In contrast to the linear regime, despite the number of spatial modes in the inner duct still being low, temporal modes are also required, and the ODEs are coupled in more ways than in the linear case. Performing a single computation for a simple duct geometry in the nonlinear case will thus take on the order of tens of minutes, meaning that the data for plots like figure~\ref{3Ds_hat_machplot.pdf} and~\ref{2Ds_hat_machplot.pdf} can take days to generate.

Moreover, even when calculating just the linear acoustic field in the outer duct, we also need to reduce the number of modes relative to a single exit-condition calculation: using the number of modes from section \ref{sec:linendcorr} would cause calculations to run very slowly. In order to compute in the outer duct in a reasonable computation time, the choices of $(\alpha_\text{max}^1,\alpha_\text{max}^2,\eta) = (10,200,1/10)$ are therefore maintained, resulting in calculations that take on the order of tens of seconds.  This is further compounded in the nonlinear case: even with this more modest number of modes, combination with the $\sim 10$ temporal modes required for an accurate nonlinear calculation is beyond the memory and time constraints of our \textsc{Matlab} code on a standard desktop computer: we therefore restrict modelling of the nonlinear regime here to the inner-duct domain.  While the emphasis in this paper is on the theoretical derivation and some proof-of-concept numerical examples to showcase possible behaviours, future work could focus on implementing the model in a more memory-efficient language than \textsc{Matlab}, or parallelising the code to run on a distributed-memory cluster, either of which would potentially make nonlinear outer-duct computations possible.

In summary, we use the numerical parameters $(\alpha^1_\mathrm{max},\alpha_\mathrm{max}^2,\eta) \!=\! (10,200,1/10)$ for the examples that follow, unless indicated otherwise.  This is shown in section \ref{sec:convergence} to give sufficiently accurate results, although higher resolutions are used in a few special cases where they are needed for accuracy, as noted in the text.

\section{The end correction}\label{sec:endcorrection}
Before we move on to giving some numerical examples using the theory above, we first consider a useful analytical application of the above analysis that allows insight to be gained into the effects of the open duct end: namely, the `end correction' problem. This is a problem in musical acoustics whereby the pressure node at the end of an open duct is not found precisely at the duct exit but rather a small distance beyond it. As a result, the length of the resonating air column differs from the physical length of the tube, and thus the instrument is out-of-tune relative to a na\"ive prediction of its pitch. Naturally, musical instrument designers (e.g.\ organ builders) are aware of this discrepancy and allow for it; however, theoretical study of the end-correction has rarely gone beyond plane waves \citep[see, e.g.][for an example with which we compare later]{snakowska2011} or dealt with nonlinearity. Here, we investigate both using the theoretical framework introduced above, although we start by reproducing the known linear plane-wave case.

\subsection{Plane wave case}
If we first consider plane waves in the linear regime, the pressure within the inner duct is given by the superposition of an incident plane wave propagating towards the exit at $s_\text{o}$ and a reflected plane wave propagating back towards the entrance with complex reflection coefficient $R$,
\begin{equation}\label{eq:endcorreqn}
    p = \Real\bigg[p_0\Big(\e^{\gamma s - \I\omega t} - |R|\e^{-\gamma s - \I\omega t + \I\arg(-R)}\Big)\bigg],
\end{equation}
with $-\I\gamma>0$ (since plane waves are always cut-on).
Assuming that a node in the pressure occurs at some  distance beyond the outlet $\hat{s}$ $ = s - s_\text{o}$, and taking the node to be a point where the incident and reflected waves are in phase and so maximally cancel, we get
\begin{align}
&\Imag\Big[\gamma\hat{s}-\I\omega t\Big] = \Imag\big[-\gamma\hat{s}-\I\omega t + \I\arg(-R)\big] &
&\Rightarrow&
    \hat{s} &= \frac{\arg(-R)}{2\Imag(\gamma)}.
\end{align}

\subsection{Linear generalisation}\label{sec:endcorrlin}
Attempting to generalise the above for non-plane-wave duct modes, we express the pressure as an expansion in the straight duct modes at a single frequency (effectively setting $a=1$ in the language of section \ref{sec:recap}):
\begin{equation}
    p = \Real\bigg[\e^{-\I\omega t}\sum_{\alpha}\psi_\alpha(\pvect{x})\left(\e^{\gamma_\alpha s}A_{\alpha}^{+} + \e^{-\gamma_\alpha s}A_{\alpha}^{-}\right)\bigg],
\end{equation}
where $\gamma_\alpha$ is the straight-duct eigenvalue $\overline{\gamma}_\alpha$ from \citet{jensenbrambley2026}, given e.g.\ in 3D by
\begin{equation}
    \gamma_\alpha = \begin{cases}
    \I\omega\sqrt{\big|1 - \lambda_\alpha^2/\omega^2R^2\big|}\quad &\Real(\omega) \geq \lambda_\alpha/R,\\
    -\omega\sqrt{\big|1 - \lambda_\alpha^2/\omega^2R^2\big|} &\Real(\omega) < \lambda_\alpha/R,
    \end{cases}
\end{equation}
and $A_\alpha^+$ and $A_\alpha^-$ are vectors of constants. The reflection matrix is given by
\begin{equation}
    \mat{R} = \left(\mat{Y} + \overline{\mat{Y}}\right)^{\!-1}\!\!\left(\overline{\mat{Y}} - \mat{Y}\right),
\end{equation}
and maps from the forward-going pressure to the backward, i.e. $\mvect{p}^- = \mat{R}\mvect{p}^+$. From the definition of the reflection matrix at the duct outlet, we may write the reflected waves in terms of the outgoing waves,
\begin{equation}
    A_\alpha^{-}\e^{-\gamma_\alpha s_\text{o}} = P_\alpha^{-}(s_\text{o}) = \sum_\beta\mat{R}_{\alpha\beta}(s_\text{o})A_\beta^{+}\e^{\gamma_\beta s_\text{o}}.
\end{equation}
Writing $\hat{s} = s - s_\text{o}$, and factoring out the $A_\beta^{+}$ vector, we have an expression equivalent to \eqref{eq:endcorreqn},
\begin{equation}
    p = \Real\bigg[\e^{-\I\omega t}\sum_{\alpha,\beta} \psi_\alpha(\pvect{x})\left(\e^{\gamma_\alpha\hat{s}}\delta_{\alpha\beta} - |\mat{R}_{\alpha\beta}(s_\text{o})|\e^{-\gamma_\alpha\hat{s} + \I\arg(-\mat{R}_{\alpha\beta}(s_\text{o}))}\right)A_\beta^{+}\e^{\gamma_\beta s_\text{o}}\bigg];
\end{equation}
since we do not have a sensible definition of the end correction for cut-off waves, we restrict ourselves to the cut-on case. Further to this, because we are comparing the arguments of entries in the reflection matrix to those of the matrix $\e^{2\gamma_\alpha\hat{s}}\delta_{\alpha\beta}$, which is diagonal, we correspondingly only consider the reflection matrix's diagonal entries. This results in a different end correction $\hat{s}_{\alpha}$ for each mode $\alpha$ that is cut-on (with $\Imag(\gamma_\alpha)\neq 0$), given by
\begin{equation}\label{eq:endcorrectiondef}
    \hat{s}_{\alpha} = \frac{\arg(-\mat{R}_{\alpha\alpha}(s_\text{o}))}{2\Imag(\gamma_\alpha)}.
\end{equation}
Given that we do not expect the oscillations in a musical instrument to be monomodal, but rather to consist of a series of harmonics, a set of modal end corrections will be informative as to how each harmonic's pitch will be affected according to the reflective properties of the instrument's outlet. This in turn will affect the timbre of the noise produced.

\subsection{Nonlinear generalisation}\label{sec:nonlin-end-theory}
It is not obvious how best to introduce nonlinearity to the end-correction problem. The reflection matrix $\mat{R}$ may be generalised to weak nonlinearity exactly as for the admittance or impedance, giving $\mvect{p}^- = \mat{R}\mvect{p}^+ + \mathcal{R}\langle\mvect{p}^+,\mvect{p}^+\rangle$, but this reflects incoming modes into \emph{different} reflected modes, and so does not \emph{directly} affect the end correction; instead, these other reflected modes feed back on the initial mode through weakly-nonlinear propagation along the inner duct.  Here, we consider a different nonlinear correction to the reflection matrix that includes both nonlinear scattering at the open end and nonlinear interactions within the duct; this alternative nonlinear reflection matrix then facilitates the calculation of a nonlinear end correction.
The linear reflection matrix at the open end, as given above in terms of the admittance, may also be obtained directly from the acoustic pressure in the duct if a single duct mode $(a,\alpha)$ is excited at the inlet, so that $P_\beta^{b+} = A_\alpha^{a+}\delta_{\alpha\beta}\delta^{ab}$. We take $a = 1$ in order to see how a source at the fundamental excites higher harmonics. In this case, at $s = s_\mathrm{o}$, we have
\begin{equation}
    \frac{\sum_{b = 1}^\infty\sum_\beta (P_\beta^{b+})^*P_\beta^{b-}}{\sum_{b = 1}^\infty\sum_\beta |P_\beta^{b+}|^2} = \frac{\sum_{b = 1}^\infty\sum_{\beta,\gamma} (P_\beta^{1+})^*\mat{R}_{\beta\gamma}^bP_\gamma^{1+}}{\sum_{b = 1}^\infty\sum_\beta |P_\beta^{1+}|^2} +O(M) = \mat{R}^1_{\alpha\alpha} + O(M),
\end{equation}
where `$*$' denotes a complex conjugate and the nonlinear reflection tensor $\mathcal{R}$ is contained within the $O(M)$ term.  We may therefore define a matricial reflection coefficient $\Tilde{\mat{R}}$ that includes both the nonlinear propagation in the duct and the nonlinear reflection at the open end by taking
\begin{align}\label{equ:nonlin-end-corr}&
    \tilde{\mat{R}}_{\alpha\alpha}^1 := \frac{\sum_{b = 1}^\infty\sum_\beta (P_\beta^{b+})^*P_\beta^{b-}}{\sum_{b = 1}^\infty\sum_\beta |P_\beta^{b+}|^2},
    &&\Rightarrow&&
    \tilde{s}_{\alpha} = \frac{\arg(-\tilde{\mat{R}}_{\alpha\alpha}^1)}{2\Imag(\gamma_\alpha^1)}.
\end{align}
where $\tilde{s}_\alpha$ is a modified, nonlinear end correction. Note that this definition only holds if the fundamental $a = 1$ is the only mode excited at the inlet.

An important difference between this calculation and the linear one is that the length and geometry of the duct now matter (since these affect the pressure at the outlet). The amount of steepening that a wave can undergo within the duct will influence its behaviour at the outlet, so the previously one-dimensional parameter space (frequency only) is now three-dimensional (frequency, Mach number and duct length). We choose to fix the frequency and study the two-dimensional relationship between Mach number and duct length for the fundamental frequency ($a=1$) plane-wave ($\alpha = 0$) coefficient $\tilde{s}_0$; the two other combinations and higher mode entries could be considered in future work. Results from this calculation will be shown in section \ref{sec:nonlinendcorr}.

\section{Results}\label{sec:results}
We now make use of the framework described above to compute some numerical examples.  These examples are intended to demonstrate a range of different behaviours.  While one motivation for this work was investigating the acoustics of musical instruments, we instead concentrate here on basic examples that aid fundamental understanding, which may or may not be relevant to musical instruments.  A thorough analysis of musical instruments in particular is left to future studies.

\subsection{Prediction of the linear end-correction coefficient}\label{sec:linendcorr}

We first consider the exit condition in isolation, without needing to solve for the acoustics either inside or outside the duct.  We do this by investigating the end correction, introduced in section~\ref{sec:endcorrection}, beginning with the linear case.  This allows comparison with existing methods, and in particular, with the Wiener--Hopf  solution~\citep{noble}, a derivation of which is given in appendix~\ref{sec:wienerhopf}.

\subsubsection{The end correction in 3D}\label{sec:3Dlinendcorr}
\begin{figure}
    \centering
    \includegraphics{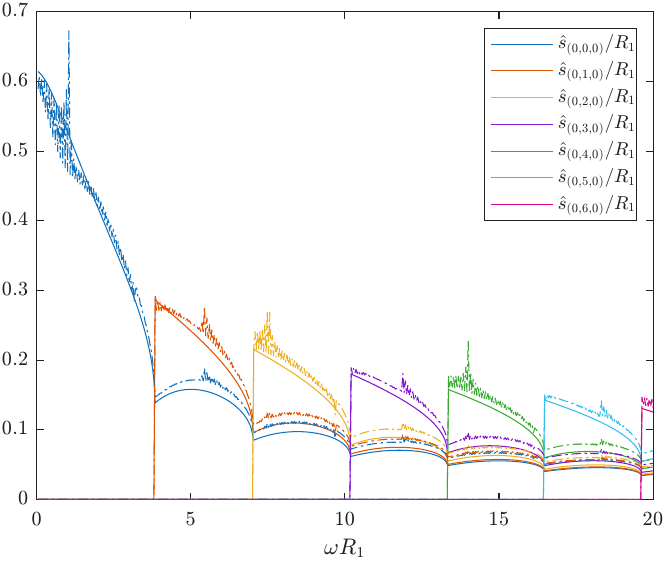}
    \caption{End-correction coefficients for a 3D duct, calculated using the Wiener--Hopf technique (solid line) and our outlet condition (dash-dotted line), for $m_\alpha=0$; 8 modes were used in the inner duct and 1200 in the outer duct, with a duct width ratio of 1/40.}
    \label{3Dendcorrectionplot.pdf}
\end{figure}

Figure~\ref{3Dendcorrectionplot.pdf}
plots the $m_\alpha = 0$ end-correction coefficients $\hat{s}_{(0,n_\alpha,\xi_\alpha)}$ for various modes as a function of nondimensionalised frequency (the Helmholtz number, $\omega R_1$) for a 3D duct exit. Here we recover the classical end correction distance of $\hat{s}_0 = \hat{s}_{(0,0,0)}$ $\approx 0.6R_1$ at low frequencies,  and then a sharp reduction as the frequency is increased before new modes cut on (in agreement with \citealt[][figure~1]{snakowska2011}, and \citealt[][figure~4]{felix2018modeling}). In contrast to \citet{snakowska2011}, modes are seen to cut-on with a nonzero end-correction, before eventually, as the frequency increases further, collapsing onto the mode-zero curve; having verified that the Wiener--Hopf solution and our own method agree on this, we conclude that this is physical and not a numerical artefact (a conclusion which also agrees with the 2D case below). It should not be concluded from this that continuous variation of $\omega R_1$ results in discontinuous variation of $\hat{s}_\alpha/R_1$: if the definition of $\hat{s}_\alpha/R_1$ could be analytically extended to account for evanescent modes, continuous results would be seen. Future work on the end correction could aim towards this.

Convergence will be discussed further in the 2D case below; taking inspiration from it in the meantime, results here are produced with a width ratio of $\eta = 1/40$, with 8 modes used in the inner duct. As the modal basis in 3D is more complicated than in 2D (and requires lots of numerical integration of Bessel function products) only $1200$ modes were used in the outer duct: a valid mode ratio, given that $\eta = 1/40$ is still considerably larger than $8/1200$. This proves a reasonable parameter choice, with relatively good agreement between the numerics and the Wiener--Hopf solution, and with the numerical solution's oscillations being of lower amplitude than for a more moderate modenumber ratio.

\subsubsection{The end correction in 2D}\label{sec:2Dlinendcorr}
\begin{figure}
    \centering
    \includegraphics{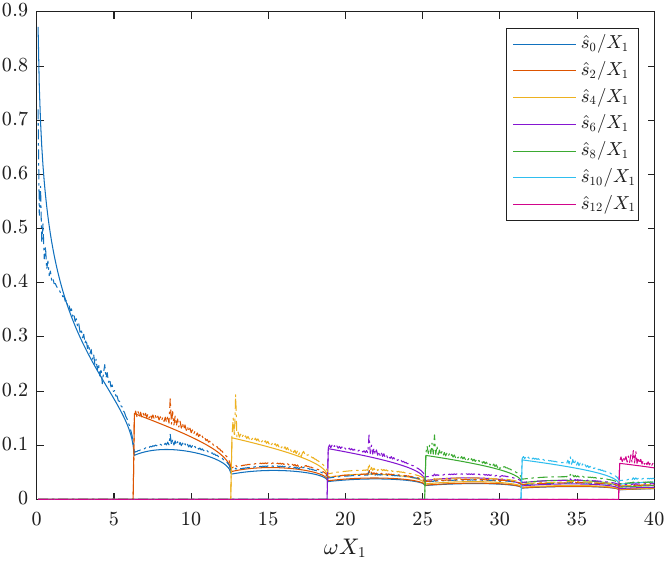}
    \caption{End-correction coefficients for a 2D duct with a symmetric source, calculated using the Wiener--Hopf technique (solid line) and our outlet condition (dash-dotted line). 13 modes were used in the inner duct and 2000 in the outer duct, with a width ratio of 1/40.}
    \label{2Dendcorrectionplot.pdf}
\end{figure}

Figure~\ref{2Dendcorrectionplot.pdf}
plots the end correction for various modes as a function of the Helmholtz number $\omega X_1$ for a 2D duct with a symmetric source. The end correction decays with increasing Helmholtz number, and its coefficients exhibit a cusp every time a new duct mode cuts on; this is again in agreement with the analytical Wiener--Hopf solution (also plotted in figure~\ref{2Dendcorrectionplot.pdf}) and also in qualitative agreement with the 3D results presented above. Here, however, in contrast to the well-known low-frequency end correction figure of $\hat{s}_0$ $\approx 0.6$$R_1$ for 3D ducts \citep{rayleigh1871}, in 2D the end correction exhibits a logarithmic singularity at low frequencies, and so is unbounded as the frequency tends to zero, or equivalently, as the inner-duct width $X_1$ tends to zero. This is not an indication that the end correction \emph{itself} diverges in practice, but rather that its convergence is sublinear with the inner-duct width $X_1$, and so it diverges when plotted using the nondimensional scaling used here. For more detail, see appendix~\ref{sec:wienerhopf}, equation~\eqref{eq:nobleexpression}.

This example is useful for validating the above modelling and its numerical implementation, at least in the linear regime, and is also useful as a yardstick for parameter choices later on. Here, at least 13 spatial modes were required in the inner duct in order to show which will go from cut-off to cut-on for the chosen frequency range. A reasonable, not-too-computationally-intensive choice for the outer duct is to use $\alpha_\text{max}^2 = 2000$ modes, since the 2000 modes will only be used for calculating admittances at one location, and we do not need to solve 2000 coupled ODEs for any propagation away from this location. According to the $\eta \geq \alpha_\text{max}^1/\alpha_\text{max}^2$ rule, this makes the most extreme possible width ratio $\eta = 13/2000 (\approx 0.0065)$. Using this value, however, our calculation noticeably overshoots the Wiener--Hopf solution to a large degree, and so we will choose our parameters more cautiously. The other extreme is when the inner and outer ducts are too similar in size and there is too much reflection from the outer duct walls. For the end-correction this manifests as large oscillations as the frequency is increased, due to modes cutting on in the outer duct. We can calculate the wavelength of the oscillations as follows: for the fundamental frequency $a = 1$, the cut-on frequency of mode $\alpha$ in the outer duct is $\alpha\pi/X_2 = \eta\alpha\pi/X_1$, so in figure \ref{2Dendcorrectionplot.pdf}, where only even modes are excited, a new outer-duct mode cuts on each time the frequency is increased by $2\eta\pi$ ($\approx 0.16$ in figure \ref{2Dendcorrectionplot.pdf}).  The amplitude of the oscillations seems to increase as this frequency increment increases, and presents a problem even for ratios as low as $\eta = 1/20$; thus, we opt for a lower ratio of $\eta = 1/40$, which is seen in figures~\ref{3Dendcorrectionplot.pdf} and~\ref{2Dendcorrectionplot.pdf} to give reasonable results.

\subsubsection{The end correction as a convergence study}\label{sec:convergence}
\begin{figure}
    \centering
    \includegraphics[width=0.7\textwidth]{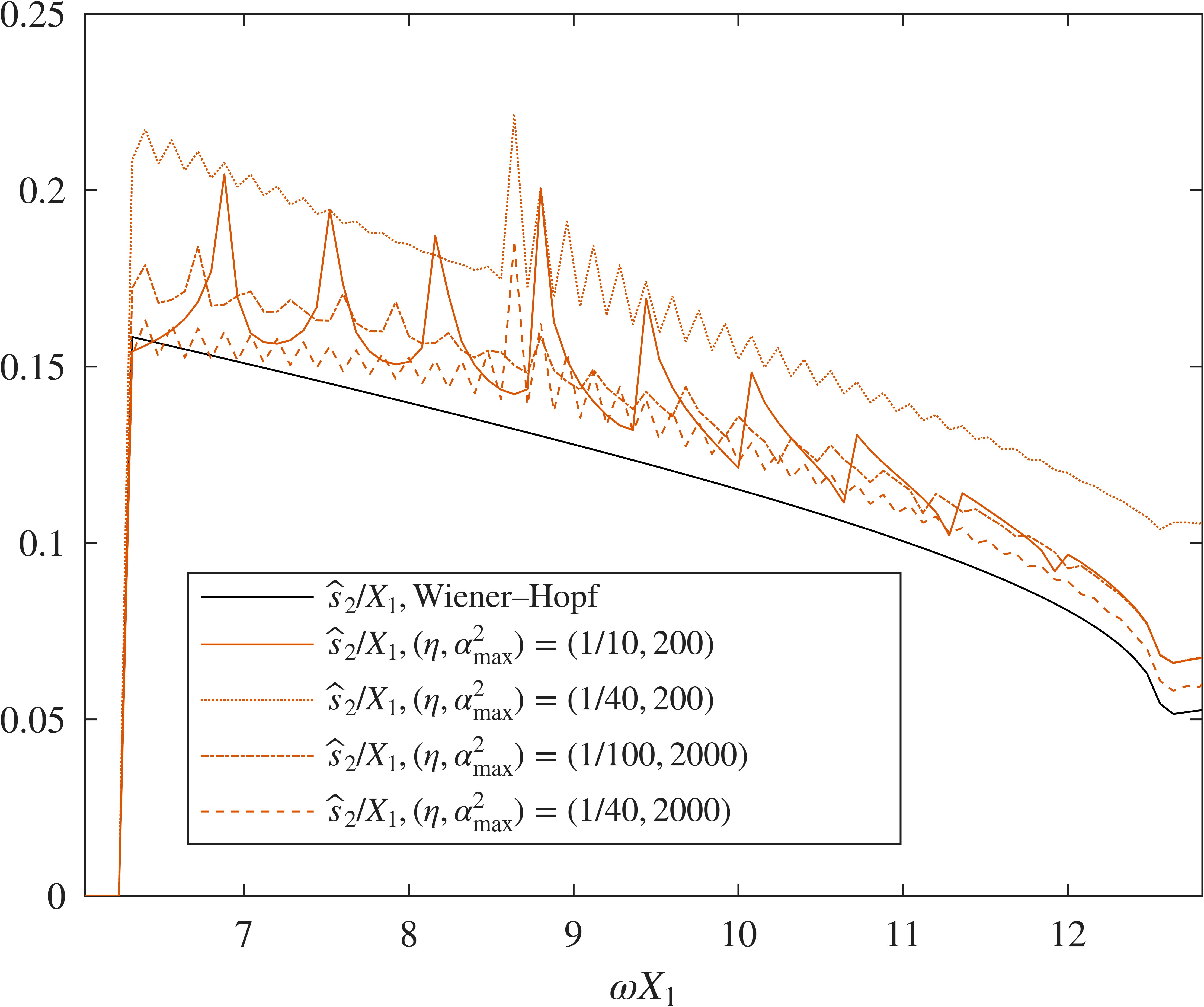}
    \caption{Comparison of different numerical parameter choices for a segment of figure \ref{2Dendcorrectionplot.pdf} between $2\pi$ and $4\pi$, with only the $\alpha^1 = 2$ mode shown. The Wiener--Hopf solution is plotted in black, and the original parameter choice $(\eta,\alpha^2_\mathrm{max}) = (1/40,2000)$ from figure \ref{2Dendcorrectionplot.pdf} is shown plotted with  an orange dashed line as before. All of these take $\alpha_\mathrm{max}^1 = 10$.}
    \label{endcorrectionconvergenceplot.pdf}
\end{figure}

To demonstrate the appropriateness of our choice of numerical parameters, we plot the end-correction coefficient as calculated for various values of $\alpha_\mathrm{max}^2$ and $\eta$ in figure \ref{endcorrectionconvergenceplot.pdf},
alongside the Wiener--Hopf solution in black. This is a zoomed-in section of figure \ref{2Dendcorrectionplot.pdf}, focussing only on Helmholtz numbers between the $\alpha = 2$ mode's cut-on and that of the next mode up (i.e. between $2\pi$ and $4\pi$).

We see that the option we chose ($\eta = 1/40$ and $\alpha_\text{max}^2 = 2000$, shown by the dashed  orange line) is the closest to the Wiener--Hopf solution. We also see that if we restrict $\alpha_\mathrm{max}$ to $200$ (while keeping $\alpha_\mathrm{max}^1 = 10$), the $\eta = 1/10$ option (solid,  orange), while oscillating much more than the $\eta = 1/40$ option (dotted,  orange), actually has an average value much closer to the Wiener--Hopf one, and never exceeds the $\eta = 1/40$ value in its inaccuracy, thus making more moderate values of $\eta$ a better choice for more moderate modenumber ratios. This is because smaller values of $\eta$ correspond to a smaller duct exit on the scale of the outer duct width, and therefore require higher resolution in the outer duct to correctly resolve. This phenomenon is also demonstrated when comparing values of $\eta$ for fixed $\alpha_\mathrm{max}^2 = 2000$: the $\eta = 1/40$ line (dashed,  orange) from figure \ref{2Dendcorrectionplot.pdf} is consistently closer to the Wiener--Hopf value than the $\eta = 1/100$ line (dot-dashed,  orange), proving the same point.

Aside from considering duct width ratios, another consideration for the outer duct is the condition on its walls. Instead of the current reflecting condition (`sound-hard') one could consider an alternative `pressure-release'  boundary. This method is tested in appendix~\ref{sec:soundsoftbc}, where it is shown that the sound-hard boundary condition provides a better match with Wiener-Hopf for the same optimised parameters as above. With this in mind, we continue with our choice of a sound-hard boundary.

\subsection{The generalised nonlinear end-correction coefficient}\label{sec:nonlinendcorr}
Since the analysis here involves not only linear but also weakly-nonlinear effects, we may investigate the effect of changing the amplitude on the position of the end correction using the results of section~\ref{sec:nonlin-end-theory}.  Two examples are provided here: a low-frequency example in 3D and an intermediate-frequency example in 2D.

\subsubsection{Low frequency in 3D}\label{sec:nonlinendcorr3D}
\begin{figure}%
    \centering%
    \includegraphics{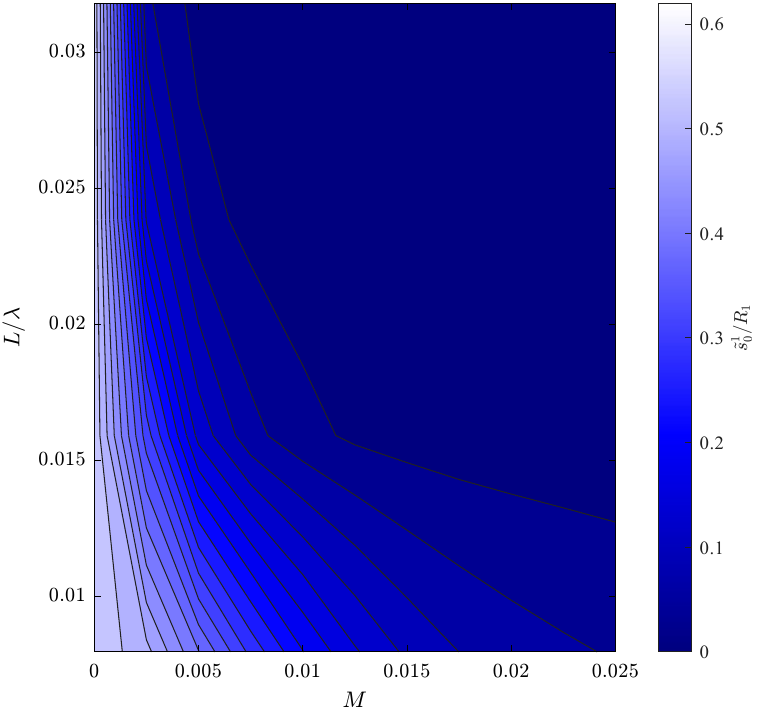}%
    \caption{The nonlinear generalised end-correction coefficient $\tilde{s}_0/R_1$ for fixed Helmholtz number $\omega R_1 = 0.05$ and a range of duct lengths (scaled by source wavelength $\lambda = 2\pi/\omega$) and Mach numbers, in 3D. Nonlinearity is shown to have a dampening effect on the discrepancy between duct length and resonant length for low frequencies, and the longer the duct the stronger this effect will be. 8 modes were used in the inner duct and 200 in the outer duct, with 10 temporal modes and a width ratio of 1/10.}%
    \label{3Ds_hat_machplot.pdf}%
\end{figure}

Figure \ref{3Ds_hat_machplot.pdf}
shows values of the generalised nonlinear end-correction coefficient $\tilde{s}_0/R_1$ (given by equation~\ref{equ:nonlin-end-corr}) for a fixed low frequency of $\omega R_1 = 0.05$, while the duct length and Mach number are varied.
In order to capture the effects of nonlinearity 10 temporal modes were used; for an accurate calculation in the inner duct this was paired with $\alpha_\text{max}^1 = 8$. Nonlinearity also requires third-rank tensors, which in the outer duct can become expensive memory-wise: as such, we used only 200 modes in the outer duct, making $\eta = 1/10$ a good intermediate width ratio choice (although the linear value of $\tilde{s}_0 \approx$ $0.6R_1$ is slightly undershot).
We see that in the linear case, the duct length plays no role as expected; once the Mach number is increased, however, the end correction quickly decays to zero. For longer ducts, the wave has steepened more by the outlet, so zero is reached for smaller Mach number. Loosely, this suggests that if an instrument is played close to its resonant frequency, higher volume means a shorter resonating length, i.e. higher frequency and a correspondingly sharper pitch.
 
\subsubsection{Intermediate frequency in 2D}\label{sec:nonlinendcorr2D}
\begin{figure}
    \centering
    \includegraphics{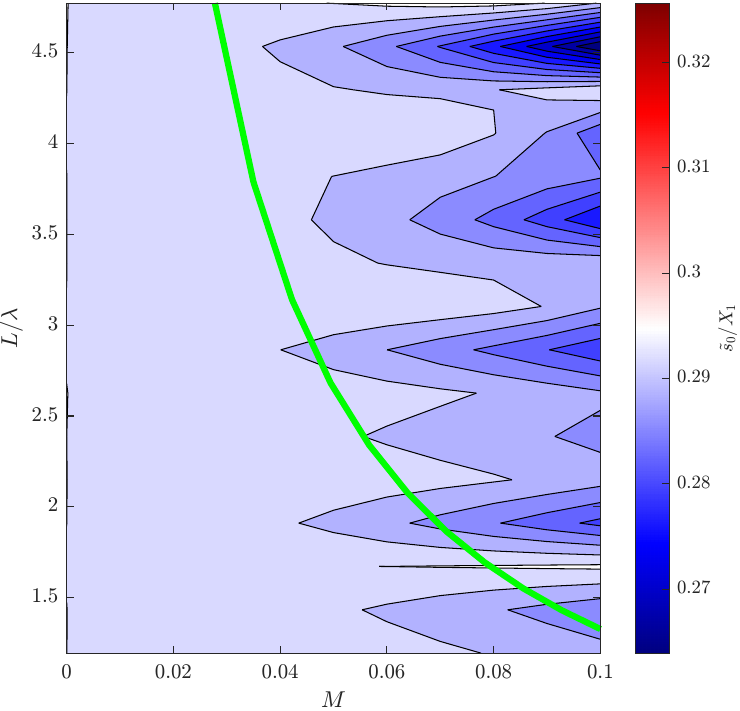}
    \caption{The nonlinear generalised end-correction coefficient $\tilde{s}_0/X_1$ for fixed Helmholtz number $\omega X_1 = 3$ and a range of duct lengths (scaled by source wavelength $\lambda = 2\pi/\omega$) and Mach numbers, in 2D. We see that for an intermediate frequency, the end correction remains the same or decreases depending on the duct length. 8 modes were used in the inner duct and 200 in the outer duct, with 10 temporal modes and a width ratio of 1/10. The shock formation distance, given by $L_\text{sf} = 1/M\beta_0\omega$, for $\beta_0$ the coefficient of nonlinearity, is shown in green (and has also been normalised by $X_1$ here).} 
    \label{2Ds_hat_machplot.pdf}
\end{figure}

Figure \ref{2Ds_hat_machplot.pdf}
shows values of the generalised nonlinear end-correction coefficient $\tilde{s}_0$ for a fixed intermediate frequency of $\omega X_1 = 3$ while the duct length and Mach number are varied. As with the previous example, here $(\alpha_\text{max}^1,\alpha_\text{max}^2,\eta) = (8,200,1/10)$. Once again, the linear value is length-independent, but this time as the Mach number is increased the end correction decays for certain duct lengths and remains constant for others. For ducts as long as three wavelengths or more, we start to see that the end correction decreases more sharply if the duct length is an odd multiple of a half-wavelength, i.e. there is a pressure node at one end and an antinode at the other. This is consistent with the fact that when a sine wave steepens into a sawtooth, the  antinodes get much closer to the half-wavelength nodes, while remaining far away from the full-wavelength nodes: thus, a small deviation from the former position results in vast pressure variation (something that might drive resistance to the end-correction phenomenon) whereas a deviation from the latter position results in roughly linear-acoustic behaviour.

\subsection{Resonances in a finite-length duct in 3D}\label{sec:resonances}

We next consider the effect of the open duct exit on the resonances of the inner duct. There are multiple ways to calculate the resonances in a duct of finite length. One method is to consider the impedance at the duct inlet. Physical instruments can have either one or two openings (described as `closed' or `open' ducts respectively). If the duct is `closed', we expect a minimum in the velocity at the inlet, corresponding to a very high impedance. Conversely, if it is open, the free-space condition at the inlet will result in a pressure minimum, i.e. a low impedance. Here, we follow~\citet[][\S4.5.1]{mctavphd} and define a scalar impedance using the rms pressure and velocity across all modes that is applicable in both linear and nonlinear regimes: 
\begin{equation}\label{equ:scalar-impedance}
    |Z_\mathrm{rms}| = \sqrt{\frac{\sum_{a = -\infty}^\infty\sum_{\alpha = 0}^\infty|P_\alpha^a|^2}{\sum_{a = -\infty}^\infty\sum_{\alpha = 0}^\infty|U_\alpha^a|^2}}.
\end{equation}
We calculate $Z_\mathrm{rms}$ at the inlet for a range of frequencies, driven by a plane-wave source at the same location.  Using this, we can infer information about the tone and harmonic series of the duct in question.

\begin{figure}
    \centering
    \includegraphics{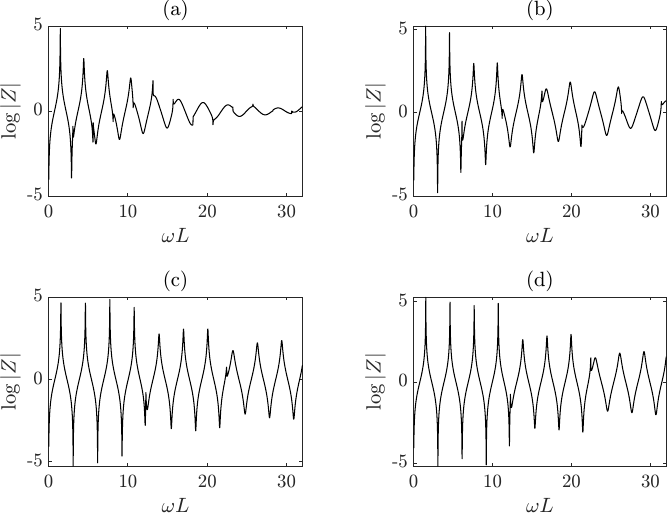}
    \caption{Inlet scalar impedance ($Z_\mathrm{rms}$ in equation~\ref{equ:scalar-impedance}) for a range of frequencies in 3D, for (a) a straight duct of aspect ratio 1/4; (b) a straight duct of aspect ratio 1/8; (c) a straight duct of aspect ratio 1/16; and (d) a duct of aspect ratio 1/16 undergoing a constant-curvature bend of angle $\pi/2$.
    8 modes were used in the inner duct, 1200 in the outer duct, and the width ratio was 1/40.}
    \label{inletimpedancetiles.pdf}
\end{figure}

\begin{figure}
    \centering
    \includegraphics{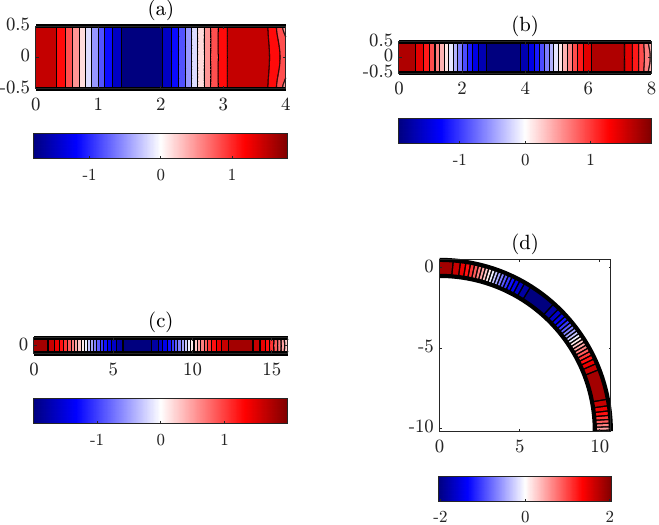}
    \caption{Linear acoustic field within a duct opening into another duct 10 times larger, in 3D, for (a) a straight duct of aspect ratio 1/4; (b) a straight duct of aspect ratio 1/8; (c) a straight duct of aspect ratio 1/16; and (d) a duct of aspect ratio 1/16 undergoing a bend of angle $\pi/2$. Each duct has a source at the inlet of frequency equal to that of the duct in question's third closed resonance (as calculated in figure \ref{inletimpedancetiles.pdf}). 8 modes were used in the inner duct, 1200 in the outer duct, and the width ratio was 1/40. An animation of this figure is given as Movie~1 in the supplementary material.}
    \label{thirdresplot.pdf}
\end{figure}

In figure \ref{inletimpedancetiles.pdf},
these calculations are performed for 3D ducts of varying aspect ratio (see figure \ref{thirdresplot.pdf} for the geometries in question), taking $(\alpha_\text{max}^1,\alpha_\text{max}^2,\eta) = (8,1200,1/40)$. The maxima occur (roughly) at odd-integer frequency multiples of the first maximum (or \emph{closed-duct fundamental}), consistent with the harmonic series of a `closed' instrument admitting oscillations of a quarter-wavelength and all odd multiples thereof. The minima, meanwhile, occur (roughly) at integer multiples of the first minimum (or \emph{open-duct fundamental)}, consistent with the harmonic series of an `open' instrument admiting integer multiples of a half-wavelength.

Also notable are the variations in the magnitude of the maxima/minima across different aspect ratios. More pronounced maxima/minima occur for more extreme aspect ratios; this agrees with the fact that broad organ pipes have a tone much more fundamental-dominated than slender ones. Between plots (c) and (d) in figure \ref{inletimpedancetiles.pdf} we are able to assess the difference that a bent duct makes: the slightly split resonance at the 8th peak in (c) is much more clearly split in (d), and peaks beyond that point have lower amplitude in (d). Figure \ref{thirdresplot.pdf}
shows the acoustics within each duct at the third closed resonant frequency in each case (corresponding to the third maximum in figure~\ref{inletimpedancetiles.pdf}). It is notable that the largely plane-wave oscillations are disrupted only at the outlet, where particularly in (a) the characteristic `cap' causing the end-correction effect may be seen.

\subsection{Linear radiation from an acoustic source}\label{sec:2Dsource}

\begin{figure}%
    \centering%
    \includegraphics{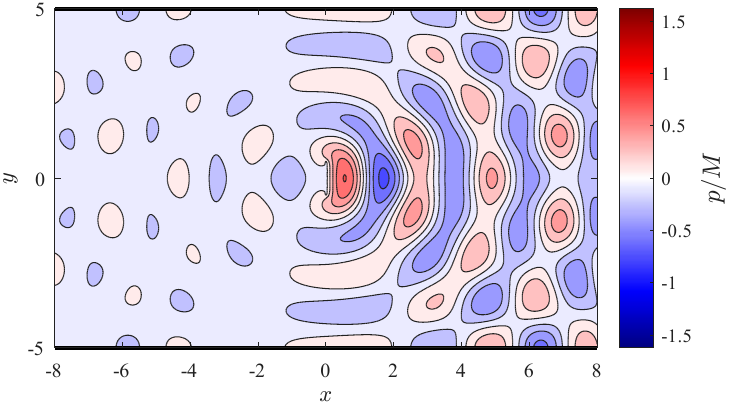}%
    \caption{Linear acoustic field emanating from a 1D acoustic source into a 2D duct 10 times larger. 10 modes were used in the inner calculation and 200 modes in the outer, with $\omega X_1 = 3$ for $X_1$ the inner width. An animation of this figure is given as Movie~2 in the supplementary material.}%
    \label{2Dsourceplot.pdf}%
    \vspace{-1ex}%
\end{figure}

Having until now considered the admittance in the inner duct without calculating in the outer duct, here we calculate the acoustic field in the outer duct without considering the inner duct's acoustics. The pressure field in the outer duct due to a plane-wave sound source located at the inner-duct exit in 2D is plotted in figure \ref{2Dsourceplot.pdf}.
In order to obey the conditions mentioned in section \ref{sec:2Dlinendcorr} while also having few enough modes to allow for solution of the governing equations in the outer duct, this figure was generated using $(\alpha_\text{max}^1,\alpha_\text{max}^2,\eta) = (10,200,1/10)$.
We see that the absorbing boundary condition on the `acoustic wormhole' surface has resulted in radiation travelling predominantly rightward from the outlet: this suggests that the absorbing condition is a reasonable workaround for the problem of the intersecting inner-duct and outer-duct geometries.  The effect of the hard outer-duct walls is clearly present to the right of the source, giving the typical interference pattern, which is also the cause of the rapid oscillations in the end correction seen in figures~\ref{3Dendcorrectionplot.pdf} and~\ref{2Dendcorrectionplot.pdf}, although the expected circular wave-fronts emanating from the source are clearly visible despite this.

\subsection{Linear radiation from a straight duct}\label{sec:straightduct}
\begin{figure}%
    \centering%
    \includegraphics{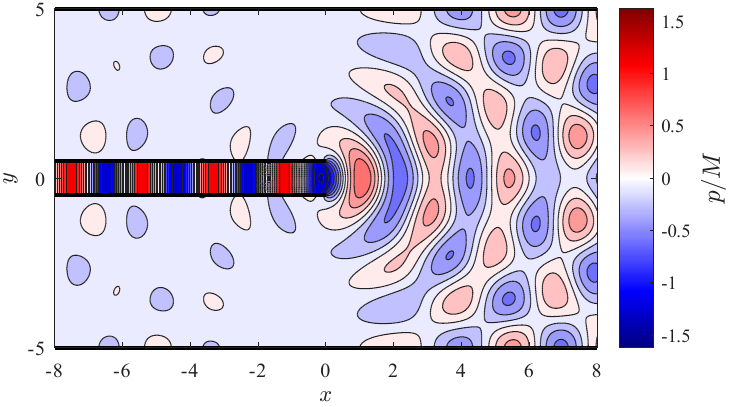}%
    \caption{Linear acoustic field emanating from a straight duct into a duct 10 times larger, in 2D. 10 modes were used in the inner calculation and 200 modes in the outer, with $\omega X_1 = 3$ for $X_1$ the inner width. An animation of this figure is given as Movie~3 in the supplementary material.}%
    \label{2Dstraightductplot.pdf}%
    \vspace{-1ex}%
\end{figure}

\begin{figure}
    \centering
    \includegraphics{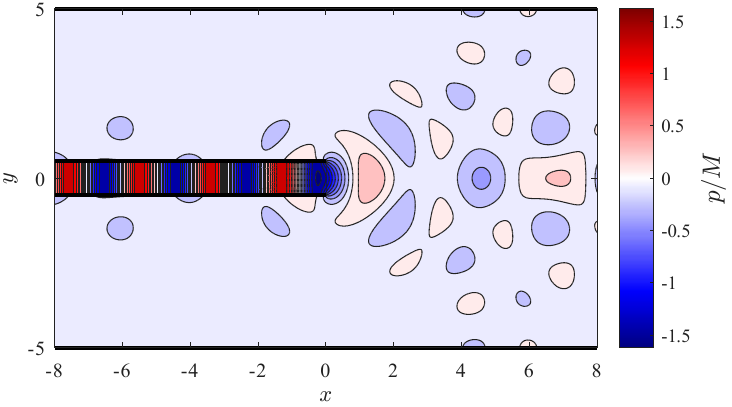}
    \caption{Linear acoustic field emanating from a straight duct into a duct 10 times larger, in 3D. 10 modes were used in the inner calculation and 200 modes in the outer, with $\omega R_1 = 3/2$ for $R_1$ the inner radius.  An animation of this figure is given as Movie~4 in the supplementary material.}
    \label{3Dstraightductplot.pdf}
\end{figure}

In contrast to the previous section, we can also choose to calculate the inner-duct acoustic field to see how sound generated by a plane-wave source within the duct subsequently exits from the duct outlet. This is shown in 2D in figure \ref{2Dstraightductplot.pdf},
and in 3D in figure \ref{3Dstraightductplot.pdf},
both with the same width ratios and numbers of modes as the previous example in figure~\ref{2Dsourceplot.pdf}.

Comparing the 2D duct (figure~\ref{2Dstraightductplot.pdf}) with a 2D source (figure~\ref{2Dsourceplot.pdf}), we unsurprisingly see great similarity. Comparing the 2D duct (figure~\ref{2Dstraightductplot.pdf}) with the 3D duct (figure~\ref{3Dstraightductplot.pdf}), it is clear that, compared with the duct interior, the radiation from the outlet in 3D has a much lower amplitude than in 2D, as the theoretical decay rate in unbounded 3D is $1/r$ with distance from the exit while it is $1/\sqrt{x}$ in 2D. However, the acoustics within the inner ducts appear to behave similarly between the two cases.

\subsection{Nonlinear acoustics within an open straight duct in 2D}\label{sec:2Dnonlin}
\begin{figure}
    \centering
    \includegraphics{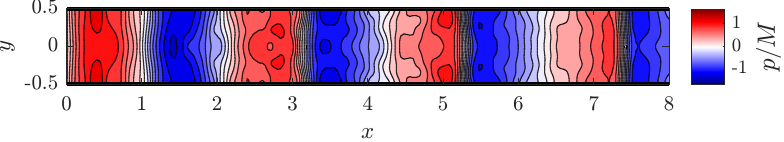}
    \caption{The nonlinear acoustic field within a straight duct, with a perturbation Mach number of $M = 0.10$, opening into a duct 10 times larger (outer duct not plotted), in 2D. 10 modes were used in the inner calculation and 200 modes in the outer, with $\omega X_1 = 3$ for $X_1$ the inner radius. An animation of this figure is given as Movie~5 in the supplementary material.}
    \label{2Dnonlin.pdf}
\end{figure}

We can also calculate the acoustic field in a straight duct with an open end in the nonlinear regime, shown in the 2D case in figure~\ref{2Dnonlin.pdf}
for a perturbation Mach number of $M=0.10$. This is calculated using the same values of $(\alpha_\text{max}^1,\alpha_\text{max}^2,\eta) = (10,200,1/10)$ as above, with $a_\text{max} = 10$.
Extending this calculation to the outer duct is computationally costly, due to the vastly greater number of spatial modes required, but we can still apply the radiation condition and see what effect it has on the acoustics within the inner duct without incurring this extra expense. An immediately noticeable physical effect of the open end that is not achieved with the idealised characteristic boundary condition from \citet{jensenbrambley2026} is that non-plane waves are excited: since the excitation at the inlet is purely planar, this nonplanarity is due to the geometry of the outlet, and nonlinear reflections back from it. In the linear case, the non-plane reflections are not significant, as seen in figure \ref{2Dstraightductplot.pdf}, but here they are very clearly visible.

\subsection{Larger-amplitude radiation from a straight duct}\label{sec:nonlinearstraight}

While the nonlinear in-duct numerics detailed earlier could also be used to calculate the weakly-nonlinear acoustic propagation outside the inner duct in theory, in practice the coupling between the modes needed in the outer duct (of which there is a large number) makes this numerically impractical.
One way to lessen the expense of the outer-duct calculation for larger-amplitude radiation is to consider only linear propagation in the outer duct, with all nonlinear effects therefore due to a steepened nonlinear outlet signal from the inner duct. This is an approximation, but due to the rapid dispersion of energy across the outer-duct domain (as would also occur in free space) as sound moves away from the outlet, linear behaviour is expected once a short distance has been covered by the sound. Given that nonlinearity has little room to develop within this short distance, a wholly linear outer-duct calculation may then be expected to approximate a weakly-nonlinear alternative fairly well. Note that this is just an approximation of the plotting in the outer duct: because we only make this approximation \textit{after} having solved the weakly-nonlinear inner-duct problem in its entirety, the quality of the weakly-nonlinear inner-duct solution is not compromised. This enables visualizing the far-field radiation pattern, and provides a useful check on the outer-duct method; in particular, it shows any reflection from the outer duct wall, and any lack of reflection from the absence of the inner duct.

\begin{figure}
    \centering
    \includegraphics{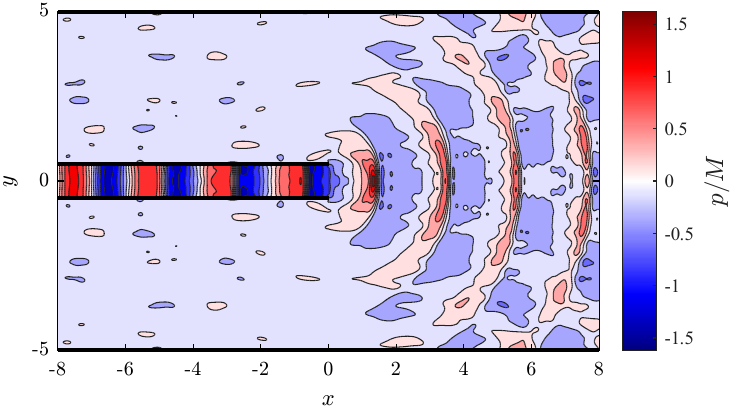}
    \caption{Nonlinear acoustic field emanating from a straight duct into a linear-regime duct 10 times larger, in 2D. 10 spatial modes were used in the inner calculation and 200 spatial modes in the outer, with $\omega X_1 = 3$ for $X_1$ the inner width. 7 temporal modes were used for perturbation Mach number $M = 0.05$. 
    An animation of this figure is given as Movie~6 in the supplementary material.
    }
    \label{2Dstraightstill.pdf}
\end{figure}

While this approximation does not reduce the number of equations to solve for the pressure, it does decouple them, which is then computationally tractable provided the number of temporal modes is reduced from 10 to 7. The result of this calculation is shown in figure~\ref{2Dstraightstill.pdf},
where the shock front is seen to widen and disperse as it exits the inner duct.

\begin{figure}
    \centering
    \includegraphics{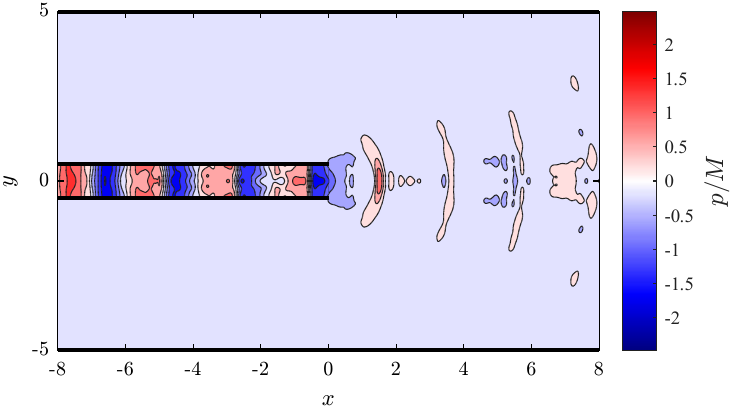}
    \caption{Nonlinear acoustic field emanating from a straight duct into a linear-regime duct 10 times larger, in 3D. 10 spatial modes were used in the inner calculation and 200 spatial modes in the outer, with $\omega X_1 = 3$ for $X_1$ the inner width. 7 temporal modes were used for perturbation Mach number $M = 0.05$. 
    An animation of this figure is given as Movie~7 in the supplementary material.
    }
    \label{3Dstraightstill.pdf}
\end{figure}

We can once more do the same thing in 3D for comparison. Figure \ref{3Dstraightstill.pdf}
shows the result (for the same parameters as the previous plot) in which we notice one major qualitative similarity with the linear case and one major qualitative difference. Similarly to the linear case, we see radiation of a much lower amplitude in the outer duct, consistent with the theoretical-decay-rate argument made in section~\ref{sec:straightduct}. However, in contrast to the almost-identical 2D and 3D inner-duct fields of the linear case, very different inner-duct fields are seen here between 2D and 3D. This is because most linear reflection from the open end is planar (with a small nonplanar amount that decays rapidly from the outlet towards the inlet) whereas the nonlinear reflection is highly nonplanar, meaning that the differences of propagation between the circular symmetry of a sliced cylindrical 3D domain and the linear symmetry of a rectangular 2D domain are much more easily visible.

\subsection{Linear radiation from a curved duct in 2D}\label{sec:2Dcurved}
\begin{figure}
    \centering
    \includegraphics{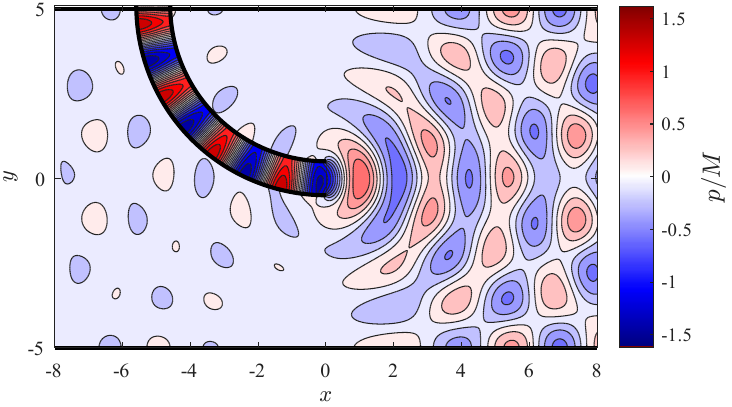}
    \caption{Linear acoustic field emanating from a curved duct into a duct 10 times larger, in 2D. 10 modes were used in the inner calculation and 200 modes in the outer, with $\omega X_1 = 3$ for $X_1$ the inner radius.  An animation of this figure is given as Movie~8 in the supplementary material.}
    \label{2Dplanarbendplot.pdf}
\end{figure}

The next case we consider is that of radiation from a curved duct into a larger straight duct, in 2D, shown in figure~\ref{2Dplanarbendplot.pdf}.
As before, this example was computed using the parameters $(\alpha_\text{max}^1,\alpha_\text{max}^2,\eta) = (10,200,1/10)$.
We can see that despite the asymmetry of the source, the radiation from the end is almost symmetric across the midline. Since symmetric and antisymmetric waves are completely uncoupled in the outer duct, this suggests that the radiation from the inner duct is predominantly symmetric, with any antisymmetric inner-duct modes being almost totally reflected from the open end.
This can be verified by examining the reflection matrix, which for this example has a dominant symmetric entry $\mat{R}_{00} = 0.053 - 0.080\I$, and a dominant antisymmetric entry $\mat{R}_{11} = -0.10 - 0.58\I$. The latter has modulus more than six times that of the former, meaning that reflections from the duct exit are indeed predominantly antisymmetric.

\subsection{Larger-amplitude radiation from a curved duct in 2D}\label{sec:nonlinearcurved}
\begin{figure}
    \centering
    \includegraphics{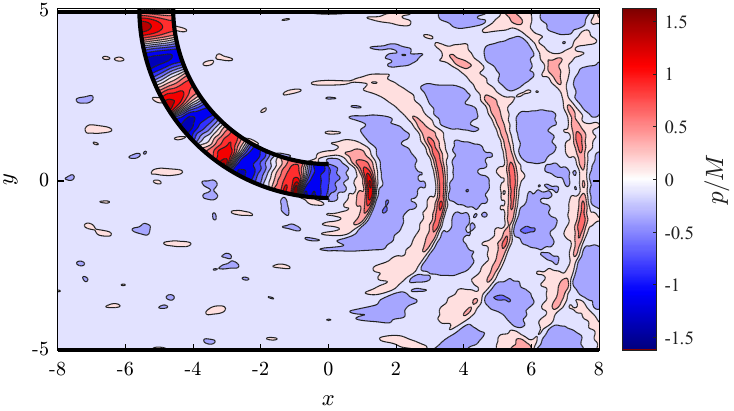}
    \caption{Nonlinear acoustic field emanating from a curved duct into a linear-regime duct 10 times larger, in 2D. 10 spatial modes were used in the inner calculation and 200 spatial modes in the outer, with $\omega X_1 = 3$ for $X_1$ the inner width. 7 temporal modes were used for perturbation Mach number $M = 0.05$. An animation of this figure is given as Movie~9 in the supplementary material.}
    \label{2Dcurvedstill.pdf}
\end{figure}

As was done with straight ducts in 2D and 3D in section~\ref{sec:nonlinearstraight}, we may also consider linearly-propagated nonlinear radiation from the curved duct geometry we just considered. Using the same parameters as before, i.e. $(\alpha_\text{max}^1,\alpha_\text{max}^2,\eta) = (10,200,1/10)$ with $a_\text{max} = 7$ and $M = 0.05$, the result shown in figure~\ref{2Dcurvedstill.pdf}
is obtained. Unlike the linear case, we see that the outer-duct radiation is now highly asymmetric. In the linear case, the peaks and troughs of the wave clearly travelled along the outside of the bend, but this did not appear to disrupt the predominantly-symmetric radiated field. Here, though, the effect of steepening seems to have had an effect, with radiation appearing to be somewhat less symmetric about the centreline $y=0$.

\subsection{Linear radiation from an exponential horn in 2D}\label{sec:2Dexphorn}
\begin{figure}
    \centering
    \includegraphics{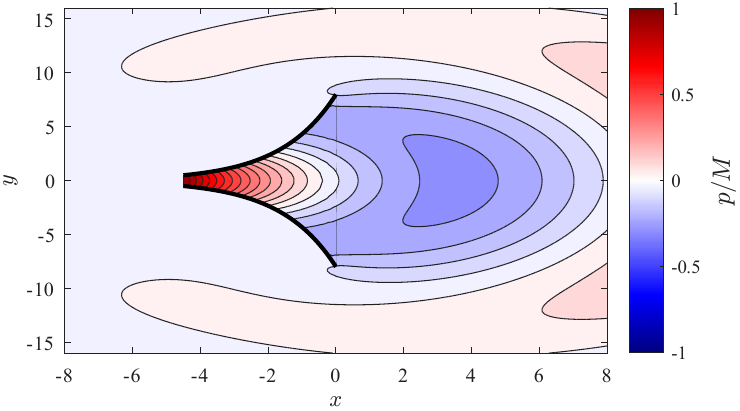}
    \caption{Linear acoustic field emanating from an exponential horn into a duct 10 times larger, in 2D (this plot shows a zoomed-in section of the outer duct, in order for the complete inner domain to be clearly visible). 40 modes were used in the inner calculation and 800 modes in the outer, with $\omega X_1 = 3$ for $X_1$ the inner radius. An animation of this figure is given as Movie~10 in the supplementary material.}
    \label{2Dexphornplot.pdf}
\end{figure}

The exponential horn is a useful object of study due to the existence of an approximate analytical solution \citep{webster}. Comparisons with this analytical solution for a duct with the characteristic admittance at the outlet were made in \citet{mctavish+brambley-2019} and \citet{jensenbrambley2026}; it was found that the approximate analytical solution's assumption of solely plane-wave propagation gave a poor quantitative agreement with the higher fidelity numerics, although the qualitative trends were correct. In figure \ref{2Dexphornplot.pdf}
the same duct is used, but this time the waves are allowed to exit from the outlet into a larger domain.
For this calculation, it was very clear that the choices of $\alpha_\text{max}^1 = 10$ and $\alpha_\text{max}^2 = 200$ used above were seriously under-resolving the acoustics at the outlet, likely due to the very steep duct width increase. As such, here the results were calculated using four times more modes while maintaining the width ratio of $\eta = 1/10$, resulting in $(\alpha_\text{max}^1,\alpha_\text{max}^2,\eta) = (40,800,1/10)$ and giving well-resolved results.
From figure~\ref{2Dexphornplot.pdf}, it is observed that this outlet shape allows for more even distribution of the radiation both behind and in front of the outlet, compared with other geometries. This is both a confirmation of this geometry's favourability for musical instrument design and also a demonstration of the importance of accurate boundary conditions, given that this effect would be completely unobservable with the previously-used characteristic outlet boundary condition \citep[see][figure 5]{mctavish+brambley-2019}.

\section{Conclusions}\label{sec:conc}

% Overview
\noindent
We have presented an exit condition in the admittance and its weakly-nonlinear counterpart for a general two- or three-dimensional duct without flow.
When solving for the acoustics either inside or outside the duct, we see in section \ref{sec:2Dnonlin} that the nonlinear reflection backward from the duct outlet results in significant deviation from the plane-wave-only predictions of the characteristic admittance \citep[as in the Blackstock validation of ][]{jensenbrambley2026}. 
We also experiment with more exotic inner-duct geometries: section \ref{sec:2Dcurved} assesses the asymmetry of radiation from a curved duct, while an exponential horn representative of a brass instrument bell is modelled in section \ref{sec:2Dexphorn}.
The condition is inspired by \citet{kemp}, \citet{felix2018modeling} and \citet{mangin2023modelling} in its use of a larger duct, concentric with the original duct, into which sound from the original duct may be radiated. However, we do not make use of Kemp's baffle-and-dipole superposition, as this does not generalize to weak nonlinearity, nor do we add the complication of both annular and cylindrical ducts as used by \citet{felix2018modeling}.  Instead, we  model the duct outlet as an acoustic discontinuity within the larger duct, which is zero-extended to the larger duct's walls. We also set the admittances in-front-of and behind the discontinuity to be positive- and negative-characteristics, respectively, and prescribe an absorbing condition on the back face of the discontinuity (termed the `acoustic wormhole'). The inner duct then has its internal acoustic field calculated using this outlet condition, and this acoustic field's resulting radiation into the outer duct may also be found. This is effectively an approximation of free space, which may be made more accurate by increasing the two ducts' width ratio. Within the inner duct, we follow \citet{jensenbrambley2026} by using an admittance-based multi-modal method to find the pressure field; in combination with this paper, a very general inner-duct geometry may be modelled in the weakly-nonlinear regime. 
The system can be solved in the smaller duct only (useful for modelling resonances), the outer duct only (for modelling radiation), both ducts (for the full acoustic field) or neither (for considering the exit condition in its own right).

% End correction
A particular application of solving for just the exit condition without considering the acoustic fields is the end-correction problem. Originally, \cite{rayleigh1871} gave an empirically-calculated figure of $0.6R$ for the difference between the actual duct length and the acoustic resonating length; this end correction has since been modelled analytically in the plane-wave case \citep{levine1948radiation} and been generalised to higher frequencies \citep{weinstein1948rigorous}, using the Wiener--Hopf technique. The model presented in this paper may also be applied to these calculations, and is validated against pre-existing analytical work on the end correction in section \ref{sec:linendcorr}. This is done both in two and three dimensions, providing both ideal parameters for future calculations and the confidence in the model's ability to accurately simulate free space.

% End correction: new in linear case
In the linear case, the end correction can be calculated as the Helmholtz number is varied. 
It is interesting to note from the comparison with the Wiener--Hopf technique in figure~\ref{3Dendcorrectionplot.pdf} that beyond very low Helmholtz numbers (i.e.\ for wavelengths far longer than the duct radius), even those corresponding to a frequency well below the cut-on frequency of the next-order spatial modes, the classical three-dimensional end correction of $0.6R$ drops off very significantly; this was also seen by \citet[][figure~1]{snakowska2011} and \citet[][figure~4]{felix2018modeling}.
In the two-dimensional case, we find, in agreement with \citet{noble}, that the plane-wave end-correction coefficient $\hat{s}_0^1/X$ tends to infinity as the Helmholtz numbers $\omega X$ tends to zero.  It should not be concluded, however, that the end correction itself becomes infinite in practice. It can be shown that the end correction tends to zero on the lengthscale of the duct length as the duct width $X$, and hence the Helmholtz number $\omega X$, tend to zero; the end correction just does so sublinearly in the duct width. Also notable in both two and three dimensions is that with increasing Helmholtz number, newly cut-on modes appear with nonzero end corrections, inducing sharp jumps in their respective end-correction coefficients. This stands in contrast to previous literature \citep{snakowska2011}, but the agreement between our numerics and the Wiener--Hopf calculation gives confidence in the correctness of these sharp jumps. Whether this is acoustically important is unclear, and is left for future study.

% End correction: new in nonlinear case
In the nonlinear case our model remains valid, which is an advantage of this model over the Wiener--Hopf technique, and the method of \citet{kemp}: in particular, in section \ref{sec:nonlinendcorr}, we calculate values of a novel \emph{nonlinear end-correction coefficient}. In contrast to the linear case, wave-steepening causes this coefficient to depend upon the duct length, meaning that there is now a three-dimensional parameter space $(M,\omega,L)$. This opens up a rich new area of study which could be approached from many directions: owing to numerical constraints, we focussed here on fixed-frequency calculations while varying duct length and Mach number. In three dimensions, an investigation of the effect of amplitude on the classical $0.6R$ result for low frequencies determined that for greater amplitude, the end-correction decays to zero; this effect is exaggerated in longer ducts, due to the increased opportunity for waves to steepen within the duct. This suggests that when playing a low note on a trumpet, blowing harder will cause the note to sharpen. In two dimensions, an intermediate frequency was investigated, and it appeared that for sufficiently long ducts, the end correction either remains stable (if the duct is an even number of half-wavelengths) or decreases (if the duct is an odd number of half-wavelengths). This suggests that when playing a high note on a trumpet, it will either sharpen or stay the same pitch, depending on how close the note is to one of the trumpet's harmonics.

% Resonances
On the topic of resonances, section \ref{sec:resonances} investigates values of the impedance at the duct inlet while varying the frequency, for various duct aspect ratios. This results in plots whose maxima or minima respectively predict the resonant frequencies of ducts closed at one end or open at both ends. Assuming the relative magnitudes of the maxima/minima predict the dominance of each resonance within the harmonic series corresponding to a particular duct geometry, shorter, fatter ducts will have a more fundamental-dominated sound, whereas longer, thinner ducts will have a greater prevalence of high harmonics. This agrees with organ pipe design: wide pipes known as \emph{flutes} have a mellow, fundamental-dominated sound, whereas narrower pipes known as \emph{strings} have brighter, high-harmonic-dominated spectra. We also investigate the effect of inner-duct curvature on inlet impedance, and plot the pressure field due to a source at the third resonance, for each geometry considered.

% Assumptions of the model
In our outer-duct model, sound from within the smaller duct appears on the right of the discontinuity at the centre of the larger duct representing the smaller duct's outlet (see figure~\ref{fig:outerductdrawings}).  This model was chosen as it allows the linear case to be generalized to weak nonlinearity, unlike other methods such as that of \citet{kemp} and the Wiener--Hopf technique, which are limited to the linear case.  However, the use of a discontinuity in the larger duct in our model gives rise to an `acoustic wormhole', requiring a choice of a ``fake'' artificial condition on the back of the discontinuity, which in reality is in the space that is occupied by the smaller duct.  Here, a perfectly absorbing boundary condition was chosen for the `acoustic wormhole' entrance, which prevents unphysical reflections from this unphysical surface.  Unfortunately, this boundary condition will absorb a small amount of acoustic energy, meaning that our model does not perfectly preserve acoustic energy (instead the energy vanishes into the `wormhole').  This appears not to pose a serious problem, though: comparing the straight duct radiation in two and three dimensions in section~\ref{sec:straightduct}, we see that the differing energy decay rates between 2D and 3D are captured by the model, and the validation in the linear regime against Wiener--Hopf results gives confidence that the energy leak is negligible.  One alternative might have been to prescribe a perfectly reflective boundary condition on this fake surface; this would have the advantage that acoustic energy is perfectly conserved, but the disadvantage that spurious reflections would be allowed.  Another possibility for future work would be to use an annular model for the outer duct behind the exit for $s<0$, although this would require a different set of duct modes and a significant complication in both mathematical and numerical complexity, so has therefore been avoided here.

% Future modifications
There are several other potential improvements that could be made to the duct exit model.  Here, the larger duct, as a proxy for free space, was nonetheless modelled with rigid perfectly-reflecting walls (or with pressure-release  walls in appendix~\ref{sec:soundsoftbc});  an outer duct with an absorbing condition on the walls, such as the perfectly-matched layers used by \cite{felix2018modeling} and \citet{mangin2023modelling}, could instead be incorporated.  While potentially difficult to implement in the weakly-nonlinear case, these methods would make extreme duct width ratios unnecessary, therefore saving on numerical costs, and would eliminate the far-field interference pattern seen within the larger duct in some of the plots presented here.  However, in practice, free space is also an idealization, and for brass instruments played within partially reflecting `rooms', the present model may well be just as accurate as a notional free space condition.
The numerical implementation of the existing model could also be made more efficient. A glaringly obvious example is the use of \textsc{Matlab} rather than a lower-level language with better memory-handling capabilities.  Similarly, as noted in section~\ref{sec:modenumberchoice}, calculating integrals of products of Bessel functions slows the 3D calculation, and these could be computed once before being cached and reused for future simulations, rather than being calculated once for each simulation.  More efficiency would enable a more comprehensive investigation of the nonlinear end-correction coefficient, calculation of the effect of the Mach number on harmonic series, and potentially even computation of the nonlinear acoustic field in the outer duct. The first of these three currently-inaccessible problems, as mentioned above, involves a three-dimensional parameter space $(M, \omega, L)$, in which we have only explored two subplanes here. With more efficient numerics, one could conduct nonlinear generalisations of figures \ref{2Dendcorrectionplot.pdf} and \ref{3Dendcorrectionplot.pdf}, or replace the length variable with frequency in figures \ref{3Ds_hat_machplot.pdf} and \ref{2Ds_hat_machplot.pdf}, as well as exploring higher Mach numbers or longer ducts than we have done here.
The infinite series of coefficients was here made finite for computational purposes by truncating the number of modes, and greater efficiency could also be gained by improving this approximation: for example, by consideration of the asymptotics of matrix entries in the admittance boundary condition as the modenumber is increased and then ``simulating'' instead of ignoring the high-mode-number coefficients.  This could result in a more accurate nonlinear exit condition for the same computational cost. It might also allow consideration of the limit in which the duct width ratio $\eta$ tends to zero: while the only comparison we attempted with the Wiener--Hopf technique here was the end-correction calculation, more direct comparisons of, for example, the pressure field in the outer duct could be performed, particularly with a more accurate free-space approximation. Another area to which the Wiener--Hopf technique has been applied is that of the duct exit with mean flow \citep{rienstra1984acoustic}: a more substantial modification to the model in this paper would be the inclusion of flow, providing in turn another area of comparison with the Wiener--Hopf technique.  While a mean flow would not be needed to model musical instruments, owing to their very low Mach number flows, it would allow, for example, the study of sound from ducted aircraft engines.  This would provide a weakly-nonlinear generalisation of the multi-modal linear work (with flow) of \citet{mangin2023modelling}.  The present study also neglected viscous effects: both linear viscous effects from friction at the duct walls and nonlinear viscous effects from vorticity and momentum shed at the duct exit.  While these are not expected to be important for musical duct acoustics resonances, adding viscous dissipation would allow the calculation of saturation amplitudes of resonances, and could allow the model to be applied to other situations involving more extreme velocities (such as Helmholtz-resonator acoustic linings and oscillations of jet nozzles) for which these effects are not negligible.

% Future application to musical instruments
As the above discussion makes clear, an obvious potential application of the model presented here is harmonic-series calculations of wind or brass instruments, as well as organ pipes. The end-correction problem is of particular interest for organ pipes, and this model provides an analytical justification for the investigation of this phenomenon's relationship to nonlinearity. It can also be used to investigate the relationship between resonant frequencies and wave amplitude (or, in musical terms, between timbre and volume). Ultimately, the combination of this paper's work with the in-duct modelling of \citet{jensenbrambley2026} will allow for analysis of the impact of both nonlinearity and duct geometry on the sound radiated from musical instruments such as trumpets and trombones.

\begin{acknowledgements}
\noindent\textbf{Supplementary material.}
Animations of figures~\ref{thirdresplot.pdf}--\ref{2Dexphornplot.pdf} are available in the supplementary movies as Movies~1--10.
\textsc{Matlab} source code to generate the results given here is also available in supplementary material.

\vspace{1ex}
\noindent\textbf{Acknowledgements.}
For the purpose of open access, the authors have applied a Creative Commons Attribution (CC
BY) licence to any Author Accepted Manuscript version arising from this submission.

\vspace{1ex}
\noindent\textbf{Funding.}
F.J.\ was funded through the Warwick Mathematics Institute Centre for Doctoral Training, and gratefully acknowledges the support of the University of Warwick and the UK Engineering and Physical Sciences Research Council (EPSRC grant EP/W523793/1).
H.T.\ gratefully acknowledges the support of the University of Warwick Undergraduate Research Support Scheme (URSS).
E.J.B.\ gratefully acknowledges the support of the UK Engineering and Physical Sciences Research Council (EPSRC grant EP/V002929/1) and the UK Research and Innovation Future Leaders' Fellowship (UKRI grant MR/V02261X/1).

\vspace{1ex}
\noindent\textbf{Declaration of interests.}
The authors report no conflict of interest.
\end{acknowledgements}

\pagebreak
\appendix
\section{The end correction calculated using the Wiener--Hopf technique}\label{sec:wienerhopf}
This appendix contains a derivation (borrowing heavily from \citet{noble}) of a solution to the problem of linear sound radiating from a duct outlet into free space. The solution is analytical in 2D and semi-analytical in 3D, and provides a helpful comparison for use in section \ref{sec:linendcorr}.
\subsection{Wiener--Hopf technique in 2D}
Here we consider the case of a semi-infinite waveguide in $s<0$ with walls at $x = \pm X/2$ containing an incident wave consisting of multiple duct modes, opening into free space at $s = 0$. We impose a forward-going pressure
\begin{equation}\label{eq:2DWHp+}
    p^+(s,x,t) = \Real\left[\sum_{\alpha = 0}^\infty \psi_\alpha(x)A_\alpha^+\exp\{-\I\omega t + \gamma_\alpha s\}\right],
\end{equation}
with $\psi_\alpha(x) = (C_\alpha/\sqrt{X})\cos\left(\lambda_\alpha\!\left[\tfrac{x}{X} + \tfrac{1}{2}\right]\right)$, where
\begin{align}
&\lambda_\alpha = \alpha\pi, &&C_\alpha = \sqrt{2 - \delta_{\alpha0}}, &&\text{and} &&\gamma_\alpha = \begin{cases}
    \I\omega\sqrt{1 - \lambda_\alpha^2/\omega^2X^2},\quad &\Real(\omega) \geq \lambda_\alpha/X,\\
    -\omega\sqrt{\lambda_\alpha^2/\omega^2X^2-1}, &\Real(\omega) < \lambda_\alpha/X.
\end{cases}
\end{align}
These are the 2D spatial modes $\psi_\alpha$ and straight-duct eigenvalues $\overline{\gamma}_\alpha$ (with the overline dropped here for clarity). The standard boundary condition used throughout this paper is $A_\alpha^+ = \I M\sqrt{X}\delta_{\alpha0}$, i.e. a piston condition of $p^+ = M\sin\omega t$ at $s = 0$ in the inner duct; equation \eqref{eq:2DWHp+} is a generalisation. We take $\omega$ to have a small positive imaginary part: this ensures that $\gamma_\alpha$ always has a negative real part, meaning that a later Fourier transform will converge.

We can express the equivalent coefficient of the \emph{total} pressure within the duct in terms of the reflection matrix $\mat{R}_{\alpha\beta} (s)$
\begin{equation}\label{eq:refmatdef}
    \begin{aligned}
        p  &= \Real\bigg[\e^{-\I\omega t}\sum_\alpha \psi_\alpha(x)\bigg(A_\alpha^+\e^{\gamma_\alpha s} + \sum_\beta\mat{R}_{\alpha\beta} (s)A_\beta^+\e^{\gamma_\beta s}\bigg)\bigg] \\
        &= \Real\bigg[\e^{-\I\omega t}\sum_{\alpha,\beta} \psi_\alpha(x)\bigg(\delta_{\alpha\beta}\e^{\gamma_\alpha s} + \bigg\{\mat{R}_{\alpha\beta} (s)\e^{\left[\gamma_\alpha  + \gamma_\beta \right]s}\bigg\}\e^{-\gamma_\alpha s}\bigg)A_\beta^+\bigg],
    \end{aligned}
\end{equation}
where we have rearranged in the manner of equation \eqref{eq:endcorreqn}. Since this is a linear problem, we may omit taking the real part and the common $\e^{-\I\omega t}$ factor in what follows (this is equivalent to working in terms of the $a = 1$ Fourier coefficient $P^1$, whose superscript we drop for simplicity). We then apply the Wiener--Hopf technique in order to calculate the reflection matrix. In the language of this method, we say that we have an \emph{incident} Fourier pressure coefficient given by
\begin{equation}
    P_\text{inc}  = \begin{cases}
        \frac{1}{2}\sum_\alpha \psi_\alpha(x)A_\alpha^+\e^{\gamma_\alpha s}\quad &|x| < \frac{X}{2}, \\
        0 &|x| > \frac{X}{2}.
    \end{cases}
\end{equation}
The total pressure $P  = P_\text{inc}  + P_\text{ref} $ (where $P_\text{ref} $ is the \emph{reflected} pressure) must satisfy the Helmholtz equation
\begin{equation}
    \frac{\partial^2P }{\partial s^2} + \frac{\partial^2P }{\partial x^2} + \omega^2P  = 0,
\end{equation}
and be continuous everywhere except at the duct walls, where it must satisfy a Neumann condition. Applying the linearity of these conditions and considering the boundary behaviour of $P_\text{inc} $,  we then have the following system for $P_\text{ref} $
\begin{subequations}
    \begin{equation}
        \left[\frac{\partial^2}{\partial s^2} + \frac{\partial^2}{\partial x^2} + \omega^2\right]P_\text{ref}  = 0,
    \end{equation}
    \begin{equation}
    \begin{dcases*}
        \frac{\partial P_\text{ref} }{\partial x}\bigg|_{x = \pm X/2} = 0 &$s<0$, \\
        \bigg[P_\text{ref} \bigg]_{x = (\pm X/2)^\mp}^{(\pm X/2)^\pm} = \frac{1}{2\sqrt{X}}\sum_\alpha C_\alpha(\mp1)^\alpha A_\alpha^+\e^{\gamma_\alpha s}\quad \bigg[\frac{\partial P_\text{ref} }{\partial x}\bigg]_{x = (\pm X/2)^\mp}^{(\pm X/2)^\pm} = 0 \quad &$s>0$.
    \end{dcases*}
    \end{equation}
\end{subequations}
We then take a Fourier transform $s \mapsto \sigma$. The wave equation transforms to
\begin{equation}\label{eq:transformedWE}
    \frac{\partial^2\hat{P}_\text{ref}}{\partial x^2} - (\sigma^2 - \omega^2)\hat{P}_\text{ref} = 0,
\end{equation}
meaning that, defining $\mu(\sigma) := \sqrt{\sigma^2 - \omega^2}$ with branch cuts chosen such that $\Real(\mu)>0$, we must have
\begin{equation}
    \hat{P}_\text{ref} = \begin{cases}
        B(\sigma)\exp(-\mu(\sigma)x) &x > X/2, \\
        C(\sigma)\cosh(\mu(\sigma)x) + D(\sigma)\sinh(\mu(\sigma)x) &x\in[-X/2,X/2], \\
        E(\sigma)\exp(\mu(\sigma)x) &x < -X/2,
    \end{cases}
\end{equation}
since the system is symmetric in $x$ and waves at infinity may only be outgoing. The Fourier transform is then split as follows 
\begin{equation}\label{eq:splitFT}
    \hat{P}_\text{ref}(\sigma,x) = \hat{P}_\text{ref}^-(\sigma,x) + \hat{P}_\text{ref}^+(\sigma,x) = \int_{s = -\infty}^0P_\text{ref}(s,x)~\e^{\I\sigma s}\intd s + \int_{s = 0}^\infty P_\text{ref}(s,x)~\e^{\I\sigma s}\intd s,
\end{equation}
and the boundary conditions become (taking a Fourier transform of the discontinuity in $P_\text{ref}$ outside of the duct, with the small positive imaginary part of $\omega$ ensuring that $\gamma_\alpha$ always has a negative real part, so the first condition on $\hat{P}_\text{ref}^+$ converges)
\begin{align}
    \frac{\partial \hat{P}_\text{ref}^-}{\partial x}\bigg|_{x = \pm X/2} = 0, &&\bigg[\hat{P}_\text{ref}^+\bigg]_{x = (\pm X/2)^\mp}^{(\pm X/2)^\pm} = \frac{\I}{2\sqrt{X}}\sum_\alpha\frac{C_\alpha(\mp1)^\alpha A_\alpha^+}{\sigma - \I\gamma_\alpha } &&\text{and} &&\bigg[\frac{\partial \hat{P}_\text{ref}^+}{\partial x}\bigg]_{x = (\pm X/2)^\mp}^{(\pm X/2)^\pm} = 0.
\end{align}
Between these conditions, the derivative must always be continuous at the boundary between the inner and outer domains; applying the continuous derivative condition to eliminate constants, we get
\begin{equation}
    \hat{P}_\text{ref} = \begin{dcases*}
        \left[\Tilde{C}(\sigma)\sinh\left(\frac{\mu X}{2}\right) + \sgn(x)\Tilde{D}(\sigma)\cosh\left(\frac{\mu X}{2}\right)\right]\exp(-\mu|x|) ~&$|x| > X/2$,\\
        -\exp\left(-\frac{\mu X}{2}\right)\left[\Tilde{C}(\sigma)\cosh(\mu x) + \Tilde{D}(\sigma)\sinh(\mu x)\right] &$|x| < X/2$.
    \end{dcases*}
\end{equation}
This results in
\begin{equation}
    \Tilde{C}(\sigma) \pm \Tilde{D}(\sigma) = \left[\hat{P}_\text{ref}\right]_{x = (\pm X/2)^\mp}^{(\pm X/2)^\pm} = \frac{\I}{2\sqrt{X}}\sum_\alpha\frac{C_\alpha(\mp1)^\alpha A_\alpha^+}{\sigma - \I\gamma_\alpha } + \left[\hat{P}_\text{ref}^-\right]_{x = (\pm X/2)^\mp}^{(\pm X/2)^\pm}.
\end{equation}
The Neumann condition reduces to 
    \begin{align}\notag
        \frac{\partial\hat{P}_\text{ref}^+}{\partial x}\bigg|_{x = \pm X/2} = \frac{\partial \hat{P}_\text{ref}}{\partial x}\bigg|_{x = \pm X/2} &= -\mu(\sigma)\e^{-\frac{\mu(\sigma) X}{2}}\left[\pm\Tilde{C}(\sigma)\sinh\left(\frac{\mu(\sigma) X}{2}\right) + \Tilde{D}(\sigma)\cosh\left(\frac{\mu(\sigma)X}{2}\right)\right] \\
        &=: \mp \mu(\sigma)^2X\Tilde{C}(\sigma)L(\sigma) - \mu(\sigma)\Tilde{D}(\sigma)K(\sigma),
    \end{align}
where $L(\sigma)$ and $K(\sigma)$ are defined \citep[equivalently to their definitions in][]{noble} for notational ease later on. Eliminating $\Tilde{C}(\sigma)$ and $\Tilde{D}(\sigma)$ yields symmetric and antisymmetric Wiener--Hopf equations
\begin{subequations}
    \begin{equation}
        \frac{\frac{\partial\hat{P}_\text{ref}^+}{\partial x}\left(\bigg|_{x = X/2} - \bigg|_{x = -X/2}\right)}{\mu(\sigma)^2XL(\sigma)} + \left[\hat{P}_\text{ref}^-\right]\left(\bigg|_{x = (X/2)^-}^{(X/2)^+} + \bigg|_{x = (-X/2)^+}^{(- X/2)^+}\right) = -\frac{\I}{\sqrt{X}}\sum_{\alpha\text{ even}}\frac{C_\alpha A_\alpha^+}{\sigma - \I\gamma_\alpha },
    \end{equation}
    \begin{equation}
        \frac{\frac{\partial\hat{P}_\text{ref}^+}{\partial x}\left(\bigg|_{x = X/2} + \bigg|_{x = -X/2}\right)}{\mu(\sigma)K(\sigma)} + \left[\hat{P}_\text{ref}^-\right]\left(\bigg|_{x = (X/2)^-}^{(X/2)^+} - \bigg|_{x = (-X/2)^+}^{(- X/2)^+}\right) = -\frac{\I}{\sqrt{X}}\sum_{\alpha\text{ odd}}\frac{C_\alpha A_\alpha^+}{\sigma - \I\gamma_\alpha }x.
    \end{equation}
\end{subequations}
If we can factorise $L(\sigma)$ and $K(\sigma)$ into two functions each, two which are analytic in the upper-half plane ($L^+(\sigma)$, $K^+(\sigma)$) and another two which are analytic in the lower-half plane ($L^-(\sigma)$, $K^-(\sigma)$), then we can arrange the Wiener--Hopf equations as
\begin{subequations}
    \begin{equation}
        \begin{aligned}
            \frac{\frac{\partial\hat{P}_\text{ref}^+}{\partial x}\left(|_{x = X/2} - |_{x = -X/2}\right)}{(\sigma + \omega)XL^+(\sigma)} &+ \frac{\I}{\sqrt{X}}\sum_{\alpha\text{ even}}\frac{C_\alpha A_\alpha^+(\I\gamma_\alpha  - \omega)L^-(\I\gamma_\alpha )}{\sigma - \I\gamma_\alpha } \\
            = &-(\sigma - \omega)L^-(\sigma)\left[\hat{P}_\text{ref}^-\right]\left(|_{x = (X/2)^-}^{(X/2)^+} + |_{x = (-X/2)^+}^{(- X/2)^+}\right) \\
            &- \frac{\I}{\sqrt{X}}\sum_{\alpha\text{ even}}\frac{C_\alpha A_\alpha^+\bigg((\sigma - \omega)L^-(\sigma) - (\I\gamma_\alpha  - \omega)L^-(\I\gamma_\alpha )\bigg)}{\sigma - \I\gamma_\alpha },
        \end{aligned}
    \end{equation}
    \begin{equation}
        \begin{aligned}
            \frac{\frac{\partial\hat{P}_\text{ref}^+}{\partial x}\left(|_{x = X/2} + |_{x = -X/2}\right)}{\sqrt{\sigma + \omega}K^+(\sigma)} &- \frac{\I}{\sqrt{X}}\sum_{\alpha\text{ odd}}\frac{C_\alpha A_\alpha^+\sqrt{\I\gamma_\alpha  - \omega}K^-(\I\gamma_\alpha )}{\sigma - \I\gamma_\alpha } \\
            = &-\sqrt{\sigma - \omega}K^-(\sigma)\left[\hat{P}_\text{ref}^-\right]\left(|_{x = (X/2)^-}^{(X/2)^+} - |_{x = (-X/2)^+}^{(- X/2)^+}\right) \\
            &+ \frac{\I}{\sqrt{X}}\sum_{\alpha\text{ odd}}\frac{C_\alpha A_\alpha^+\bigg(\sqrt{\sigma - \omega}K^-(\sigma) - \sqrt{\I\gamma_\alpha  - \omega}K^-(\I\gamma_\alpha )\bigg)}{\sigma - \I\gamma_\alpha }.
        \end{aligned}
    \end{equation}
\end{subequations}
In each case, the left-hand side is analytic in the upper-half plane and the right-hand side is analytic in the lower-half plane. Therefore, both sides are analytic everywhere, and so we follow \citet{noble} in setting them to zero (corresponding physically to a choice of the smoothest possible solution at the scattering point, i.e. the outlet corners). As such, we determine $\Tilde{C}(\sigma)$ and $\Tilde{D}(\sigma)$ to be
\begin{subequations}
    \begin{equation}
        \Tilde{C}(\sigma) = \frac{\I\exp\left(\frac{\mu(\sigma)X}{2}\right)(\sigma + \omega)L^+(\sigma)}{4\sqrt{X}\mu(\sigma)\sinh\left(\frac{\mu(\sigma)X}{2}\right)}\sum_{\alpha\text{ even}}\frac{C_\alpha A_\alpha^+(\I\gamma_\alpha  - \omega)L^-(\I\gamma_\alpha )}{\sigma - \I\gamma_\alpha },
    \end{equation}
    \begin{equation}
        \Tilde{D}(\sigma) = -\frac{\I\exp\left(\frac{\mu(\sigma)X}{2}\right)\sqrt{\sigma + \omega}K^+(\sigma)}{2\sqrt{X}\mu(\sigma)\cosh\left(\frac{\mu(\sigma)X}{2}\right)}\sum_{\alpha\text{ odd}}\frac{C_\alpha A_\alpha^+\sqrt{\I\gamma_\alpha  - \omega}K^-(\I\gamma_\alpha )}{\sigma - \I\gamma_\alpha }.
    \end{equation}
\end{subequations}
Substituting these formulae into the expression for $\hat{P}_\text{ref} $ and inverting, we find
\begin{equation}
    P_\text{ref}  = \begin{dcases*}
        \frac{\I}{4\pi\sqrt{X}}\int_{\sigma = -\infty}^\infty\frac{\e^{-\mu(\sigma)\left(|x| - X/2\right) - \I\sigma s}}{\mu(\sigma)}\Bigg[\sum_{\alpha\text{ even}}\frac{C_\alpha A_\alpha^+(\sigma + \omega)(\I\gamma_\alpha  - \omega)XL^+(\sigma)L^-(\I\gamma_\alpha )}{2(\sigma - \I\gamma_\alpha )} \\
        \qquad\qquad\qquad- \sgn(x)\sum_{\alpha\text{ odd}}\frac{C_\alpha A_\alpha^+\sqrt{(\sigma + \omega)(\I\gamma_\alpha  - \omega)}K^+(\sigma)K^-(\I\gamma_\alpha )}{\sigma - \I\gamma_\alpha }\Bigg]\intd\sigma \qquad\quad |x| > X/2,\\[3ex]
        \frac{\I}{4\pi\sqrt{X}}\int_{\sigma = -\infty}^\infty\e^{-\I\sigma s}\Bigg[-\sum_{\alpha\text{ even}}\frac{C_\alpha A_\alpha^+(\sigma + \omega)(\I\gamma_\alpha  - \omega)XL^+(\sigma)L^-(\I\gamma_\alpha )\cosh(\mu(\sigma)x)}{2(\sigma - \I\gamma_\alpha )\mu(\sigma)\sinh\left(\frac{\mu(\sigma)X}{2}\right)} \\
        \qquad\qquad\qquad+ \sum_{\alpha\text{ odd}}\frac{C_\alpha A_\alpha^+\sqrt{(\sigma + \omega)(\I\gamma_\alpha  - \omega)}K^+(\sigma)K^-(\I\gamma_\alpha )\frac{\sinh\mu(\sigma)x}{\mu(\sigma)}}{(\sigma - \I\gamma_\alpha )\cosh(\frac{\mu(\sigma)X}{2})}\Bigg]\intd\sigma \qquad |x| < X/2.
    \end{dcases*}
\end{equation}
In order to calculate these integrals, the contour of integration can be deformed either into the lower or upper-half plane, depending on the sign of $s$. There are two possible signs of $|x| - X/2$, each with two possible signs of $s$, so altogether there are four cases.

We know that $L^+(\sigma)$ and $K^+(\sigma)$ are analytic in the upper-half plane. In the lower-half plane they must contain all of the singularities of $L(\sigma)$ and $K(\sigma)$, i.e. a branch cut downward from $\sigma = -\omega$. For $|x| > X/2$, we have a branch cut in the upper-half-plane anyway due to the factors of $\mu$, meaning that this integral will be deformed around a branch cut for both $s > 0$ and $s < 0$ (cases 1 and 2). For $|x| < X/2$ and $s > 0$, the integral may only be deformed into the lower-half plane, where there is a branch cut (case 3). However, for $|x| < X/2$ and $s < 0$, because the series expansion of the integrand contains only $\mu^2$ (not $\mu$), and $L^+$ and $K^+$ are analytic in the upper-half plane, we can deform upwards and encounter only poles, whose residues become duct modes (case 4, corresponding to the duct interior). This has the following result
\begin{equation}\label{2DPref}
        P_\text{ref}  = -\frac{1}{2X}\Bigg(\sum_{\substack{\alpha\text{ even},\\ \beta\text{ even}}}\!\!\frac{C_\alpha A_\alpha^+\e^{-\gamma_\beta s} L^+(-\I\gamma_\beta )L^-(\I\gamma_\alpha )C_\beta \psi_\beta(x)}{(\gamma_\alpha  + \gamma_\beta )\gamma_\beta }
        + \sum_{\substack{\alpha\text{ odd},\\ \beta\text{ odd}}}\!\!\frac{C_\alpha A_\alpha^+\e^{-\gamma_\beta s} K^+(-\I\gamma_\beta )K^-(\I\gamma_\alpha )C_\beta \psi_\beta(x)}{(\gamma_\alpha  + \gamma_\beta )\gamma_\beta }\Bigg).
\end{equation}
This is substituted into the total pressure with the $\alpha$ and $\beta$ labels flipped, and upon comparison with equation \eqref{eq:refmatdef} we get the following expression
\begin{equation}
    \mat{R}_{\alpha\beta} (s) = \begin{dcases*}
        -\frac{C_\alpha C_\beta L^+(-\I\gamma_\alpha )L^-(\I\gamma_\beta )(\gamma_0  + \gamma_\alpha )(\gamma_0  + \gamma_\beta )}{2\gamma_\alpha (\gamma_\alpha  + \gamma_\beta )X\e^{(\gamma_\alpha  + \gamma_\beta )s}} \qquad\qquad&$\alpha$ even, $\beta$ even, \\
        -\frac{C_\alpha C_\beta K^+(-\I\gamma_\alpha )K^-(\I\gamma_\beta )\sqrt{-\I(\gamma_0  + \gamma_\alpha )}\sqrt{\I(\gamma_0  + \gamma_\beta )}}{\gamma_\alpha (\gamma_\alpha  + \gamma_\beta )X\e^{(\gamma_\alpha  + \gamma_\beta )s}} &$\alpha$ odd, $\beta$ odd, \\
        0 &otherwise,
    \end{dcases*}
\end{equation}
so then
\begin{equation}
    \mat{R}_{\alpha\alpha} (0) = \begin{dcases*}
        -\frac{C_\alpha^2 L^+(-\I\gamma_\alpha )L^-(\I\gamma_\alpha )(\gamma_0  + \gamma_\alpha )^2}{4(\gamma_\alpha )^2X} &$\alpha$ even,\\
        -\frac{\I C_\alpha^2 K^+(-\I\gamma_\alpha )K^-(\I\gamma_\alpha )|\gamma_0  + \gamma_\alpha |}{2(\gamma_\alpha )^2X} &$\alpha$ odd.
    \end{dcases*}
\end{equation}

\subsubsection{Further analytical expressions}
A more precise expression may be found for the reflection matrix: for particular ranges of $\omega$, this can lead to elegant closed-form expressions for the end correction itself.  From \citet{noble}, we have analytical expressions for $L^+$ and $K^+$ as follows
\begin{equation}
      \begin{Bmatrix} L^+(\sigma) \\ K^+(\sigma) \end{Bmatrix}=\exp\!\left(\frac{\I\sigma X}{2\pi}\!\left(\!1 - C + \log\frac{\{4,1\}\pi}{\omega X}\right) - \frac{\sigma X}{4} - \frac{\mu(\sigma)X}{2\pi}\arccos\frac{\sigma}{\omega}\right)
      \prod_{\mathclap{\substack{\beta > 0,\\\{\text{even,odd}\}}}}^\infty\left(\frac{X}{\beta\pi}\right)\!\left(-\gamma_\beta  - \I\sigma\right)\exp\!\left(\frac{\I\sigma X}{\beta\pi}\right),
\end{equation}
with $L^-(\sigma) = L^+(-\sigma)$ and $K^-(\sigma) = K^+(-\sigma)$, and $C$ being the Euler-Mascheroni constant. We then have
\begin{equation}
    \arg\left(-\mat{R}_{\alpha\alpha} (0)\right) = \begin{cases}
        \arg(L^+(-\I\gamma_\alpha ))^2 &\alpha\text{ even}, \\
        \arg(K^+(-\I\gamma_\alpha ))^2 - \frac{\pi}{2} \quad&\alpha\text{ odd}.
    \end{cases}
\end{equation}
$(L^+(-\I\gamma_\alpha ))^2$ may be written in a way that makes taking the argument easier, i.e.
\begin{equation}
    \begin{aligned}
        (L^+(-\I\gamma_\alpha ))^2 = &\exp\left(\frac{\I|\gamma_\alpha |X}{\pi}\left(1 - C + \log\frac{4\pi}{\omega X}\right) - \frac{|\gamma_\alpha |X}{2} - \I\alpha\arccos\frac{|\gamma_\alpha |}{\omega}\right) \\
        &\times\prod_{\substack{\beta > 0,\\ \text{even}}}^{2\lfloor\frac{\omega X}{2\pi}\rfloor}\left\{\left(\frac{X}{\beta\pi}\right)^2\e^{-\I\pi}\left(|\gamma_\beta | + |\gamma_\alpha |\right)^2\exp\left(\frac{2\I|\gamma_\alpha X}{\beta\pi}\right)\right\} \\
        &\times\prod_{\substack{\beta > 2\lfloor\frac{\omega X}{2\pi}\rfloor,\\ \text{even}}}^\infty\left\{\left(\frac{X}{\beta\pi}\right)^2\left(|\gamma_\beta | - \I|\gamma_\alpha |\right)^2\exp\left(\frac{2\I|\gamma_\alpha X}{\beta\pi}\right)\right\},
    \end{aligned}
\end{equation}
so that after a bit of manipulation we find the expression for even modes to be
\begin{align}
  \arg(-\mat{R}_{\alpha\alpha} (0)) &=
        - \alpha\left(\frac{\pi}{2} - \arccos\frac{|\gamma_\alpha |}{\omega}\right) - \pi\left(\left\lfloor\frac{\omega X}{2\pi}\right\rfloor - \frac{\alpha}{2}\right)
\\\notag&\!\!\!\!
        + \frac{|\gamma_\alpha |X}{\pi}\left\{1 - C + \log\frac{4\pi}{\omega X} + H_{\lfloor\omega X/2\pi\rfloor} \vphantom{+ \sum_{n = \lfloor\omega X/2\pi\rfloor+1}^\infty\left[\frac{1}{n} - \frac{2\pi}{|\gamma_\alpha |X}\arctan\frac{|\gamma_\alpha |X}{2\pi n\sqrt{1 - \frac{\omega^2X^2}{4n^2\pi^2}}}\right]}
  + \!\!\!\!\sum_{n = \lfloor\omega X/2\pi\rfloor+1}^\infty\left[\frac{1}{n} - \frac{2\pi}{|\gamma_\alpha |X}\arctan\frac{|\gamma_\alpha |X}{2\pi n\sqrt{1 - \frac{\omega^2X^2}{4n^2\pi^2}}}\right]\right\}.
\end{align}
Doing the same for odd modes, we get
    \begin{align}
        \arg(-\mat{R}_{\alpha\alpha} (0)) &= \frac{|\gamma_\alpha |X}{\pi}\left\{1 - C + \log{\frac{\pi}{\omega X}} + 2H_{2\left\lfloor\omega X/2\pi + 1/2\right\rfloor} - H_{\lfloor\omega X/2\pi + 1/2\rfloor} \vphantom{\frac{|\gamma_\alpha |X}{2(n  - \frac{1}{2})\pi\sqrt{1 - \frac{\omega^2X^2}{4(n - 1/2)^2X^2}}}}\right. \\\notag
        &\left.\qquad+ \sum_{n = \lfloor\omega X/2\pi + 1/2\rfloor + 1}\left[\frac{1}{n - \frac{1}{2}} - \frac{2\pi}{|\gamma_\alpha |X}\arctan\frac{|\gamma_\alpha |X}{2(n  - \frac{1}{2})\pi\sqrt{1 - \frac{\omega^2X^2}{4(n - 1/2)^2X^2}}}\right]\right\}\\\notag
        &\qquad- \alpha\left(\frac{\pi}{2} - \arccos\frac{|\gamma_\alpha |}{\omega}\right) - \pi\left(\left\lfloor\frac{\omega X}{2\pi} + \frac{1}{2}\right\rfloor - \frac{\alpha + 1}{2}\right),
    \end{align}
where $H_n$ is the $n^\text{th}$ harmonic number. For $\omega \in (\overline{\omega}_\alpha ,\overline{\omega}_{\alpha+2} )$, if $\alpha$ is even we have
\begin{align}
  \frac{\hat{s}_\alpha }{X} =&
    - \frac{\alpha}{2|\gamma_\alpha |X}\!\left(\frac{\pi}{2} - \arccos\frac{|\gamma_\alpha |}{\omega}\right)
\\\notag&
    + \frac{1}{2\pi}\!\left\{\!1 - C + \log\frac{4\pi}{\omega X} + H_{\alpha/2} + \!\!\!\!\sum_{n = \alpha/2+1}^\infty\!\left[\frac{1}{n} - \frac{2\pi}{|\gamma_\alpha |X}\arctan\frac{|\gamma_\alpha |X}{2\pi n\sqrt{1 - \frac{\omega^2X^2}{4n^2\pi^2}}}\right]\!\right\},
\end{align}
and if $\alpha$ is odd we have
\begin{align}
  \frac{\hat{s}_\alpha }{X} =
  - \frac{\alpha}{2|\gamma_\alpha |X}\left(\frac{\pi}{2} - \arccos\frac{|\gamma_\alpha |}{\omega}\right)
  &+ \frac{1}{2\pi}\left\{1 - C + \log{\frac{\pi}{\omega X}} + 2H_{\alpha + 1} - H_{(\alpha+1)/2} \vphantom{\frac{|\gamma_\alpha |X}{2(n  - \frac{1}{2})\pi\sqrt{1 - \frac{\omega^2X^2}{4(n - 1/2)^2X^2}}}}
  \right.\\\notag&\left.
        + \!\!\!\!\sum_{n = (\alpha+1)/2 + 1}\left[\frac{1}{n - \frac{1}{2}} - \frac{2\pi}{|\gamma_\alpha |X}\arctan\frac{|\gamma_\alpha |X}{2(n  - \frac{1}{2})\pi\sqrt{1 - \frac{\omega^2X^2}{4(n - 1/2)^2X^2}}}\right]\right\},
\end{align}
and in particular if $\omega < \overline{\omega}_1 $, we can use $\arcsin(z) = \arctan(z/\sqrt{1 - z^2})$ to get
\begin{equation}\label{eq:nobleexpression}
    \frac{\hat{s}_0 }{X} = \frac{1}{2\pi}\left\{1 - C + \log\frac{4\pi}{\omega X} + \sum_{n = 1}^\infty\left[\frac{1}{n} - \frac{2\pi}{\omega X}\arcsin\frac{\omega X}{2\pi n}\right]\right\},
\end{equation}
as derived in \citet{noble}. Note that this gives a logarithmic singularity as $\omega X \to 0$, but that $\hat{s}_0  \sim X\log\frac{1}{X}$ as $X \to 0$: this means the end correction \emph{does} tend to zero, just sublinearly with duct width.

\subsection{Wiener--Hopf technique in 3D}
In order to prescribe an incident pressure of the same form as in 2D, we set our incident pressure to be
\begin{equation}
    P_\text{inc}  = \begin{cases}
        \frac{1}{2}\sum_\alpha \psi_\alpha(r,\theta) A_\alpha^+\exp(\gamma_\alpha s) &r < R, \\
        0 &r > R.
    \end{cases}
\end{equation}
The spatial modes in 3D are $\psi_\alpha(r,\theta) = (C_\alpha/\sqrt{\pi} R) \J_{m_\alpha}\!(\lambda_\alpha r/R)\cos(m_\alpha\theta - \xi_\alpha\pi/2)$, where modenumbers, constants and eigenvalues are defined by 
\begin{subequations}
\begin{align}
    &\lambda_\alpha = j_{m_\alpha n_\alpha}',
    && C_\alpha = 
    \begin{cases}
        \big|\J_{0}(\lambda_\alpha)\big|^{-1},&m_\alpha = 0,\\
        \bigg(\sqrt{\frac{1}{2}\left[1 - \frac{{m_\alpha}^2}{{\lambda_\alpha}^2}\right]} ~\big|\J_{m_\alpha}(\lambda_\alpha)\big|\bigg)^{-1},&m_\alpha \neq 0,
    \end{cases} \\ &\xi_\alpha \in \{0,1\} \qquad \text{and} \qquad &&\gamma_\alpha = \begin{cases}
    \I\omega\sqrt{1 - \lambda_\alpha^2/\omega^2R^2},\quad &\Real(\omega) \geq \lambda_\alpha/R,\\
    -\omega\sqrt{\lambda_\alpha^2/\omega^2R^2-1}, &\Real(\omega) < \lambda_\alpha/R,
\end{cases}
\end{align}
\end{subequations}
where $j_{m_\alpha n_\alpha}'$ is the $n_\alpha$-th zero of the Bessel function of order $m_\alpha$. Once again, taking $A_\alpha^+ = \I M\sqrt{\pi}R\delta_{\alpha0}$ results in a piston condition of $p^+ = M\sin\omega t$ at $s = 0$ in the inner duct. Taking a Fourier transform and solving the Helmholtz equation results in a solution for the reflected pressure (dropping the superscript once more)
\begin{equation}
    \hat{P}_\text{ref} = \begin{cases}
        \sum_{m = 0}^\infty \Ib_m(\mu(\sigma) r)\left[A_m(\sigma)\cos m\theta + B_m(\sigma)\sin m\theta\right] &r < R,\\
        \sum_{m = 0}^\infty \K_m(\mu(\sigma)r)\left[C_m(\sigma)\cos m\theta + D_m(\sigma)\sin m\theta\right] &r > R,
    \end{cases}
\end{equation}
where again we have $\mu^2 := \sigma^2 - \omega^2$. In writing down this solution, we have already applied the conditions of periodicity in $\theta$, regularity at $r = 0$ and decay in the far field. Defining the split Fourier transform as before, we then have duct wall and continuity conditions
\begin{equation}
    \begin{dcases*}
        \frac{\partial\hat{P}_\text{ref}^-}{\partial r}\bigg|_{r = R} = 0, \\
        \left[\hat{P}_\text{ref}^+\right]_{R^-}^{R^+} = \frac{\I}{2\sqrt{\pi}R}\sum_\alpha\frac{C_\alpha\J_{m_\alpha}(\lambda_\alpha)\cos\left(m_\alpha\theta - \frac{\xi_\alpha\pi}{2}\right)A_\alpha^+}{\sigma - \I\gamma_\alpha },\quad\left[\frac{\partial\hat{P}_\text{ref}^+}{\partial r}\right]_{R^-}^{R^+} = 0.
    \end{dcases*}
\end{equation}
Applying the continuous derivative condition reduces the number of constants to two, giving
\begin{equation}
    \hat{P}_\text{ref} = \begin{dcases*}
        \sum_{m = 0}^\infty \frac{\Ib_m(\mu(\sigma) r)}{\Ib_m'(\mu(\sigma) R)}\left[\widetilde{A}_m(\sigma)\cos m\theta + \widetilde{B}_m(\sigma)\sin m\theta\right] &$r < R$,\\
        \sum_{m = 0}^\infty \frac{\K_m(\mu(\sigma)r)}{\K_m'(\mu(\sigma)R)}\left[\widetilde{A}_m(\sigma)\cos m\theta + \widetilde{B}_m(\sigma)\sin m\theta\right] &$r > R$.
    \end{dcases*}
\end{equation}
We can apply the jump condition to this expression, getting
\begin{equation}
    \left[\hat{P}_\text{ref}\right]_{R^-}^{R^+} = \sum_{m = 0}^\infty\left(\frac{\K_m(\mu(\sigma)R)}{\K_m'(\mu(\sigma)R)} - \frac{\Ib_m(\mu(\sigma)R)}{\Ib_m'(\mu(\sigma)R)}\right)\left(\widetilde{A}_m(\sigma)\cos m\theta + \widetilde{B}_m(\sigma)\sin m\theta\right),
\end{equation}
where the bracketed quantity on the left may be re-expressed in terms of the Wronskian $W\left\{\K_m(z),\Ib_m(z)\right\} = 1/z$, i.e.
\begin{equation}
    \begin{aligned}
        \frac{\K_m(\mu(\sigma)R)}{\K_m'(\mu(\sigma)R)} - \frac{\Ib_m(\mu(\sigma)R)}{\Ib_m'(\mu(\sigma)R)} &= \frac{W\left\{\K_m(\mu(\sigma)R),\Ib_m(\mu(\sigma)R)\right\}}{\K_m'(\mu(\sigma)R)\Ib_m'(\mu(\sigma)R)} \\
        &= \frac{1}{\mu(\sigma)R\K_m'(\mu(\sigma)R)\Ib_m'(\mu(\sigma)R)} =: -\frac{2}{L_m(\sigma)}.
    \end{aligned}
\end{equation}
Here we have defined $L_m(\sigma)$ as we did with $L(\sigma)$ and $K(\sigma)$ in the 2D case. Integrating the jump condition over a period in $\theta$ allows us to find $\widetilde{A}_m(\sigma)$ and $\widetilde{B}_m(\sigma)$
\begin{equation}
    -\frac{2}{L_m(\sigma)}\begin{Bmatrix}
            \widetilde{A}_m(\sigma) \\ \widetilde{B}_m(\sigma)
        \end{Bmatrix} = \left[\frac{1}{\pi}\int_0^{2\pi}\left[\hat{P}_\text{ref}^-\right]_{R^-}^{R^+}\begin{Bmatrix}\cos m\theta \\ \sin m\theta\end{Bmatrix}\mathrm{d}\theta + \frac{\I}{2\sqrt{\pi}R}\sum_{\substack{\alpha:m_\alpha = m,\\ \xi_\alpha = \{0, 1\}}}\frac{C_\alpha\J_{m_\alpha}(\lambda_\alpha)A_\alpha^+}{\sigma - \I\gamma_\alpha }\right].
\end{equation}
The Fourier transform of the continuity condition here was previously a sum of modes, but the orthogonality of these modes has been exploited through integration over a period, hence the new limits on the sum. We may also find $\widetilde{A}_m(\sigma)$ and $\widetilde{B}_m(\sigma)$ in terms of the derivative on the boundary, by noting that we have
\begin{equation}
    \frac{\partial\hat{P}_\text{ref}^+}{\partial r}\bigg|_{r = R} = \frac{\partial\hat{P}_\text{ref}}{\partial r}\bigg|_{r = R} = \mu(\sigma)\sum_{m = 0}^\infty\left(\widetilde{A}_m(\sigma)\cos m\theta + \widetilde{B}_m(\sigma)\sin m\theta\right),
\end{equation}
and integrating similarly. Eliminating $\widetilde{A}_m(\sigma)$ and $\widetilde{B}_m(\sigma)$ results in
\begin{multline}
  -\frac{L_m(\sigma)}{2}\left[\frac{1}{\pi}\int_0^{2\pi}\left[\hat{P}_\text{ref}^-\right]_{R^-}^{R^+}\!\begin{Bmatrix}\cos m\theta \\ \sin m\theta\end{Bmatrix}\mathrm{d}\theta + \frac{\I}{2\sqrt{\pi}R}\sum_{\substack{\alpha:m_\alpha = m,\\ \xi_\alpha = \{0, 1\}}}\!\!\!\!\frac{C_\alpha\J_{m_\alpha}(\lambda_\alpha)A_\alpha^+}{\sigma - \I\gamma_\alpha }\right]
  \\
        = \frac{1}{\pi\mu(\sigma)}\int_0^{2\pi}\!\left[\frac{\partial\hat{P}_\text{ref}^+}{\partial r}\right]_{r = R}\!\!\begin{Bmatrix}
            \cos m\theta \\ \sin m\theta
        \end{Bmatrix}\intd\theta.
\end{multline}
Taking the decomposition $L_m(\sigma) = L_m^+(\sigma)L_m^-(\sigma)$, we can write this as an equality between an UHP-analytic function and a LHP-analytic one
\begin{align}\notag
      &\frac{\mu^-(\sigma)L_m^-(\sigma)R}{\pi}\int_0^{2\pi}\left[\hat{P}_\text{ref}^-\right]_{R^-}^{R^+}\begin{Bmatrix}
        \cos m\theta \\ \sin m\theta
      \end{Bmatrix}\intd\theta
%      \\&\qquad\qquad\qquad
      + \frac{\I}{2\sqrt{\pi}R}\!\sum_{\substack{\alpha:m_\alpha = m, \\ \xi_\alpha = \{0, 1\}}}\!\!\!\!\!\frac{C_\alpha \J_{m_\alpha}(\lambda_\alpha)A_\alpha^+\left[\mu^-(\sigma)L_m^-(\sigma) - \mu^-(\I\gamma_\alpha )L_m^-(\I\gamma_\alpha )\right]}{\sigma - \I\gamma_\alpha }
      \\&\quad =
      -\frac{2}{\pi\mu^+(\sigma)L_m^+(\sigma)}\int_0^{2\pi}\left[\frac{\partial\hat{P}_\text{ref}^+}{\partial r}\right]_{r = R}\!\begin{Bmatrix}
        \cos m\theta \\ \sin m\theta
      \end{Bmatrix}\intd\theta
%      \\&\qquad\qquad\qquad
      - \frac{\I}{2\sqrt{\pi}R}\sum_{\substack{\alpha:m_\alpha = m,\\ \xi_\alpha = \{0, 1\}}}\!\!\!\!\frac{C_\alpha\J_{m_\alpha}(\lambda_\alpha)A_\alpha^+\mu^-(\I\gamma_\alpha )L_m^-(\I\gamma_\alpha )}{\sigma - \I\gamma_\alpha }.
\end{align}
Assuming that both sides are equal to zero as before, we can eliminate $\left[\partial\hat{P}_\text{ref}^+/\partial r\right]_{r = R}$ and $\left[\hat{P}_\text{ref}^-\right]_{R^-}^{R^+}$ to find expressions for $\widetilde{A}_m(\sigma)$ and $\widetilde{B}_m(\sigma)$; these are substituted into $\hat{P}_\text{ref}$ to give an inner-duct result (in which we have turned the addition of $\widetilde{A}_m(\sigma)$ and $\widetilde{B}_m(\sigma)$ into a sum over $\xi$)
\begin{equation}
        \hat{P}_\text{ref} = -\!\sum_{m = 0}^\infty\frac{\I\mu^+(\sigma)L_m^+(\sigma)\Ib_m(\mu r)}{4\sqrt{\pi}R\mu(\sigma)\Ib_m'(\mu R)}\!\sum_{\xi = 0} \!\sum_{\substack{\alpha:\\m_\alpha = m, \\ \xi_\alpha = \xi}}\!\!\!\frac{C_\alpha\J_{m_\alpha}(\lambda_\alpha)A_\alpha^+\mu^-(\I\gamma_\alpha )L_m^-(\I\gamma_\alpha )\cos \left(m_\alpha\theta - \frac{\xi\pi}{2}\right)}{\sigma - \I\gamma_\alpha },
\end{equation}
and an outer-duct result with $\K_m$ in place of $\Ib_m$. A power-series calculation verifies that only factors of $\mu^2$ appear in the inner-duct expression for $s<0$ (as before) so we can calculate the residue at each pole, getting
    \begin{align}
        \I~\res_{\sigma = -\I\gamma_\beta }\hat{P}_\text{ref}\e^{-\I\sigma s} =& -\frac{\e^{-\gamma_\beta s}\mu^+(-\I\gamma_\beta )L_{m_\beta}^+(-\I\gamma_\beta )C_\beta \J_{m_\beta}(\lambda_\beta)\psi_\beta}{4(2 - \delta_{m_\beta 0})\gamma_\beta } \\\notag
        &\quad\times\sum_{\xi = 0} \frac{C_\beta\J_{m_\beta}(\lambda_\beta r/R)\cos\left(m_\beta\theta - \frac{\xi\pi}{2}\right)}{\pi R}\sum_{\substack{\alpha:\\m_\alpha = m_\beta \\ \xi_\alpha = \xi}}\frac{C_\alpha \J_{m_\alpha}(\lambda_\alpha)A_\alpha^+\mu^-(\I\gamma_\alpha )L_{m_\beta}^-(\I\gamma_\alpha )}{\gamma_\alpha  + \gamma_\beta },
    \end{align}
where the sum over $\xi$ accounts for the fact that modes with $m \neq 0$ are degenerate with multiplicity 2. Since we have $P_\text{ref}  = 2\pi\I\sum_\beta\res_{\sigma = -\I\gamma_\beta }\hat{P}_\text{ref}\e^{-\I\sigma s}/2\pi$, this results in a sum of all duct modes, which can be rephrased in the manner of equation \eqref{2DPref}: the reflection matrix is then calculated to be
\begin{equation}
    \mat{R}_{\alpha\beta} (s) = -\frac{C_\alpha C_\beta\J_{m_\alpha}(\lambda_\alpha)\J_{m_\beta}(\lambda_\beta)\mu^+(-\I\gamma_\alpha )\mu^-(\I\gamma_\beta )L_{m_\alpha}^+(-\I\gamma_\alpha )L_{m_\beta}^-(\I\gamma_\beta )}{2(2 - \delta_{m_\alpha0})\gamma_\alpha (\gamma_\alpha  + \gamma_\beta )\exp\left((\gamma_\alpha  + \gamma_\beta )s\right)}\delta_{m_\alpha m_\beta}\delta_{\xi_\alpha\xi_\beta},
\end{equation}
so we have
\begin{equation}
    \mat{R}_{\alpha\alpha} (0) = -\frac{\I C_\alpha^2|\J_{m_\alpha}(\lambda_\alpha)|^2\left|\gamma_0  + \gamma_\alpha \right|L_{m_\alpha}^+(-\I\gamma_\alpha )L_{m_\beta}^-(\I\gamma_\beta )}{4(2 - \delta_{m_\alpha0})\left|\gamma_\alpha \right|^2}.
\end{equation}
\subsubsection{Multiplicative decomposition}
We wish to multiplicatively decompose $L_m(\sigma)$. The form of $L_m$ was chosen because of the asymptotic expansions
\begin{align}
    \Ib_m'(\mu R) \sim \frac{e^{\mu R}}{\sqrt{2\pi\mu R}}\left[1 + O\left(\frac{1}{\mu R}\right)\right] &&\text{and} &&\K_m'(\mu R) \sim -\sqrt{\frac{\pi}{2\mu R}}e^{-\mu R}\left[1 + O\left(\frac{1}{\mu R}\right)\right],
\end{align}
as $\mu \rightarrow \infty$. From this we have $L_m(\sigma) \rightarrow 1$ as $\sigma \rightarrow \pm\infty$, which is useful because it means that $\log(L_m(\sigma))$ will converge at these limits. Because $\mu(\sigma) := \sqrt{\sigma^2 - \omega^2}$ will necessarily involve branch cuts, we will need a multiplicative decomposition of $\mu(\sigma)$. We write
\begin{equation}
    \mu(\sigma) = \mu^+(\sigma)\mu^-(\sigma) := \sqrt{-\I(\sigma + \omega)}\sqrt{\I(\sigma - \omega)},
\end{equation}
so that $\mu^+(\sigma)$ has a branch cut vertically downwards from $\sigma = -\omega$, and $\mu^-(\sigma)$ has a branch cut vertically upwards from $\sigma = \omega$. Therefore $L_m(\sigma)$ is analytic, and also nonzero, in the strip between $\Imag(\sigma) = -\Imag(\omega)$ and $\Imag(\sigma) = \Imag(\omega)$. Thus if we take $\log(L_m(\sigma))$, this function is analytic in the same region. Cauchy's integral formula then says that for $Y \in \mathbb{R}^+$ and $\delta < \Imag(\omega)$ we can write
\begin{equation}
    2\pi\I L_m(\sigma) = \left(\int_{-Y - \I\delta}^{Y - \I\delta} + \int_{Y - \I\delta}^{Y + \I\delta} + \int_{Y + \I\delta}^{-Y + \I\delta} + \int_{-Y + \I\delta}^{-Y - \I\delta}\right)
    \frac{\log(L_m(\Tilde{\sigma}))}{\Tilde{\sigma} - \sigma}\intd\Tilde{\sigma}.
\end{equation}
Taking $Y \rightarrow \infty$ and assuming the end integrals converge, we then have
\begin{equation}
    2\pi\I\log(L_m(\sigma)) = \int_{-\infty - \I\delta}^{\infty - \I\delta}\frac{\log(L_m(\Tilde{\sigma}))}{\Tilde{\sigma} - \sigma}\intd\Tilde{\sigma} - \int_{-\infty + \I\delta}^{\infty + \I\delta}\frac{\log(L_m(\Tilde{\sigma}))}{\Tilde{\sigma} - \sigma}\intd\Tilde{\sigma}.
\end{equation}
The first integral is analytic for $\Imag(\sigma) > -\delta$ and the second integral is analytic for $\Imag(\sigma) < \delta$. Therefore we write
\begin{equation}
    \log(L_m(\sigma)) = \log(L_m^+(\sigma)) + \log(L_m^-(\sigma)),
\end{equation}
and have the definitions
\begin{equation}
L_m^\pm(\sigma) = \exp\left(\pm\frac{1}{2\pi\I}\int_{-\infty}^\infty\frac{\log L_m(\Tilde{\sigma})}{\Tilde{\sigma} - \sigma}\intd\Tilde{\sigma}\right).
\end{equation}
It is evident from these definitions that $L_m^-(\sigma) = L_m^+(-\sigma)$, therefore
\begin{equation}
        \arg(-\mat{R}_{\alpha\alpha} (0)) = 2\arg(L_{m_\alpha}^+(-\I\gamma_\alpha )) + \frac{\pi}{2}
        = -\frac{1}{\pi}\Real\left(\int_{-\infty}^\infty\frac{\log L_m(\Tilde{\sigma})}{\Tilde{\sigma} - \sigma}\intd\Tilde{\sigma}\right) + \frac{\pi}{2}.
\end{equation}
While we require $\omega$ to have a small positive imaginary part for the above calculation to work, we would ideally like it to be real. Thus we define a factorising contour $\mathcal{C}_\sigma$ with equation
\begin{align}&
    \sigma = \omega\left(\xi - \I h\frac{4(\xi/q)}{3 + (\xi/q)^4}\right),
    &&\xi\in\mathbb{R},
    \end{align}
which ensures we can bend around the point $\omega$ on the real axis, as well as any cut-on poles that now lie on it. In our calculations, we took $h = 0.01$ and $q = 1$, with $\omega$ having an imaginary part of $\I\times10^{-5}$ added to it. 

\section{Pressure-release walls in the outer duct}\label{sec:soundsoftbc}
\begin{figure}
    \centering
    \includegraphics{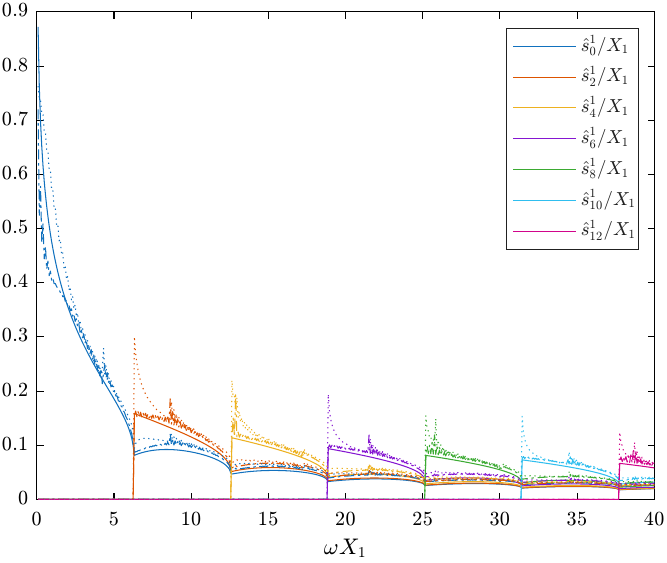}
    \caption{End-correction coefficients for a 2D duct with a symmetric source, calculated using the Wiener--Hopf technique (solid line), the sound-hard outlet condition (dash-dotted line) and the pressure-release outlet condition (dotted line). 13 modes were used in the inner duct and 2000 in the outer duct, with a width ratio of 1/40.}
    \label{comparison_equalparams.pdf}
\end{figure}

As well as considering an outer duct with zero-velocity  walls (i.e. a \textit{sound-hard} boundary condition), we can also compute the outlet condition with a \textit{pressure-release} boundary, i.e. one where the pressure is zero rather than the velocity. Considering the 2D case, this simply amounts to the use of an alternative basis in the outer duct, i.e. $\{\varphi_\alpha^2(x)\}_{\alpha=1}^\infty$ with $\varphi_\alpha^2 = 0$ on $x = \pm X/2~\forall~\alpha$. With the same governing equation (Helmholtz) and normalisation condition as the original duct modes, this results in 
\begin{align}
    \varphi_\alpha^2(x) = \frac{1}{\sqrt{2X}}\sin\left(\frac{\lambda_\alpha(x + X/2)}{X}\right)
\end{align}
where $\lambda_\alpha = \alpha\pi$ as before. Note that there is no zero mode in this case (as this would fail to satisfy normalisation) so modes begin at $\alpha = 1$ instead. Carrying this through to the definition of an alternative $\tilde{\mat{F}}_{\alpha\beta} := \int_{S_1}\psi_\alpha^1\varphi_\beta^2\intd S$, we find
\begin{equation}
    \tilde{\mat{F}}_{\alpha\beta} = \begin{cases}
        \frac{C_\alpha\beta\eta^{3/2}}{\sqrt{2}\pi(\alpha^2 - \beta^2\eta^2)}\left((-1)^\alpha\cos\left(\frac{(1+\eta)\beta\pi}{2}\right) - \cos\left(\frac{(1 - \eta)\beta\pi}{2}\right)\right), & \alpha \neq \eta\beta, \\
        \frac{C_\alpha\sqrt{\eta}}{4\sqrt{2}\alpha\pi}\left(\cos\left(\frac{(1-1/\eta)\alpha\pi}{2}\right) - (-1)^\alpha\cos\left(\frac{(1+1/\eta)\alpha\pi}{2}\right)\right), & \alpha = \eta\beta.
    \end{cases}
\end{equation}
This $\tilde{\mat{F}}$ is then used to calculate an admittance condition for the inner duct, resulting in a reflection matrix, which in turn is used to form a linearly-generalised end-correction coefficient as in section \ref{sec:endcorrlin}. Plotting the end-correction coefficients then results in figure~\ref{comparison_equalparams.pdf}. 
As can be seen, the pressure-release boundary condition is fairly consistently further away from the Wiener-Hopf solution than the sound-hard one, and also exhibits unphysically sharp jumps at every mode cut-on. From this, we conclude that the sound-hard boundary condition is a better approximation to free space (as calculated by Wiener-Hopf).

\urlstyle{rm}
\bibliography{biblio}

@string(JFM="J.~Fluid Mech")

@string(JSV="J.~Sound Vib")

@string(JASA="J.~Acoust. Soc. Am")

@string(AIAAJ="{AIAA}~J")

@string{PNAS="Proc. Natl Acad. Sci"}

@article{mctavish+brambley-2019,
    author="J. P. McTavish and E. J. Brambley",
    year="2019",
    title="Nonlinear Sound Propagation in Two-Dimensional Curved Ducts: A Multimodal Approach",
    journal=JFM,
    volume="875",
    pages="411--447",
    doi="10.1017/jfm.2019.497"
}

@phdthesis{mctavphd,
    title       = "{N}onlinear acoustics in a general waveguide",
    author      = "J. P. McTavish",
    school      = "University of Cambridge",
    year        = "2018",
    doi         = "10.17863/CAM.35714"
}

@article{felix1,
    title   =   "Sound propagation in rigid bends: A multimodal approach",
    author  =   "F\'elix, S. and Pagneux, V.",
    journal =   JASA,
    volume  =   "110",
    pages   =   "1329--1337",
    year    =   "2001",
    DOI     =   "10.1121/1.1391249",
}

@article{felix2,
    title   =   "Multimodal analysis of acoustic propagation in three-dimensional bends",
    author  =   "S. F\'elix and V. Pagneux",
    journal =   "Wave Motion",
    volume  =   "36",
    pages   =   "157--168",
    year    =   "2002",
    DOI     =   "10.1016/S0165-2125(02)00009-4",
}

@article{webster,
    title   =   "Acoustical impedance and the theory of horns and of the phonograph",
    author  =   "A. G. Webster",
    journal =   PNAS,
    volume  =   "5",
    number  =   "7",
    pages   =   "275--282",
    year    =   "1919",
    doi     =   "10.1073/pnas.5.7.275"
}

@inproceedings{kemp,
    title   =   "Pressure fields in the vicinity of brass musical instrument bells measured using a two dimensional grid array and comparison with multimodal models",
    author  =   "J. A. Kemp and A. Lopez-Carromero and M. Campbell",
    booktitle = "Proc.~ICSV24",
    year    =   "2017",
    url     =   "https://hdl.handle.net/10023/12128",
}

@article{fernando,
    title   =   "Nonlinear waves and shocks in a rigid acoustical guide",
    author  =   "R. Fernando and Y. Druon",
    journal =   JASA,
    volume  =   "129",
    pages   =   "604--615",
    year    =   "2011",
    DOI     =   "10.1121/1.3531799",
}

@article{hirschberg,
    title   =   "Shock waves in trombones",
    author  =   "A. Hirschberg and J. Gilbert and R. Msallam and A. P. J. Wijnands",
    journal =   JASA,
    volume  =   "99",
    pages   =   "1754--1758",
    year    =   "1996",
    DOI     =   "10.1121/1.414698",
}

@article{rendon2010,
    author = {Rend\'on, P. L. and Ordu\~na-Bustamante, F. and Narezo, D. and P\'erez-L\'opez, A. and Sorrentini, J.},
    title = {Nonlinear progressive waves in a slide trombone resonator},
    journal = JASA,
    volume = {127},
    number = {2},
    pages = {1096--1103},
    year = {2010},
    doi = {10.1121/1.3277221},
}

@article{gilbert2008,
  TITLE = {A simulation tool for brassiness studies},
  AUTHOR = {Gilbert, J. and Menguy, L. and Campbell, M.},
  JOURNAL = JASA,
  VOLUME = {123},
  NUMBER = {4},
  PAGES = {1854--1857},
  YEAR = {2008},
  doi={10.1121/1.2872342}
}

@article{arnold2012,
    author = {Myers, A. and Pyle, R. W., Jr. and Gilbert, J. and Campbell, D. M. and Chick, J. P. and Logie, S.},
    title = {Effects of nonlinear sound propagation on the characteristic timbres of brass instruments},
    journal = JASA,
    volume = {131},
    number = {1},
    pages = {678--688},
    year = {2012},
    doi = {10.1121/1.3651093},
}

@article{pandya2003,
  title={Schlieren imaging of shock waves from a trumpet},
  author={Pandya, B. H. and Settles, G. S. and Miller, J. D.},
  journal=JASA,
  volume={114},
  number={6},
  pages={3363--3367},
  year={2003},
  DOI={10.1121/1.1628682}
}

@article{jensenbrambley2026,
      title={Multimodal nonlinear acoustics in two- and three-dimensional curved ducts}, 
      author={F. Jensen and E. J. Brambley},
      journal=JFM,
      year={2026},
      volume={1030},
      pages={A19},
      doi={10.1017/jfm.2026.11220},
}

@book{noble,
    author="B. Noble",
    year="1958",
    title="Methods based on the {W}iener--{H}opf Technique for the Solution of Partial Differential Equations",
    publisher="Pergamon",
    isbn="0828403325",
}

@article{snakowska2011,
  title={Open end correction for arbitrary mode propagating in a cylindrical acoustic waveguide},
  author={Snakowska, A. and Jurkiewicz, J. and Smolik, D.},
  journal={{A}cta {P}hysica {P}olonica {A}},
  volume={120},
  number={4},
  pages={736--739},
  year={2011},
  publisher={{P}olska {A}kademia {N}auk. {I}nstytut {F}izyki {PAN}},
  doi={10.12693/APhysPolA.120.736}
}

@article{levine1948radiation,
  title={On the radiation of sound from an unflanged circular pipe},
  author={Levine, H. and Schwinger, J.},
  journal={Phys.~Rev.},
  volume={73},
  number={4},
  pages={383},
  year={1948},
  publisher={APS},
  doi={10.1103/PhysRev.73.383}
}

@article{rayleigh1871,
  title={V.~{O}n the theory of resonance},
  author={Rayleigh, {Lord}},
  journal={Phil. Trans. R.~Soc. Lond.},
  volume={161},
  pages={77--118},
  year={1871},
  publisher={The Royal Society London},
  doi={10.1098/rstl.1871.0006}
}

@article{weinstein1948rigorous,
  title={Rigorous solution of the problem of an open-ended parallel-plate waveguide},
  author={Weinstein, L. A.},
  journal={Izv. Akad. Nauk, Ser. Fiz},
  volume={12},
  pages={144--165},
  year={1948}
}

@article{rienstra1984acoustic,
  title={Acoustic radiation from a semi-infinite annular duct in a uniform subsonic mean flow},
  author={Rienstra, S. W.},
  journal=JSV,
  volume={94},
  number={2},
  pages={267--288},
  year={1984},
  publisher={Elsevier},
  doi={10.1016/S0022-460X(84)80036-X}
}

@article{felix2018modeling,
    author = {F\'elix, S. and Doc, J.-B. and Boucher, M. A.},
    title = {Modeling of the multimodal radiation from an open-ended waveguide},
    journal = JASA,
    volume = {143},
    number = {6},
    pages = {3520--3528},
    year = {2018},
    doi = {10.1121/1.5041268},
}

@phdthesis{mangin2023modelling,
    author= {B. Mangin},
    title= {Modelling acoustic propagation in modern turbofan intakes using a multimodal method},
    URL = {https://theses.hal.science/tel-04355151},
  NUMBER = {2023LEMA1020},
  SCHOOL = {Le {M}ans Universit{\'e}},
  YEAR = {2023},
  eprint = {tel-04355151},
  arxive = {HAL},
}

@article{cummings1983,
    author="A. Cummings and W. Eversman",
    title="High Amplitude Acoustic Transmission through Duct Terminations: Theory",
    journal=JSV,
    volume="91",
    pages="503--518",
    year="1983",
    doi="10.1016/0022-460X(83)90829-5",
    }

@article{zinn1970,
    author="B. T. Zinn",
    title="A Theoretical Study of Non-linear Damping by {H}elmholtz Resonators",
    journal=JSV,
    volume="13",
    pages="347--356",
    year="1970",
    doi="10.1016/S0022-460X(70)80023-2",
}

@article{disselhorst1980,
    author="Disselhorst, J. H. M. and van Wijngaarden, L.",
    title="Flow in the Exit of Open Pipes During Acoustic Resonance",
    journal=JFM,
    volume="99",
    pages="293--319",
    year="1980",
    doi="10.1017/S0022112080000626",
    }

@article{auregan2008,
    author="Y. Aur\'egan and M. Leroux",
    year="2008",
    title="Experimental Evidence of an Instability over an Impedance Wall in a Duct with Flow",
    journal=JSV,
    volume="317",
    pages="432--439",
    doi="10.1016/j.jsv.2008.04.020"
}

@article{rienstra2018,
    author = {Rienstra, S. W. and Singh, D. K.},
    title = {Nonlinear Asymptotic Impedance Model for a {H}elmholtz Resonator of Finite Depth},
    journal = AIAAJ,
    volume = {56},
    number = {5},
    pages = {1792--1802},
    year = {2018},
    doi = {10.2514/1.J055882},
}
\end{document}